\documentclass[twocolumn]{aastex631}

\shorttitle{CMZoom IV}
\shortauthors{Hatchfield et al.}

\graphicspath{{./}{figures/}}

\begin{document}

\title{CMZoom IV. Incipient High-Mass Star Formation Throughout the Central Molecular Zone\footnote{\copyright 2023, all rights reserved, valid until copyright transferred to publisher}}

\author[0000-0003-0946-4365]{H Perry Hatchfield}
\affiliation{Jet Propulsion Laboratory, California Institute of Technology, 4800 Oak Grove Drive, Pasadena, CA, 91109, USA}
\affiliation{University of Connecticut, Department of Physics, 196A Auditorium Road, Unit 3046, Storrs, CT 06269 USA}

\author[0000-0002-6073-9320]{Cara Battersby}
\affiliation{University of Connecticut, Department of Physics, 196A Auditorium Road, Unit 3046, Storrs, CT 06269 USA}
\affiliation{Center for Astrophysics $|$ Harvard \& Smithsonian, MS-78, 60 Garden St., Cambridge, MA 02138 USA}

\author[0000-0003-0410-4504]{Ashley~T.~Barnes}
\affiliation{Argelander-Institut f\"{u}r Astronomie, Universit\"{a}t Bonn, Auf dem H\"{u}gel 71, 53121, Bonn, DE}

\author[0000-0002-4013-6469]{Natalie Butterfield}
\affiliation{National Radio Astronomy Observatory, 520 Edgemont Road, Charlottesville, VA 22903, USA}

\author[0000-0001-6431-9633]{Adam Ginsburg}
\affiliation{University of Florida Department of Astronomy, Bryant Space Science Center, Gainesville, FL, 32611, USA}

\author[0000-0001-9656-7682]{Jonathan D. Henshaw}
\affiliation{Astrophysics Research Institute, Liverpool John Moores University, 146 Brownlow Hill, Liverpool L3 5RF, UK}
\affiliation{Max-Planck-Institute for Astronomy, Koenigstuhl 17, 69117 Heidelberg, Germany}

\author[0000-0001-6353-0170]{Steven N. Longmore}
\affiliation{Astrophysics Research Institute, Liverpool John Moores University, 146 Brownlow Hill, Liverpool L3 5RF, UK}
\affiliation{Cosmic Origins Of Life (COOL) Research DAO, coolresearch.io}

\author[0000-0003-2619-9305]{Xing Lu}
\affiliation{Shanghai Astronomical Observatory, Chinese Academy of Sciences, 80 Nandan Road, Shanghai 200030, People’s Republic of China}

\author[0000-0002-8502-6431]{Brian Svoboda}
\affiliation{National Radio Astronomy Observatory, PO Box O, Socorro, NM 87801 USA}

\author[0000-0001-7330-8856]{Daniel Walker}
\affiliation{UK ALMA Regional Centre Node, Jodrell Bank Centre for Astrophysics, The University of Manchester, Manchester M13 9PL, UK}
\affiliation{University of Connecticut, Department of Physics, 196A Auditorium Road, Unit 3046, Storrs, CT 06269 USA}

\author{Daniel Callanan}
\affiliation{Astrophysics Research Institute, Liverpool John Moores University, 146 Brownlow Hill, Liverpool L3 5RF, UK}
\affiliation{Center for Astrophysics $|$ Harvard \& Smithsonian, MS-78, 60 Garden St., Cambridge, MA 02138 USA}

\author[0000-0001-8782-1992]{Elisabeth A. C. Mills}
\affiliation{Department of Physics and Astronomy, University of Kansas, 1251 Wescoe Hall Drive, Lawrence, KS 66045, USA}

\author[0000-0001-6947-5846]{Luis C. Ho}
\affiliation{Kavli Institute for Astronomy and Astrophysics, Peking University, Beijing 100871, China}
\affiliation{Department of Astronomy, School of Physics, Peking University, Beijing 100871, China}

\author[0000-0002-5094-6393]{Jens Kauffmann}
\affiliation{Haystack Observatory, Massachusetts Institute of Technology, 99 Millstone Road, Westford, MA 01886, USA}

\author[0000-0002-8804-0212]{J. M. Diederik Kruijssen}
\affiliation{Cosmic Origins Of Life (COOL) Research DAO, coolresearch.io}

\author{J{\"u}rgen Ott}
\affiliation{National Radio Astronomy Observatory, 1003 Lopezville Rd., Socorro, NM 87801, USA}

\author[0000-0003-2133-4862]{Thushara Pillai}
\affiliation{Boston University Astronomy Department, 725 Commonwealth Avenue, Boston, MA 02215, USA}

\author[0000-0003-2384-6589]{Qizhou Zhang}
\affiliation{Center for Astrophysics $|$ Harvard \& Smithsonian, MS-42, 60 Garden St., Cambridge, MA 02138 USA}

\begin{abstract}
In this work, we constrain the star-forming properties of all possible sites of incipient high-mass star formation in the Milky Way's Galactic Center. We identify dense structures using the \textit{CMZoom} 1.3mm dust continuum catalog of objects with typical radii of $\sim$0.1pc, and measure their association with tracers of high-mass star formation. We incorporate compact emission at 8, 21, 24, 25, and 70$\mu$m from MSX, Spitzer, Herschel, and SOFIA, catalogued young stellar objects, and water and methanol masers to characterize each source. We find an incipient star formation rate (SFR) for the CMZ of $\sim$0.08 M$_\odot$yr$^{-1}$ over the next few 10$^5$yr. We calculate upper and lower limits on the CMZ's incipient SFR of $\sim$0.45 M$_\odot$yr$^{-1}$ and $\sim$0.05 M$_\odot$yr$^{-1}$ respectively, spanning between roughly equal to and several times greater than other estimates of CMZ's recent SFR. Despite substantial uncertainties, our results suggest the incipient SFR in the CMZ may be higher than previously estimated. We find that the prevalence of star formation tracers does not correlate with source volume density, but instead $\gtrsim$75\% of high-mass star formation is found in regions above a column density ratio (N$_{\rm SMA}$/N$_{\rm Herschel}$) of $\sim$1.5. Finally, we highlight the detection of ``atoll sources'', a reoccurring morphology of cold dust encircling evolved infrared sources, possibly representing HII regions in the process of destroying their envelopes.

\end{abstract}



\section{Introduction} \label{sec:intro}
Understanding how the process of star formation varies with environment demands the detailed study of molecular cloud evolution under diverse physical conditions. Star-forming regions in the solar neighborhood provide an extremely detailed look at the formation of stars in one relatively narrow window of physical parameter space. The interstellar medium occupying the radially innermost few hundred parsecs of the Milky Way provides an excellent opportunity to observe the formation of stars within molecular clouds subject to a more extreme environment than that of the Galaxy's disk while still being nearby enough for modern observatories to resolve the fine details of the star formation process. The substantial reservoir of gas surrounding the Galactic Center at a radius of about 100pc known as the Central Molecular Zone (CMZ) hosts more than 3$\times$10$^7$ M$_\odot$ of molecular hydrogen \citep[e.g. ][]{dahmen_molecular_1998, ferriere_spatial_2007}, largely contained in giant molecular clouds with volume densities commonly exceeding 10$^{4}$ cm$^{-3}$ \citep[e.g.][]{mills_dense_2018, guesten_new_1982}, high kinetic temperatures \citep[50-100, e.g.][]{ginsburg_dense_2016, krieger_survey_2017}, broad turbulent velocity dispersions ($\sim$10 km s$^{-1}$), and complex three-dimensional morphologies (e.g.\ \citealt{shetty_linewidthsize_2012, henshaw_investigating_2016, kauffmann_galactic_2017a, henshaw_brick_2019}). These conditions bear some resemblance to the properties of z$\sim$1-3 galaxies near the apex of cosmic star formation, and might be used to further our understanding of how star formation occurs in the distant reaches of the universe \citep[e.g.][]{kruijssen_comparing_2013}. 

Over the last decade, studies measuring the star formation rate (SFR) within the Galactic Center have revealed a global dearth of recent star formation relative to the substantial dense gas content \citep[e.g.][]{immer_recent_2012, longmore_variations_2013, barnes_star_2017, lu_census_2019}. There are, however, plentiful signposts of vigorous past and ongoing high and low-mass star formation throughout the CMZ \citep[e.g.][]{ginsburg_distributed_2018, walker_Star_2018, walker_star_2021, lu_ALMA_2020,lu_ALMA_2021}. The composition of the Nuclear Stellar Disk implies periods of much higher SFR within the past $\sim$1 Gyr \citep[][]{nogueras-lara_early_2020}, and the presence of the Arches and Quintuplet star clusters, thought to have been formed within the last $\sim$6 Myr \citep[e.g.][]{figer_massive_2002,hosek_unusual_2019}, suggests that the CMZ has a prolific and irregular star formation history. Some theoretical studies bolster these claims, showing that computational models of the Milky Way's CMZ display highly episodic star formation histories on similar timescales \citep[e.g. ][]{kruijssen_what_2014, krumholz_dynamical_2015, krumholz_dynamical_2017, torrey_instability_2017, armillotta_life_2019, orr_fiery_2021}. 

There are several young massive clusters (with total masses $\gtrsim 10^4$M$_\odot$) actively forming throughout the CMZ, most notably in Sgr B2, Dust Ridge Clouds C and E, and Sgr C \citep[e.g.][]{schmiedeke_physical_2016, ginsburg_distributed_2018, walker_comparing_2016, walker_Star_2018, lu_census_2019, lu_star_2019, barnes_young_2019,lu_ALMA_2021}. However, these clouds are only a subset of a population of massive molecular clouds with extremely high densities, many of which appear to have few if any signatures of high-mass star formation \citep[e.g.][]{longmore_variations_2013, lu_census_2019, walker_star_2021, williams_initial_2022}. In order to understand the global deficiency of active star formation in the CMZ relative to its total dense gas mass, we must consider the complete sample of giant molecular clouds throughout the CMZ. 

\textit{CMZoom} \citep[][]{battersby_cmzoom_2020} is the first survey at 1.3mm to map all high-density (N(H$_2) > 10^{23}$ cm$^{-2}$) gas in the CMZ with sufficient resolution and sensitivity to uncover their dense gas substructure on sub-parsec scales. The survey's catalog of compact sources, constructed from the Submillimeter Array (SMA)'s 1.3mm dust continuum emission, characterized all possible sites of ongoing and incipient high-mass star formation, placing an upper limit on present-day star formation potential of the Galactic Center \citep[][]{hatchfield_cmzoom_2020}. However, from the \textit{CMZoom}'s 1.3mm dust continuum alone, it is impossible to determine which of these compact submillimeter sources are actively forming high-mass protostars and which remain quiescent. 

In this paper, we perform a detailed analysis of each \textit{CMZoom} source, incorporating previous multi-wavelength observations and catalogs of star formation tracers from the literature to characterize the properties of each possible site of deeply embedded, incipient high-mass star formation in the Galactic Center (excluding Sgr A* and Sgr B2). In Section \ref{sec:data_catalog}, we briefly summarize the details of the \textit{CMZoom} survey and describe the design of the \textit{CMZoom} catalog. In Section \ref{sec:method}, we describe our method for comparing \textit{CMZoom} objects with compact far-infrared sources and previous catalogs of young stellar object (YSO) candidates and masers. In Section \ref{sec:results}, we present our results for which \textit{CMZoom} clouds host ongoing massive star formation, and how the physical properties of sources correlate with the indicators of active star formation. In Section \ref{sec:discussion}, we discuss the implications of these results, focusing on the scaling relationships and star formation rates presented in the previous section. Finally, in Section \ref{sec:summary}, we summarize the results of this work and list our key conclusions.

\section{Summary of Data and Catalog Design} \label{sec:data_catalog}
The \textit{CMZoom} survey is an SMA large program, surveying all dense gas (N(H$_2)>$10$^{23}$ cm$^{-2}$) in the CMZ. Using the compact and subcompact configurations of the SMA, the 1.3mm continuum emission achieved a typical resolution of $\sim$3'' ($\sim$0.1pc), revealing a wealth of differing morphologies and complexities of dust structure in the 35 regions and $\sim$240 square arcmin ($\sim$1360 pc$^2$) area surveyed. In addition to the continuum, \textit{CMZoom} surveyed emission from notable transitions of key molecular species, such as CO(2-1) and its isotopologue companions, the para-H$_2$CO triplet transitions ($2_{0,3}-1_{0,2}$, $2_{2,2}-1_{2,1}$, $2_{2,1}-1_{2,0}$), SiO(5-4) and others, presented in \citet{callanan_cmzoom_2023}. While the vast majority of regions observed by \textit{CMZoom} have been confirmed to be near the Galactic Center using a variety of methods including their spectral properties, a handful of regions (4 of 36) cannot be definitively localized to the CMZ and may represent foreground emission. These sources (G0.393, G0.212, G1.670, and G359.137) have been flagged as possible foreground regions, though they have been included in the following analysis for completeness.

\begin{figure*}
\begin{center}
\includegraphics[trim = 0mm 0mm 0mm 0mm, width = .99 \textwidth]{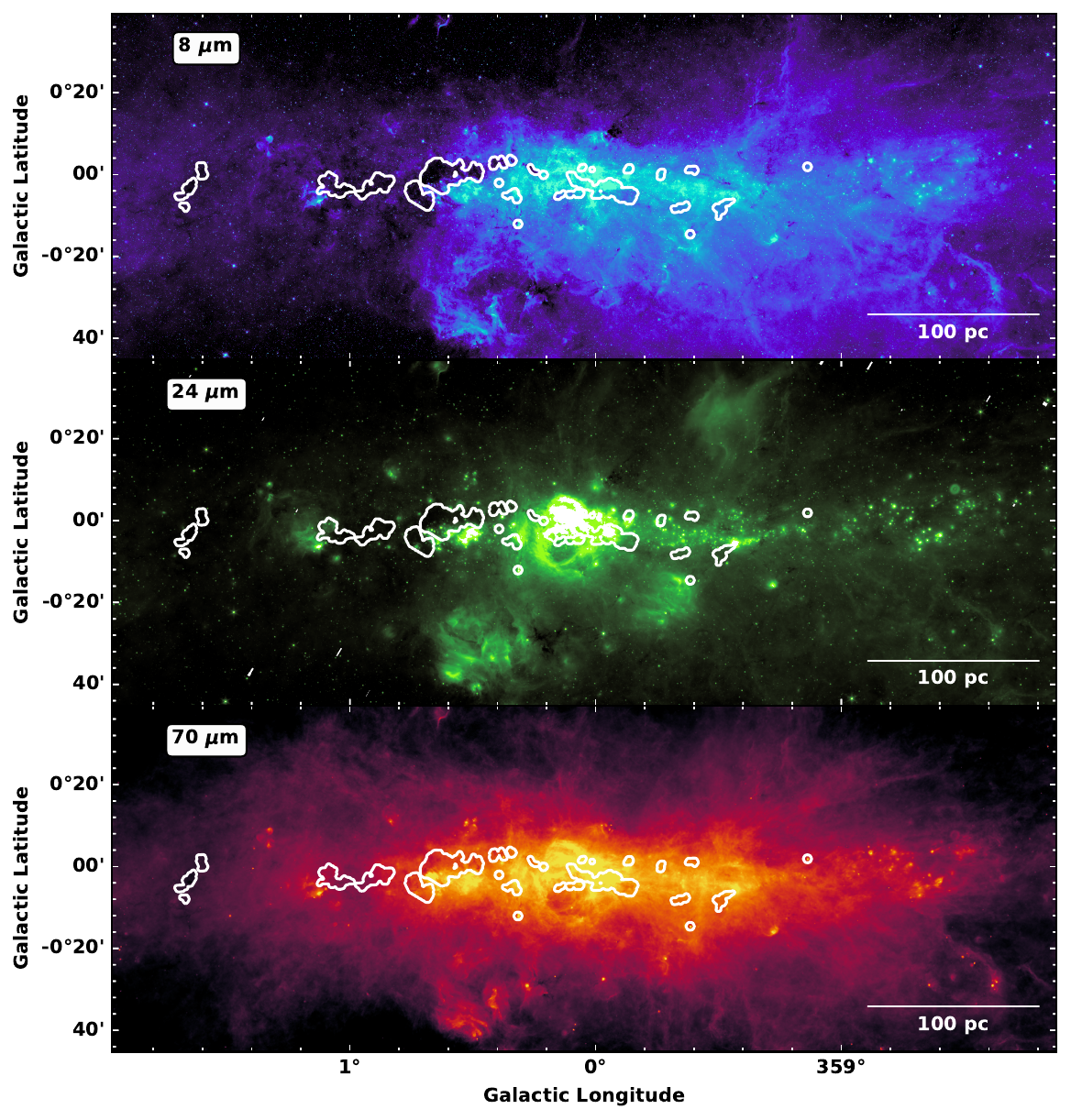}
\end{center}
\caption{Mosaics of the innermost four degrees of the CMZ constructed from 8$\mu$m (B), 24$\mu$m (G) and 70$\mu$m (R) continuum from GLIMPSE \citep[top,][]{benjamin_glimpse_2003}, MIPSGAL \citep[middle,][]{carey_mipsgal_2009}, and Herschel's Hi-GAL survey \citep[bottom,][]{molinari_hi-gal_2010}, respectively. Contours in white are overlaid showing the \textit{CMZoom} survey footprint in each panel. }
\label{fig:rgb_mosaic}
\end{figure*}

The dust continuum catalog, detailed in \cite{hatchfield_cmzoom_2020}, was designed to characterize the properties of the compact dust emission structures throughout the CMZ's dense gas content as completely as possible using a pruned dendrogram. We produce a hierarchical decomposition of the flux represented by a dendrogram (implemented using {\sc astrodendro}\footnote{\href{http://www.dendrograms.org/}{http://www.dendrograms.org/}}), a tree-like plot where the highest level, brightest structures are identified  as ``leaves''. The significance of each of these leaves is determined by a global estimate of the noise, $\sigma_{\rm global}$, with a minimum structure boundary value of 3$\sigma$ and a minimum leaf boundary significance of 1$\sigma_{\rm global}$. The dendrogram algorithm also considers a minimum pixel size for leaves, which we have chosen to be half the typical beam size. There is no enforced maximum leaf size, and the shapes and sizes of resulting leaves are entirely determined by the significance and extent of local maxima in the flux map and the dendrogram parameters described above. The dynamic range and variable noise across the surveyed regions necessitated the ``pruning'' of this initial dendrogram, in which leaves were removed from the final catalog if they did not meet a local noise threshold in the mean flux and peak flux. Local root-means-squared (RMS) noise estimates were constructed from the SMA residual maps. The only leaves that remain in the catalog are those which have a peak flux of 6$\sigma$ above the local RMS noise, and a mean flux 2$\sigma$ above the local RMS. The resulting catalog is uniquely reproducible from the input dust continuum mosaic and residuals. The interpretation of these leaves is complicated by their physical scale and the SMA's physical resolution at the Galactic Center, as the emission from these leaves is likely due to a combination of unresolved protostellar sources and resolved dusty envelopes. ALMA observations of a subset of the \textit{CMZoom} leaves have revealed a population of multiple protostellar sources that are unresolved by the SMA, so in this work we interpret the star-forming sources in the catalog as protoclusters containing one or more high-mass protostars \citep[e.g. ][]{ginsburg_distributed_2018, barnes_young_2019, walker_star_2021}.

The flux and mass completeness of this catalog were determined using synthetic interferometric observations of clouds of point-source-like dust sources, and replicating the imaging pipeline used to process the observed dust continuum maps of the \textit{CMZoom} data products. \citet{hatchfield_cmzoom_2020} uses these synthetic observations to estimate that more than $95\%$ of dust continuum sources with masses above 80 M$_\odot$ have been recovered by this cataloging procedure and are characterized in the high robustness version of the \textit{CMZoom} catalog (assuming an average dust temperature of at least 20K across the entire dendrogram leaf). Using this mass completeness, given an initial mass function (IMF, e.g.\ \citealt{kroupa_variation_2001}) and an assumed star formation efficiency (SFE) for star-forming structures on scales resolved by the SMA, this mass completeness limit suggests that virtually all (more than 95$\%$ for ) of possible sites of deeply embedded high-mass star formation within the CMZ are included within the \textit{CMZoom} catalog. While the \textit{CMZoom} survey does not observe the entire contiguous spatial range of the CMZ, it does sample all regions above a column density of 10$^{23}$ N(H$_2$) cm$^2$, and therefore is almost certain to capture the deeply embedded high-mass star formation present in this extreme part of the Galaxy \citep[][]{battersby_cmzoom_2020}. It must also be noted that the catalog does not include star-forming sources that are sufficiently evolved to destroy their envelopes, though such sources account for a large portion of the star formation tracers throughout the CMZ considered in the other work cited throughout. The present analysis aims to characterize an earlier stage of star formation, during which the young protosclusters are still embedded in their cold envelopes.

\section{Method} \label{sec:method}
\subsection{Selection and Identification of Star Formation Tracers}\label{sec:method:sf_tracers}
By comparing positions and properties of \textit{CMZoom} catalog leaves with the positions of a variety of previously catalogued tracers of active or recent star formation, we determine the incipient star formation rate for each source in the \textit{CMZoom} Survey. The presence of certain maser species are known to trace recent star formation within molecular clouds. In particular, emission at 6.7 GHz from methanol (CH$_3$OH) masers indicates ongoing high-mass (M $>$ 8M$_\odot$) star formation \citep[e.g.][]{minier_protostellar_2003, xu_high-sensitivity_2008}. Water (H$_2$O) masers are commonly associated with both high and low-mass star-forming regions, but may also indicate the presence of evolved sources on the asymptotic branch of AGB stars, Mira variables, and possibly other astrophysical phenomena \citep[e.g.][]{hinkle_infrared_1979,forster_oh_1999,miranda_water-maser_2001}, but such sources will be much dimmer and can be excluded based on their emission in other wavebands \citep{longmore_candidate_2013}. We consider the coincidence of methanol or H$_2$O masers with \textit{CMZoom} catalog objects as an indication that those leaves are actively forming high-mass stars. We incorporate the 6.67 GHz emission from CH$_3$OH masers catalogued using the Parkes telescope by \citealt{caswell_6-ghz_2010} and the J=6$_{1,6}$-5$_{2,3}$ transition of H$_2$O masers catalogued using the Australia Telescope Compact Array (ATCA) by \citealt{walsh_accurate_2014}. If a \textit{CMZoom} catalog leaf's contour overlaps with one of the masers from either of these catalogs within the reported astrometric uncertainty (see Table \ref{tab:tracer_summary}), it is considered associated and the leaf is considered to have a robust signature of active star formation. While deeper and higher resolution maser observations are available for a subset of the \textit{CMZoom} area \citep[e.g.][]{lu_deeply_2015,lu_census_2019}, we do not include the analysis of these data here in order to preserve a consistent approach across all regions in the survey.

We must consider that these masers may not be present or detectable at all sites and evolutionary stages of incipient star formation, so we also consider the presence of mid and far-infrared continuum emission from the dusty envelopes of young protoclusters. The presence of point-like far-infrared emission between 20 and 70$\mu$m associated with a compact submillimeter dust source is a signature of early high-mass star formation \citep[e.g.][]{yusef-zadeh_massive_2008}, since the spectral energy distributions (SEDs) of dust surrounding Class 0 and I protostars peak in the far-infrared. While some populations of evolved stars may display similar far-infrared profiles, potentially masquerading as high-mass YSOs, these compact far-infrared sources that are associated with cold, dense submillimeter gas structures are much less likely to be interlopers \citep[e.g.][]{schultheis_near-ir_2003,an_massive_2011,koepferl_main-sequence_2015}.

Past surveys have catalogued point-like infrared sources according to a variety of other criteria, identifying a large number of YSO candidates throughout the Galactic Center. Here, we cross-correlate the locations of 24\micron and 70\micron point-like sources identified by \citet{gutermuth_24_2015} and \citet{molinari_hi-gal_2016}, respectively. The 24\micron point source catalog presented by \citet{gutermuth_24_2015} using data from the MIPSGAL Survey \citep[][]{carey_mipsgal_2009} is constructed using PhotVis, a modification of the DAOFIND algorithm \citep[][]{gutermuth_spitzer_2008, stetson_daophot_1987}. \textit{CMZoom} sources with a leaf contour overlapping within one MIPSGAL beam (as used in generating the catalog, 6.25'' FWHM) are considered associated with the YSO candidate and thus are considered robustly star-forming. The 70\micron point source catalog generated from the Hi-Gal survey \citep[][]{molinari_hi-gal_2010, molinari_hi-gal_2016} is generated using the CuTex algorithm, which fits 2D Gaussians to perform photometry by considering the curvature of the flux map \citep[][]{molinari_source_2011}. Any \textit{CMZoom} sources whose leaf contour overlaps within one Hi-Gal beam FWHM ($\sim$5'') of these sources are considered as associated.

We perform the same general procedure for several previous catalogs of YSO candidates throughout the survey region. \citet{an_massive_2011} identify YSO candidates from the infrared point source catalog presented in \citet{ramirez_point_2008} using spectra from 5-35$\mu$m from the Infrared Spectrograph (IRS) of the Spitzer space telescope. \citet{yusef-zadeh_star_2009} use 24 and 70$\mu$m observations with Spitzer/MIPS, supplemented by other far-IR, submillimeter and radio observations to identify a large sample of YSOs in the Galactic Center, though this sample is likely contaminated with a significant fraction of evolved stars interacting with surrounding dust \citep[][]{koepferl_main-sequence_2015}. Lastly, we consider the YSO catalog produced by \citet{immer_recent_2012} by analyzing SEDS observed with IRS toward point sources identified using ISOGAL \citep{schuller_explanatory_2003, omont_isogal_2003}. For each of these previous YSO catalogs, we determine their association with the \textit{CMZoom} sources using the beam FWHM from the corresponding instruments as the corresponding astrometric uncertainty. Leaves with contours overlapping within the positional uncertainty of these YSOs are flagged as ``robustly star-forming''. 

\begin{figure*}[htbp]
\begin{center}
\includegraphics[trim = 0mm 0mm 0mm 0mm,clip, width = .8 \textwidth]{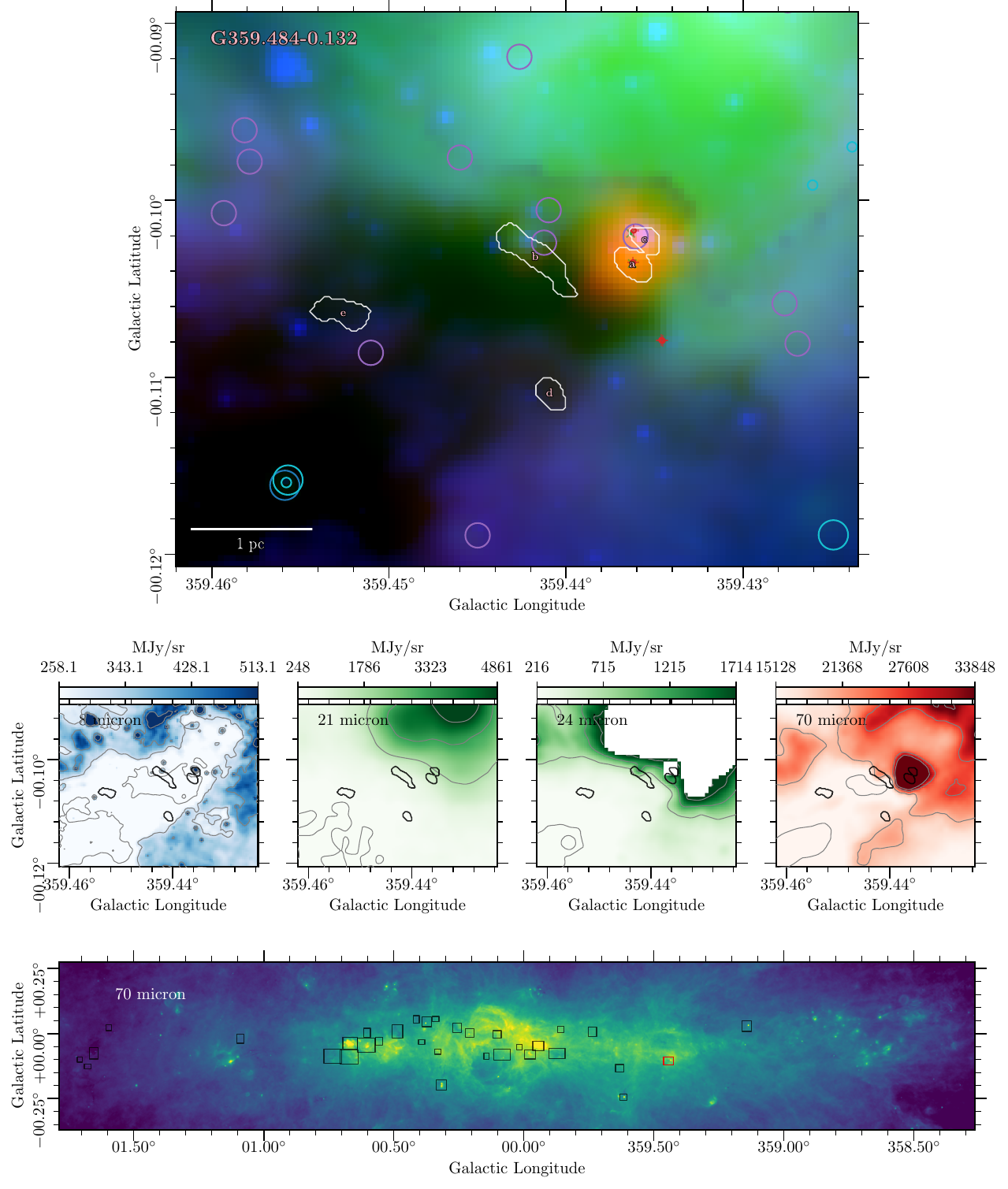}
\end{center}
\caption{An example (Sgr C) of the three-color images used in the by-eye search for compact emission signifying the early onset of star formation associated with \textit{CMZoom} catalog objects. The three colors used in the top panel are 8$\mu$m (blue, GLIMPSE, from \citet{benjamin_glimpse_2003}, 24$\mu$m (green, MIPSGAL, from \citealt{carey_mipsgal_2009}), and 70$\mu$m (red, Hi-Gal, from \citet{molinari_hi-gal_2010}). These are each shown individually in the bottom four panels, along with the 21$\mu$m emission from MSX \citep[][]{egan_midcourse_2003}. Overlaid on the composite three-color image are white contours outlining the \textit{CMZoom} leaves, along with cyan circles demarcating YSO candidates from a compilation of those identified by \citet{yusef-zadeh_star_2009}, \citet{an_massive_2011}, and \citet{immer_recent_2012}, purple circles indicating the 70$\mu$m point sources catalogued by \citet{molinari_hi-gal_2010}, and darker blue circles representing the point sources identified by \citet{gutermuth_24_2015}. The radial size of these circles corresponds to the FWHM condition used do determine plausible association with \textit{CMZoom} leaves. The bottom panel shows a 70$\mu$m mosaic of the Galactic Center with each \textit{CMZoom} region shown as black box, with the specific region from the above panels highlighted in red. In this case, leaves a, b, and c are labeled have robust signatures of high-mass star formation, and have been flagged as ``robustly star-forming'' in the catalog, while leaves d and e are labeled ``robustly quiescent''. The full sample of clouds is presented in Appendix \ref{app:rgb_gallery}.}
\label{fig:rgb_example}
\end{figure*}

\subsection{Cataloging of Compact Far-Infrared Sources By-Eye}
\label{sec:method:by-eye}
To supplement the previously noted sources in the far IR, we performed a by-eye search for point-like emission in the 8$\mu$m, 21$\mu$m, 24$\mu$m and 70$\mu$m dust continuum associated with each CMZoom robust catalog source. This combination of determining association with star formation tracers algorithmically as well as by-eye is helpful because the far-IR dust continuum landscape of the Galactic Center is very complex, with highly variable background levels and noise properties, significant potential for confusion with foreground sources, and bright extended emission that is not necessarily associated with sites of early star formation. Performing a by-eye survey allows us to handle ambiguous cases that may have been missed or mischaracterized in previous studies on an individual basis while using independent selection criteria. 

\begin{figure}
\begin{center}
\includegraphics[trim = 0mm 0mm 0mm 0mm, width = .45 \textwidth]{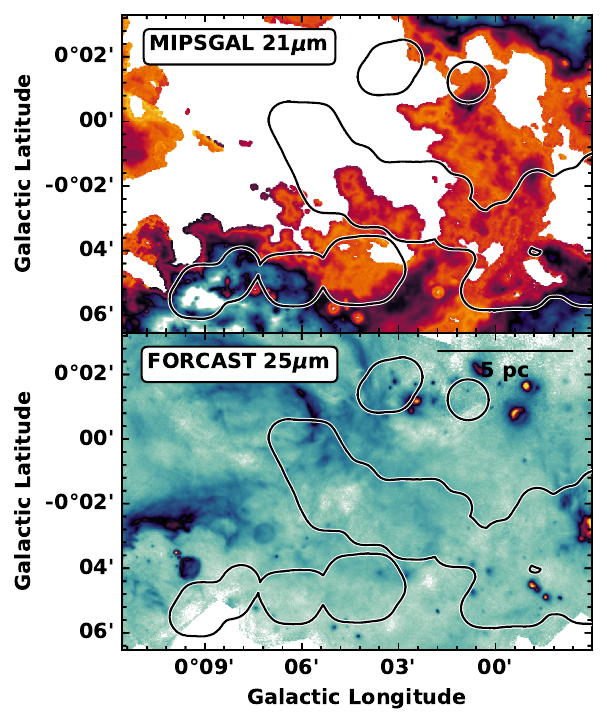}
\end{center}
\caption{A comparison of the MIPSGAL (top panel) and SOFIA FORCAST (bottom panel) observations toward a subset of \textit{CMZoom} fields near the Galactic Nucleus, including ``The Three Little Pigs'' and ``The 50 km s$^{-1}$ Cloud''.  The saturated regions in MIPSGAL are resolved with FORCAST, though the FORCAST Galactic Center Survey does not cover all regions observed by \textit{CMZoom}, and so could not be applied to the entire catalog sample. }
\label{fig:mips_comp}
\end{figure}

The goal of this search was to robustly identify the presence or absence of any signature of active high-mass star formation. To further mitigate the inherent biases of human error in performing this by-eye cataloging, eight team members (initials HH, CB, AB, NB, SL, XL, BS, DW)  performed the task independently. Each team member, using locally-scaled RGB maps (R: 70$\mu$m, G: 21$\mu$m and 24$\mu$m and B: 8$\mu$m, e.g.\ Fig \ref{fig:rgb_example}), designated either a ``robust detection'', ``robust non-detection'', or ``ambiguous detection'' for each star formation tracer, for each of the 285 high-robustness catalog sources. A final determination of ``robustly star-forming'', ``robustly-non-starforming'' (or, equivalently, ``robustly quiescent'', and ``ambiguous star-forming state'' was then decided considering the compilation of the independent results. Leaves with significant consensus (at least 6 of the 7 team members in agreement) for a certain designation for 70$\mu$m, 21$\mu$m or 24$\mu$m point-like sources were considered to be robustly star-forming or robustly quiescent, corresponding to the consensus. Leaves for which significant disagreement was recorded ($\geq 2$ disagreeing designations) were re-inspected to determine if there was a discernible cause for the discrepancy, or if the leaf should be designated as ``ambiguous''. The most prevalent cases for ambiguity are marginally compact sources embedded in or near bright, structured extended emission. While it is possible that these ambiguous sources are actively star-forming, their properties cannot be easily distinguished from their envelope emission, and higher resolution and sensitivity measurements or more sophisticated fitting techniques \citep[e.g.][]{immer_recent_2012} would be necessary to arrive at a more definitive designation. Therefore, such leaves are designated as ``ambiguously star-forming'' in the following analysis.

\subsection{Sources of Uncertainty in the By-Eye Analysis}\label{sec:by-eye_unc}

In identifying the association of each high-mass star formation tracer with the catalog leaves, there are several opportunities for ambiguity and uncertainty that must be addressed. While the majority of sources designated as ``robustly star-forming'' or ``robustly quiescent'' had near-unanimous agreement about the presence or absence of each star formation tracer, ambiguous edge-cases were re-inspected by eye by only one team member, compiling the previous results and reaching a final designation based on the relevant context. It is clear that this final designation for the cases handled in this way might be biased, but such edge-cases largely fell into two categories. Firstly, for regions with bright but diffuse emission suffusing the leaf contours and their immediate surroundings with non-point-source-like substructure, some team members marked leaves as having compact far-IR emission. Secondly, some leaf contours overlap with, but are not centered on a bright, far-IR compact source, and some team members marked such leaves as non-star-forming. The final designation for leaves in both cases was ``ambiguously star-forming'' unless other information was available to better constrain their properties. 

The by-eye component of the star formation tracer cataloging does not significantly alter the statistics of the final populations. Only 11 \textit{CMZoom} leaves achieved a ``robustly star-forming'' designation without association with a previously-catalogued star formation tracer, of which 10 are in the immediate proximity, though not overlapping with, a previously-catalogued tracer. The remaining source, G0.619+0.012i, hosts a clear 8$\mu$m point source and a much more subtle 24-25$\mu$m point source, visible in both MIPSGAL and SOFIA FORCAST continuum observations. Because of the low signal-to-noise value for this far-IR source, it is possible previous automatic cataloging algorithms have missed it, and the presence of a cold 1.3mm continuum envelope better highlights its likely-protostellar nature. While many of the more ``ambiguously star-forming'' sources may correspond to previously not-catalogued YSO candidates, characterizing their nature is left for future efforts using a more complete set of observations.

\subsection{Accounting for Saturated Pixels with SOFIA FORCAST Data}

Several regions in the \textit{CMZoom} field are saturated in the 24\micron emission map from Spitzer/MIPS (see section 5.2 in \citet{carey_mipsgal_2009}). Because these saturated sources in the MIPSGAL field are typically associated with compact, bright far-infrared sources, some are particularly relevant to the present study. We consider lower resolution 21\micron maps from MSX \citep[][]{egan_midcourse_2003} in cases of saturation in the MIPSGAL maps. In the most complex regions of the CMZ's warm dust component, the additional spatial resolution gained by considering more recent observations from other instruments is critical for distinguishing the star-forming properties of the relevant \textit{CMZoom} catalog objects.

To that end, we use data from the FORCAST Galactic Center Legacy Survey \citep[][]{hankins_sofia/forcast_2019, hankins_sofiaforcast_2020} to characterize the compact 25.3\micron emission associated with the 34 \textit{CMZoom} catalog objects for which Spitzer/MIPS data are saturated. An example of such a region is shown in Figure \ref{fig:mips_comp}. Using these data, we confirm whether the \textit{CMZoom} sources host signatures of high-mass star formation. Several key \textit{CMZoom} clouds are not covered by the SOFIA FORCAST Galactic Center Survey (most notably the 1.6 degree cloud complex, the 50\,km\,s$^{-1}$ cloud, and the Dust Ridge clouds), so this analysis is not available for all leaves in the catalog. This may lead to a potential bias in the identified star-forming sources, as it is possible that higher-resolution measurements with greater sensitivity to far-infrared point-sources from SOFIA's FORCAST instrument or future observatories would reveal even more compact emission within seemingly quiescent \textit{CMZoom} catalog leaves. Future analysis will be expanded to include a more detailed characterization of the \textit{CMZoom} sources using the data available from the FORCAST Galactic Center Legacy Survey.

\subsection{Summary of Method for Star Formation Designation}

For the sake of clarity, here we concisely summarize the procedure for identifying which of the \textit{CMZoom} sources are ``robustly star-forming'', ``robustly quiescent'', and ``ambiguously star-forming''. 
A source is considered ``robustly star-forming'' if it has:
\begin{itemize}
    \item A previously identified YSO within its leaf contour, within the uncertainty of the YSO's position
    \item A previously catalogued H$_2$O or CH$_3$OH maser within its leaf contour, up to the maser's positional uncertainty
    \item A previously catalogued 24$\mu$m or 70$\mu$m point source overlapping the leaf contour, within the beam of the corresponding observation
    \item A by-eye determined point-like source of 21$\mu$m, 24$\mu$m, 25$\mu$m, or 70$\mu$m emission overlapping with the leaf contour. 
\end{itemize}
A source is designated as ``robustly quiescent'' if it has none of the above signatures of high-mass star formation. Finally, a source is considered ``ambiguously star-forming'' if any of the above criteria are ambiguously valid, or if the by-eye designation was clouded by bright and diffuse emission surrounding the source.

\section{Results} \label{sec:results}
\subsection{The Star-Forming Properties of CMZoom Sources}
The expansion of the \textit{cmzoom} catalog produced via the methods described in the previous section has been made available online on the \textit{CMZoom} dataverse, along with the rest of the survey's data products\footnote{\url{https://dataverse.harvard.edu/dataverse/cmzoom}}. With this compiled catalog of deeply embedded star formation throughout the CMZ, we can consider the physical properties of the \textit{CMZoom} leaves in the context of their star-forming status. The properties we consider here are those reported in the CMZoom survey's high-robustness catalog\footnote{This catalog is on the CMZoom Dataverse at \dataset[DOI: 10.7910/DVN/RDE1CH]{https://doi.org/10.7910/DVN/RDE1CH}}, and their calculation is fully explained in \cite{hatchfield_cmzoom_2020}, though we briefly review several key definitions here.

\begin{table*}\label{tab:tracer_summary}
\centering
\caption{A brief summary of which data are used as tracers of incipient and ongoing star formation in this work. }
\begin{tabular}{lll}
\hline\hline
Star Formation Tracer & Astrometic Uncertainty & Citation \\
\hline
8$\mu$m point-like emission & -- & \citet{benjamin_glimpse_2003} \\

21$\mu$m point-like emission & -- & \citet{egan_midcourse_2003} \\

24$\mu$m point-like emission & -- & \citet{carey_mipsgal_2009} \\

70$\mu$m point-like emission & -- & \citet{molinari_hi-gal_2010,molinari_source_2011} \\

Prev. catalogued 70$\mu$m point sources & 5'' & \citet{molinari_hi-gal_2016} \\ 

Prev. catalogued 24$\mu$m point sources & 6'' & \citet{gutermuth_24_2015} \\

H$_2$O Masers (J = 6$_{1,6}$ - 5 $_{2,3}$) & 1'' & \citet{walsh_accurate_2014} \\

CH$_3$OH Masers (6.7GHz) & 0.6'' & \citet{caswell_6-ghz_2010} \\

Prev. catalogued YSOs & 2-6'' & \citet{an_massive_2011, immer_recent_2012}

\end{tabular}
\end{table*}

The compact dust sources resolved by CMZoom are defined as the leaves of the dendrogram produced by the \texttt{astrodendro} algorithm applied to the entire survey mosaic, and do not necessarily display elliptical boundary contours. We consider the effective radius of each source to be $R_{\rm eff}\equiv(N_{\rm pix}A_{\rm pix}/\pi)^{1/2}$, with $N_{\rm pix}$ as the number of pixels associated with a given dendrogram leaf and $A_{\rm pix}$ being the area of a single pixel. The mass of each catalog source is defined as 

\begin{equation}
M_{\rm leaf}=\frac{d^{2}S_{\nu}R_{\rm gd}}{\kappa_{\nu}B_{\nu}(T_{\rm d})}, 
\label{eq:mass}
\end{equation}
where $d$ is the distance to the Galactic Center \citep[8.2kpc, ][]{the_gravity_collaboration_geometric_2019}, $S_\nu$ is the leaf's integrated flux,  $R_{\rm gd}$ is the gas-to-dust ratio (taken to be 100, e.g.\ \citealt{battersby_Characterizing_2011}), $\kappa_{\nu}$ is the dust opacity at frequency $\nu$ and $B_{\nu}(T_{\rm d})$ is the Planck function evaluated for dust temperature $T_{\rm d}$, as determined from \textit{Herschel} dust emission modeling (\citealt{mills_Origins_2017}). We also derive source-scale column densities from the \textit{CMZoom} data, calculated for each catalog leaf as
\begin{equation}\label{eq:column}
N_{\rm H_{2}}=\frac{F^{\rm peak}_{\rm \nu}R_{\rm gd}}{\mu_{\rm H_{2}}m_{\rm H}\kappa_{\nu}B_{\rm \nu}(T_{\rm d})},
\end{equation}
using the same definitions as above, with $F^{\rm peak}_{\rm \nu}$ as the leaf's peak flux density in Jy beam$^{-1}$, $\mu_{\rm H_2}$ as the mean atomic weight (2.8, e.g.\ \citealt{kauffmann_MAMBO_2008}), and $m_H$ as the mass of hydrogen. 

Using the total leaf mass $M_{\rm leaf}$ defined above in equation \ref{eq:mass}, we can estimate the average number density of each leaf by assuming that each leaf's mass is distributed uniformly with spherical symmetry inside a radius equal to $R_{\rm eff}$. This simplifying assumption yields 
\begin{equation}\label{eq:n}
    n(H_2) = \frac{3M}{4\pi R_{\rm eq}^3 \mu_{\rm H_2} m_{\rm H} }.
\end{equation}

from which we can also construct the object's free-fall time, 

\begin{equation} \label{eq:tff}
t_{\rm ff}=\bigg(\frac{3\pi}{32G\mu_{\rm H_{2}}m_{\rm H} n_{\rm H_{2}}}\bigg)^{1/2},
\end{equation}

where $G$ is the gravitational constant and the remaining variables and constants are the same as defined above. The cloud-scale column density used in the following analysis is similarly derived from modeling the spectral energy distribution of dust measured by \textit{Herschel}, which will be presented in Battersby et al. in prep. The spatial resolution of this \textit{Herschel}-derived column density is 36'', or about 1.4pc at the distance of the Galactic Center.

\begin{figure*}
\begin{center}
\includegraphics[trim = 0mm 0mm 0mm 0mm, width = .9 \textwidth]{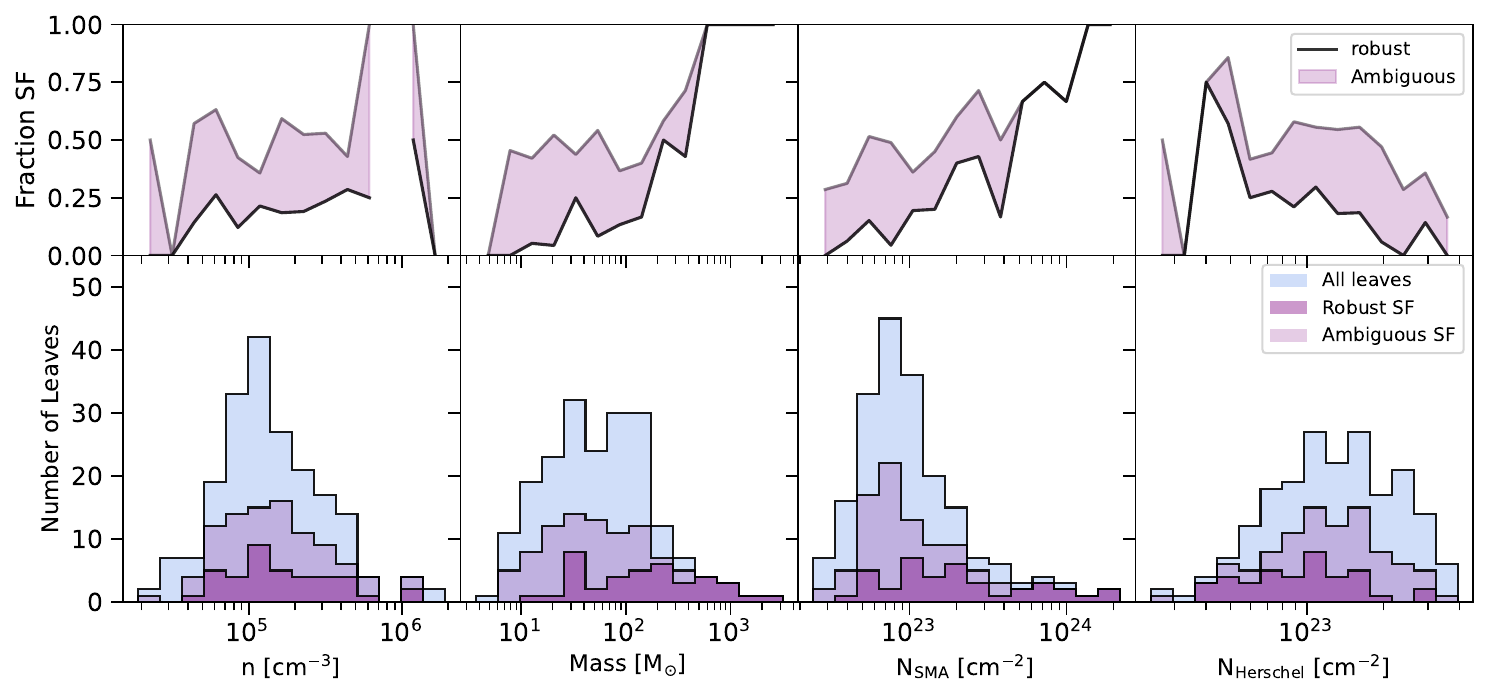}
\end{center}
\caption{ The physical properties of the robustly star-forming CMZoom catalog objects compared with the properties of all catalog objects (from left to right: volume number density of H$_2$, total leaf mass, peak column density of H$_2$ resolved by the SMA, and the cloud-scale column density in H$_2$ as measured by Herschel). Each bottom panel shows a binning of the CMZoom catalog's distribution for the respective property. Each panel shows three histograms, the dark purple histogram shows the distribution of leaves associated with robust tracers of active star formation, the pale purple histogram shows the distribution of leaves with either robust or ambiguous star formation signatures, and the distribution of all leaves is shown in light blue. The solid black line in the upper panels for each physical property shows the ratio between the histogram counts for the distributions of star-forming leaves for each bin, representing the fraction of leaves at that quantity's value that are robustly forming stars. A region is shown above each of these representing how this fraction increases if we include the ambiguous star formation tracers as well.}
\label{fig:prophist1}
\end{figure*}

\begin{figure*}
\begin{center}
\includegraphics[trim = 0mm 0mm 0mm 0mm, width = .9 \textwidth]{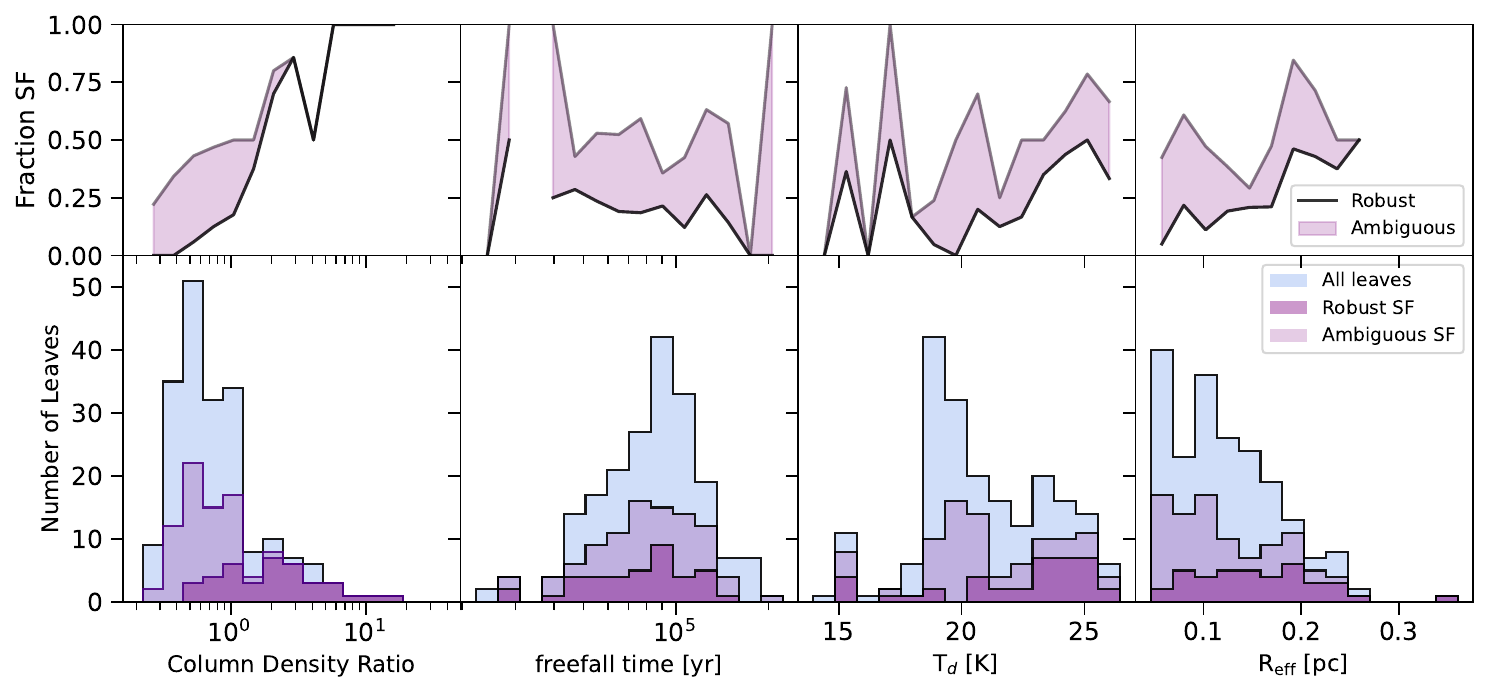}
\end{center}
\caption{A continuation of Figure \ref{fig:prophist1}. The physical properties of the robustly star-forming CMZoom catalog objects compared with the properties of all catalog objects (from left to right: the column density ratio (N$_{\rm SMA}$ / N$_{\rm Herschel}$), the leaf free-fall time, the local dust temperature, and the leaf effective radius). Each bottom panel shows a binning of the CMZoom catalog's distribution in the respective physical properties. Each panel shows three histograms, the dark purple histogram shows the distribution of leaves associated with robust tracers of active star formation, the pale purple histogram shows the distribution of leaves with either robust or ambiguous star formation signatures, and the distribution of all leaves is shown in light blue. The solid black line in the upper panels for each physical property shows the ratio between the histogram counts for the distributions of star-forming leaves for each bin, representing the fraction of leaves at that quantity's value that are robustly forming stars. A region is shown above each of these representing how this fraction increases if we include the ambiguous star formation tracers as well.}
\label{fig:prophist2}
\end{figure*}

In Figures \ref{fig:prophist1} and \ref{fig:prophist2}, we consider how the fraction of leaves with indications of active star formation correlates with the catalog properties. For this analysis, we exclude the regions surrounding Sgr B2 and Sgr A$^*$. We exclude the former since the region hosts a particularly massive, more-evolved protocluster for which significantly more accurate estimates of star formation rates and high-resolution YSO counting and modeling are available (for example, see \citealt{schmiedeke_physical_2016} and \citealt{ginsburg_distributed_2018}). We exclude the region surrounding Sagittarius A and the circumnuclear disk due to the highly non-Gaussian noise and contamination from synchrotron emission described in more detail in \citealt{battersby_cmzoom_2020}. Excluding these regions, we find a total of 39 leaves with a robust signature of active star formation, 57 leaves with ambiguous signs of active star formation, and 104 leaves with a robust non-detection of active star formation. These results are displayed for only robustly star-forming sources in Table \ref{tab:cloud_SFR} and for robustly and ambiguously star-forming sources in Table \ref{tab:cloud_SFR_2}.

\subsection{Star Formation Rates of CMZ clouds}\label{sec:results:sfrs}
With a robust estimate of the star formation status of the CMZoom catalog sources, we can derive a new estimate for the SFR of each cloud observed by \textit{CMZoom}. \citealt{hatchfield_cmzoom_2020} calculated an upper limit on the SFR of the CMZ, a star formation potential, in the range of 0.08 - 2.20 M$_\odot$ yr$^{-1}$ (or  0.04 - 0.47 M$_\odot$ yr$^{-1}$ excluding Sgr B2)  depending on the assumed star formation efficiency. This method makes the assumption that each of the catalogued objects will collapse to form stars on their free-fall timescales, which is likely to overestimate the resulting star formation rates. By considering which of these sources are associated with active star formation tracers, we can provide a more accurate estimate of the incipient star formation rate of clouds in the CMZ. 

To calculate the incipient star formation rates for the \textit{CMZoom} clouds, we follow a modified version of the procedure used in \citealt{hatchfield_cmzoom_2020} for each catalog leaf with a robust star formation tracer. We translate the properties from the catalog into star formation rates for each of the \textit{CMZoom} clouds as the summation of the individual leaves' star formation rates, using

\begin{equation}\label{eq:sfr_cloud}
    \text{SFR}_{\rm cloud} = \frac{\epsilon_{\rm SF}}{\xi_{\rm IMF}} \sum_i \frac{M_{i}}{t_{ff,i}},
\end{equation}
where the sum runs over the $i$ leaves in the cloud, $M_i$ and $t_{ff,i}$ are respectively the mass and free-fall time of the $i$th star-forming leaf, $\epsilon_{\rm SF}$ is the star formation efficiency per freefall time, and $\xi_{\rm IMF}$ is a correction to account for low-mass star formation to which the \textit{CMZoom} catalog and the tracers used in this work are not sensitive. In this equation, we make several simplifying assumptions. Firstly, we must choose an average star formation efficiency with which these structures are to form stars over the course of a free-fall time. We assume that each leaf is forming stars with an efficiency 
\begin{equation}
\epsilon_{\rm SF} = 0.25\pm0.15
\end{equation}
over the course of one of the leaf's free-fall time, $t_{ff}$. The star formation efficiency per free-fall time has been investigated on the scales of individual protostellar cores ($\sim20-50\%$, e.g.\ \citet{walker_Star_2018, barnes_young_2019}) as well as on cloud scales ($\sim$1-10\%, e.g.\ \citet{leroy_cloud-scale_2017, barnes_star_2017, utomo_star_2018, krumholz_star_2019, lu_star_2019, grudic_dynamics_2022, chevance_life_2022}), but on the intermediate scales investigated in this work, it is unclear what a realistic value might be. The range of values assumed above is intended to be broad enough to encompass a variety of sensible possibilities in agreement with the values estimated for bounding physical scales. 

The high-mass star-forming sources characterized here only provide a partially complete picture of the CMZ's incipient star formation. There is substantial evidence for a lower-mass star-forming population in both the most prolific clouds, such as Sgr B2 \citep{ginsburg_distributed_2018} as well as the supposedly most star formation-deficient clouds, like The Brick \citep{walker_star_2021}. These lower mass protostellar populations are not likely to be detectable using the method employed in this work. While the \textit{CMZoom} catalog is complete to more than 95\% of all possible sites of incipient high-mass star formation (the robust catalog is complete to $>95\%$ of sources with total gas mass $>80$M$_\odot$ for an assumed dust temperature  $>$20K, \citet{hatchfield_cmzoom_2020}), it is not sensitive to the population of more isolated low-mass protostars that have been detected in some of the Galactic Center's molecular clouds. 

To account for the missing star-forming mass associated with the lower-mass star protostellar objects in the CMZ, we extrapolate from the stellar Initial Mass Function, or IMF \citep[][]{kroupa_variation_2001} using the IMF python module (https://github.com/keflavich/imf). By drawing 1000 samples from the IMF of a cluster with a mass of $>100$ M$_\odot$ (sufficiently above the \textit{CMZoom} catalogs' completeness), we find that an average of $\xi_{\rm IMF} \approx 20\%$ of the cluster's star mass is ultimately contained in high-mass protostars\footnote{The high-mass stars' contribution to the total fraction of star mass in a cluster is approximately the same for clusters with mass above 100M$_\odot$, but may be smaller for low-mass clusters for which the high mass tail of the IMF is not well sampled.}. This may be an overestimate for sources that contain significant further fragmentation that includes a significant number of lower-mass protostars, but if the flux detected with the SMA is largely due to dust associated with gas around high-mass protostars, then it is not likely to be a significant overestimate. Future observations capable of measuring the masses of a more complete sample of the CMZ's protostellar population are necessary to better approximate this value.

The uncertainties for these values are dominated by our assumptions in $\epsilon_{\rm SF}$, $\xi_{\rm IMF}$, as well as the systematic uncertainties in our assumed value of the gas-to-dust ratio, dust grain properties and the random uncertainty from noise in the SMA's flux measurements. The uncertainties in mass, dust temperature, and free-fall time are further detailed in \citealt{hatchfield_cmzoom_2020}. The masses reported in the catalog use dust temperatures calculated from Herschel SED fitting (Battersby et al. in prep.), and therefore apply to a spatial scale of 36'', significantly larger than the size of the \textit{CMZoom} catalog objects. It is likely that dust temperatures would be systematically higher for sites of active star formation on smaller spatial scales, so the star formation rates calculated using the Herschel temperatures may be too high. This is because an assumed lower temperature in Equation \ref{eq:mass} implies a higher mass (and similarly a shorter freefall time from Equations \ref{eq:n} and \ref{eq:tff}), leading to a significantly elevated SFR using Equation \ref{eq:sfr_cloud}. Dust temperatures on the scales of individual cores in the CMZ may be higher, $57 - 220$K or greater \citep{walker_Star_2018}, so we expect the temperatures on the intermediate scales probed by \textit{CMZoom} to be respectively intermediate to the Herschel-derived values and the core temperatures, though where they fall between these boundary values is difficult to constrain further without higher resolution observations in far-infrared and submillimeter bands. The values of the star formation rates for each cloud are reported in Table \ref{tab:cloud_SFR} assuming both the Herschel-derived temperatures and an intermediate dust temperature of 50K.

\begin{table*}
\centering
\caption{The summary of the star-forming properties for each cloud in the CMZ considering only robustly star-forming sources. The first three columns include a coordinate name, colloquial name and cloud mask index for each cloud. The column ``SFR$_{\rm Herschel}$'' represents the star formation rate calculated in equation \ref{eq:sfr_cloud}, assuming a dust temperature from Herschel, as reported in the catalog, with uncertainties due to star formation efficiency and random error in the mass estimation. The column ``SFR$_{\rm 50K}$'' represents the star formation rate calculated in equation \ref{eq:sfr_cloud}, assuming a dust temperature of 50K, with uncertainties due to star formation efficiency and random error in the mass estimation.The column ``N$_{\rm SF}$'' lists the number of robustly star-forming sources within each \textit{CMZoom} cloud and the final column N$_{\rm SF}$/N$_{\rm tot}$ lists the fraction of sources with robust star formation tracers in each cloud. For clouds without star formation tracers, and for Sgr B2, these numbers are omitted and replaced with a dash. }

\begin{tabular}{lllllll}
\hline\hline
 Cloud ID & Colloquial Name & \textit{CMZoom} & SFR$_{\rm Herschel}$ & SFR$_{\rm 50K}$ & N$_{\rm SF}$ & N$_{\rm SF}$/N$_{\rm tot}$ \\
 & & mask number & 10$^{-3}$ M$_\odot$yr$^{-1}$ & 10$^{-3}$ M$_\odot$yr$^{-1}$ & \\
\hline
G1.683-0.089 &  & 1 & -- & -- & 0 & -- \\
G1.670-0.130 &  & 2 & 3.2$\pm$1.9 & 0.4$\pm$0.2 & 2 & 0.33 \\
G1.651-0.050 &  & 3 & 1.5$\pm$0.9 & 0.2$\pm$0.1 & 1 & 0.50 \\
G1.602+0.018 &  & 4 & 4.0$\pm$2.4 & 0.4$\pm$0.3 & 1 & 0.20 \\
G1.085-0.027 &  & 5 & 6.5$\pm$3.9 & 1.0$\pm$0.6 & 2 & 0.67 \\
G1.038-0.074 &  & 6 & -- & -- & 0 & -- \\
G0.891-0.048 &  & 7 & -- & -- & 0 & -- \\
G0.714-0.100 &  & 8 & -- & -- & 0 & -- \\
G0.699-0.028 & Sgr B2 & 9 & -- & -- & 0 & -- \\
G0.619+0.012 & Sgr B2 SE & 10 & 30.0$\pm$18.0 & 8.1$\pm$4.9 & 7 & 0.70 \\
G0.489+0.010 & Dust Ridge Clouds E\&F & 11 & 15.9$\pm$9.5 & 2.8$\pm$1.7 & 1 & 0.08 \\
G0.412+0.052 & Dust Ridge Cloud D & 12 & -- & -- & 0 & -- \\
G0.393-0.034 &  & 13 & -- & -- & 0 & -- \\
G0.380+0.050 & Dust Ridge Cloud C & 14 & 37.9$\pm$22.8 & 10.0$\pm$6.0 & 1 & 0.12 \\
G0.340+0.055 & Dust Ridge Cloud B & 15 & -- & -- & 0 & -- \\
G0.326-0.085 & The Sailfish & 16 & -- & -- & 0 & -- \\
G0.316-0.201 &  & 17 & -- & -- & 0 & -- \\
G0.253+0.016 & The Brick & 18 & -- & -- & 0 & -- \\
G0.212-0.001 &  & 19 & 5.0$\pm$3.0 & 1.6$\pm$0.9 & 3 & 0.60 \\
G0.145-0.086 & The Straw Cloud & 20 & -- & -- & 0 & -- \\
G0.106-0.082 & The Sticks Cloud & 21 & -- & -- & 0 & -- \\
G0.070-0.035 &  & 22 & -- & -- & 0 & -- \\
G0.068-0.075 & The Stone Cloud & 23 & 2.2$\pm$1.3 & 0.6$\pm$0.4 & 1 & 0.08 \\
G0.054+0.027 &  & 24 & -- & -- & 0 & -- \\
G0.014+0.021 &  & 25 & -- & -- & 0 & -- \\
G0.001-0.058 & The 20 km/s cloud & 26 & 4.0$\pm$2.4 & 1.2$\pm$0.7 & 5 & 0.26 \\
G359.948-0.052 & Sgr A*, circumnuclear disk & 27 & -- & -- & 0 & -- \\
G359.889-0.093 & The 50 km/s cloud & 28 & 32.3$\pm$19.4 & 5.9$\pm$3.5 & 1 & 0.05 \\
G359.865+0.022 &  & 29 & -- & -- & 0 & -- \\
G359.734+0.002 &  & 30 & -- & -- & 0 & -- \\
G359.648-0.133 &  & 31 & -- & -- & 0 & -- \\
G359.611+0.018 &  & 32 & -- & -- & 0 & -- \\
G359.615-0.243 &  & 33 & -- & -- & 0 & -- \\
G359.484-0.132 & Sgr C & 34 & 65.4$\pm$39.2 & 15.4$\pm$9.2 & 3 & 0.60 \\
G359.137+0.031 &  & 35 & -- & -- & 0 & -- \\
\hline\hline
\end{tabular}

\label{tab:cloud_SFR}
\end{table*}

\begin{table*}
\centering
\caption{A version of Table \ref{tab:cloud_SFR} including ambiguously star-forming sources in addition to robustly star-forming sources. The first three columns include a coordinate name, colloquial name and cloud mask index for each cloud. The column ``SFR$_{\rm Herschel}$'' represents the star formation rate calculated in equation \ref{eq:sfr_cloud}, assuming a dust temperature from Herschel, as reported in the catalog, with uncertainties due to star formation efficiency and random error in the mass estimation. The column ``SFR$_{\rm 50K}$'' represents the star formation rate calculated in equation \ref{eq:sfr_cloud}, assuming a dust temperature of 50K, with uncertainties due to star formation efficiency and random error in the mass estimation.The column ``N$_{\rm SF}$'' lists the number of robustly star-forming sources within each \textit{CMZoom} cloud and the final column N$_{\rm SF}$/N$_{\rm tot}$ lists the fraction of sources with robust star formation tracers in each cloud. For clouds without star formation tracers, and for Sgr B2, these numbers are omitted and replaced with a dash. }

\begin{tabular}{lllllll}
\hline\hline
 Cloud ID & Colloquial Name & \textit{CMZoom} & SFR$_{\rm Herschel}$ & SFR$_{\rm 50K}$ & N$_{\rm SF}$ & N$_{\rm SF}$/N$_{\rm tot}$ \\
 & & mask number & 10$^{-3}$ M$_\odot$yr$^{-1}$ & 10$^{-3}$ M$_\odot$yr$^{-1}$ & \\
\hline
G1.683-0.089 &  & 1 & -- & -- & 0 & -- \\
G1.670-0.130 &  & 2 & 3.9$\pm$2.4 & 0.5$\pm$0.3 & 4 & 0.67 \\
G1.651-0.050 &  & 3 & 1.5$\pm$0.9 & 0.2$\pm$0.1 & 1 & 0.50 \\
G1.602+0.018 &  & 4 & 5.0$\pm$3.0 & 0.5$\pm$0.3 & 3 & 0.60 \\
G1.085-0.027 &  & 5 & 7.1$\pm$4.2 & 1.1$\pm$0.6 & 3 & 1.00 \\
G1.038-0.074 &  & 6 & -- & -- & 0 & -- \\
G0.891-0.048 &  & 7 & -- & -- & 0 & -- \\
G0.714-0.100 &  & 8 & 13.7$\pm$8.2 & 2.7$\pm$1.6 & 11 & 0.44 \\
G0.699-0.028 & Sgr B2 & 9 & -- & -- & 0 & -- \\
G0.619+0.012 & Sgr B2 SE & 10 & 32.0$\pm$19.2 & 8.6$\pm$5.2 & 8 & 0.80 \\
G0.489+0.010 & Dust Ridge Clouds E\&F & 11 & 16.9$\pm$10.1 & 2.9$\pm$1.8 & 3 & 0.23 \\
G0.412+0.052 & Dust Ridge Cloud D & 12 & 2.7$\pm$1.6 & 0.5$\pm$0.3 & 2 & 0.17 \\
G0.393-0.034 &  & 13 & -- & -- & 0 & -- \\
G0.380+0.050 & Dust Ridge Cloud C & 14 & 38.7$\pm$23.2 & 10.2$\pm$6.1 & 3 & 0.38 \\
G0.340+0.055 & Dust Ridge Cloud B & 15 & -- & -- & 0 & -- \\
G0.326-0.085 & The Sailfish & 16 & 0.2$\pm$0.1 & 0.1$\pm$0.0 & 2 & 1.00 \\
G0.316-0.201 &  & 17 & -- & -- & 0 & -- \\
G0.253+0.016 & The Brick & 18 & 2.3$\pm$1.4 & 0.4$\pm$0.3 & 2 & 0.18 \\
G0.212-0.001 &  & 19 & 5.3$\pm$3.2 & 1.6$\pm$1.0 & 4 & 0.80 \\
G0.145-0.086 & The Straw Cloud & 20 & 0.4$\pm$0.3 & 0.1$\pm$0.1 & 1 & 0.50 \\
G0.106-0.082 & The Sticks Cloud & 21 & -- & -- & 0 & -- \\
G0.070-0.035 &  & 22 & 1.7$\pm$1.0 & 0.5$\pm$0.3 & 3 & 0.60 \\
G0.068-0.075 & The Stone Cloud & 23 & 6.4$\pm$3.8 & 1.5$\pm$0.9 & 12 & 1.00 \\
G0.054+0.027 &  & 24 & -- & -- & 0 & -- \\
G0.014+0.021 &  & 25 & -- & -- & 0 & -- \\
G0.001-0.058 & The 20 km/s cloud & 26 & 6.3$\pm$3.8 & 1.8$\pm$1.1 & 7 & 0.37 \\
G359.948-0.052 & Sgr A*, circumnuclear disk & 27 & -- & -- & 0 & -- \\
G359.889-0.093 & The 50 km/s cloud & 28 & 50.6$\pm$30.4 & 9.2$\pm$5.5 & 9 & 0.43 \\
G359.865+0.022 &  & 29 & 0.1$\pm$0.1 & 0.0$\pm$0.0 & 1 & 0.50 \\
G359.734+0.002 &  & 30 & 0.1$\pm$0.1 & 0.0$\pm$0.0 & 1 & 0.33 \\
G359.648-0.133 &  & 31 & -- & -- & 0 & -- \\
G359.611+0.018 &  & 32 & -- & -- & 0 & -- \\
G359.615-0.243 &  & 33 & -- & -- & 0 & -- \\
G359.484-0.132 & Sgr C & 34 & 65.4$\pm$39.2 & 15.4$\pm$9.2 & 3 & 0.60 \\
G359.137+0.031 &  & 35 & -- & -- & 0 & -- \\
\hline\hline
\end{tabular}

\label{tab:cloud_SFR_2}
\end{table*}

The star formation rates of each CMZoom cloud are presented in  Figure \ref{fig:sfr_bounds}. Due to the considerable uncertainties in the masses of the \textit{CMZoom} sources within the Sgr B2 complex and the circumnuclear field surrounding Sgr A*, we exclude these two regions from our analysis (for more details on the issues with the interpretation of these regions see section 5.1 of \citet{hatchfield_cmzoom_2020} and section 5.4 of \citet{battersby_cmzoom_2020}). While low-mass star formation is likely to occur throughout CMZ clouds both with and without star formation tracers used in this work, we cannot know their star-forming properties. For the excluded clouds and those with no high-mass star formation indicators, we do not report a star formation rate, listing `---' instead in Tables \ref{tab:cloud_SFR} and \ref{tab:cloud_SFR_2}.

\section{Discussion} \label{sec:discussion}

\subsection{Number Density, Column Density Ratio, and their Relationship to Star Formation}\label{sec:discussion:n_od}
Some theories and empirical models of star formation suggest the existence of a volume density threshold \citep[e.g.][]{krumholz_minimum_2008, kauffmann_mass-size_2010, lada_star_2010, lada_star_2012, padoan_star_2014}, below which molecular gas cannot efficiently collapse and form new stars. Above this density threshold, some of these theories and models predict an increase in the number of YSOs in the cloud as a scaling relation, their YSO count and therefore star formation rate increasing with density. The densities of all \textit{CMZoom} sources appear to be well above the surface and volume density thresholds suggested for solar neighborhood clouds \citep[e.g.][]{, krumholz_minimum_2008, lada_star_2010,clark_column_2014}. There is also evidence accumulating for the existence of an environmentally dependent density threshold for star formation, which is needed to explain the distribution of star formation observed in CMZ clouds \citep[e.g.][]{kauffmann_galactic_2013, kruijssen_what_2014, rathborne_g0.253+0.016:_2014, kauffmann_galactic_2017b, walker_Star_2018, barnes_young_2019, lu_census_2019, lu_star_2019, battersby_cmzoom_2020}. If there exists a universal density threshold above which the amount of star formation increases in proportion to the dense gas mass, we should expect to see a positively correlated scaling relationship, or even a step function relating the density properties of the catalog objects and the prevalence of ongoing star formation. If instead there should be an environmentally dependent threshold, we might expect to see very little star formation below some surface or volume density, and increasingly more star formation above that threshold. Neither of these situations appears to be the case for the \textit{CMZoom} catalog objects. 

In some cases, we do see correlations within the \textit{CMZoom} sample. Figures \ref{fig:prophist1} and \ref{fig:prophist2} show the fraction of star-forming leaves across the range of physical properties derived in \citealt{hatchfield_cmzoom_2020}. Three leaf properties do appear to increase significantly along with the prevalence of robust star formation indicators: the total leaf's mass, leaf's peak source-scale column density, and the leaf's column density ratio (the ratio of the source-scale peak column density to the cloud-scale column density, $N_{\rm SMA} / N_{\rm Herschel}$). The uptick in frequency of star-forming leaves that occurs for both mass and peak source-scale column density is largely due to a very small number of leaves in high-mass and high-column density bins, corresponding to the handful of already well-known and well-studied young massive clusters in the CMZ (Dust Ridge Cloud C, Sgr C). 

\begin{figure*}[htbp]
\begin{center}
\includegraphics[trim = 0mm 0mm 0mm 0mm, width = .7 \textwidth]{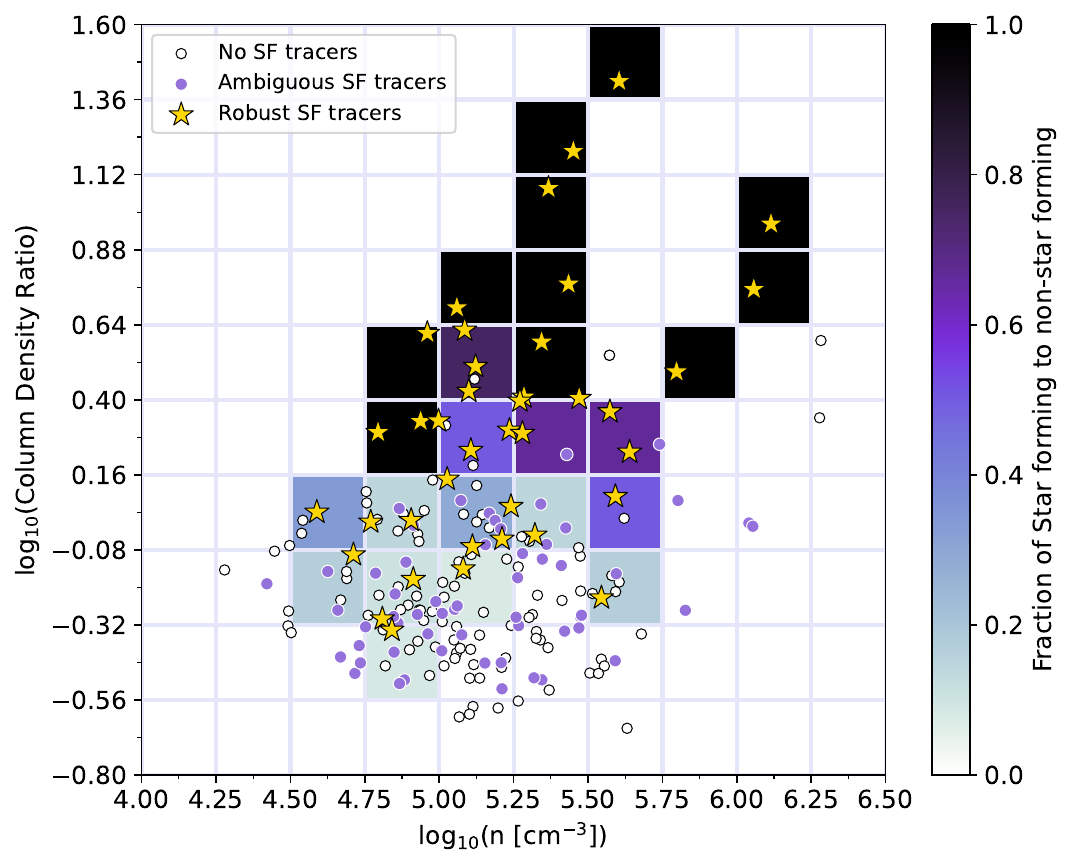}
\end{center}
\caption{The distribution of robustly star-forming, robustly quiescent, and ambiguous \textit{CMZoom} catalog leaves as a function of $n$(H$_2$), the volume number density of H$_2$, and the column density ratio, defined as the ratio of the SMA-derived source-scale column density to the Herschel-derived cloud-scale column density for each source. The colorscale background represents the ratio of star-forming leaves to all leaves in each bin. While leaves with signatures of active star formation occur at both low and high number densities and column density ratios, a sharp uptick in the fraction of star-forming sources occurs at a column density ratio threshold of $\sim$ 1.5 (or log$_{10}$(Column Density Ratio) $\approx$ 0.17). There is no obvious scaling between volume density and active star formation.}
\label{fig:scatterhist_odn}
\end{figure*}

The sharp rise in the prevalence of robustly star-forming sources that occurs above a column density ratio of $\sim 1.5$ N$_{\rm SMA} / $N$_{\rm Herschel}$ comes from a significantly larger sample of sources (N$>$30), including a combination of well-studied regions and previously understudied clouds. Therefore, column density ratio seems to correlate most significantly with the presence of active star formation. The correlation of column density ratio with the indicators of active star formation is not necessarily surprising \citep[e.g. ][]{kruijssen_what_2014}, as the conditions for active star formation demand the presence of local, gravitationally bound overdensities. Figure \ref{fig:scatterhist_odn} shows the stark contrast between distribution of star formation within leaves as a function of volume density and of column density ratio. While star formation does occur in sources across the range of volume densities and column density ratios probed by \textit{CMZoom}, the vast majority (over $\sim$75\%) of leaves above a column density ratio of $\sim1.5$ have tracers of star formation. It is possible that this number represents a threshold in the relative surface density, above which turbulent fluctuations are increasingly likely to produce substructures that are dense enough to form stars. While star formation can occur in sources with low column density ratio, any substructure in excess of this column density ratio threshold is extremely likely to be gravitationally bound and actively be forming stars. However, from this study alone it is not possible to determine if this threshold is dependent on the observational parameters of the data used to construct the column density ratio estimate. It is therefore unclear if this can be generalized to a dense gas fraction threshold on similar spatial scales, and further analysis will elucidate its relevance to other surveys of dense substructure in molecular clouds. 

Notably, there is no corresponding correlation of star formation tracer prevalence with the volume density of these sources. The lack of such a correlation with density begs the question - why are stars not preferentially forming in the highest density substructures? It is possible that the assumptions that we use to derive the leaf volume densities are overly-simplistic and fail to accurately represent the importance of the ISM's three-dimensional complexity. For instance, we assume that the mass derived from each dendrogram leaf's integrated flux is spherically distributed, isotropic in both the plane-of-the-sky and along the line of sight. While leaves in the catalog are not uniformly circular, equation \ref{eq:n} uses the effective radius of each leaf to calculate an average volume density. Densities on the physical scales of the \textit{CMZoom} are certain to have non-uniformities, and the unresolved substructure of the leaves complicates their interpretation.

\citealt{clark_column_2014} used hydrodynamic simulations of star-forming clouds to evaluate the efficacy of translating between measured column density and the true volume density of star-forming regions to evaluate star formation density thresholds, finding that the mass of clouds above a given column density does not translate straightforwardly to a mass above a corresponding measured volume density. The average density of a clump-scale object has an inconvenient degeneracy. When averaging over an entire structure composed of unresolved quiescent or star-forming substructure, the mean volume density may represent a log-normal distribution due to the turbulent velocity and density field (as is predicted in the substantial theoretical literature of turbulent molecular clouds, e.g. \citealt{burkhart_self-gravitating_2019, chen_anatomy_2018}) or it may represent a more diffuse envelope embedding a collection of gravitationally bound cores at very high volume density. Since we cannot distinguish between these scenarios with the mean volume density, one which may either represent a quiescent, turbulent structure or an actively star-forming object, we should not expect mean volume density to consistently correlate with star formation tracers. In this way of thinking, the peak SMA column density and the column density ratio are better suited to distinguish between a dense, but unbound structure and an actively star-forming protocluster. This helps to explain why the column density ratio, in the bottom left panel of Figure \ref{fig:prophist2} shows the most robust correlation with active star formation tracers.

It is also possible that the densities measured are an adequate probe of the true volume densities of the sources, and the lack of correlation between volume density and tracers of active star formation is representative of the physical timescales relevant to star formation. The timescales required for protostars to exhibit the star formation tracers used in this work (far infrared radiation from heated gas envelopes and masers) are not well constrained. The flux from which we measure the volume density is not necessarily from material that will form stars, but perhaps also represents gas that is actively escaping the process of star formation. Simultaneously, many of the dense sources that do not exhibit any indication of star formation may truly be pre-stellar gas structures, so the ``robustly non-star-forming'' sources may be contributing to future star formation in a way that is not well-constrained by this procedure. Higher resolution observations and virial analysis \citep[e.g.][]{myers_virial_2022} provide more insight into what component of the gas in these structures should be treated as star-forming or quiescent, though very few sources seem to be entirely gravitationally bound according to recent sprectral analysis presented by \citet{callanan_cmzoom_2023}. 

\begin{figure*}[h]
\begin{center}
\includegraphics[trim = 0mm 0mm 0mm 0mm, width = .99 \textwidth]{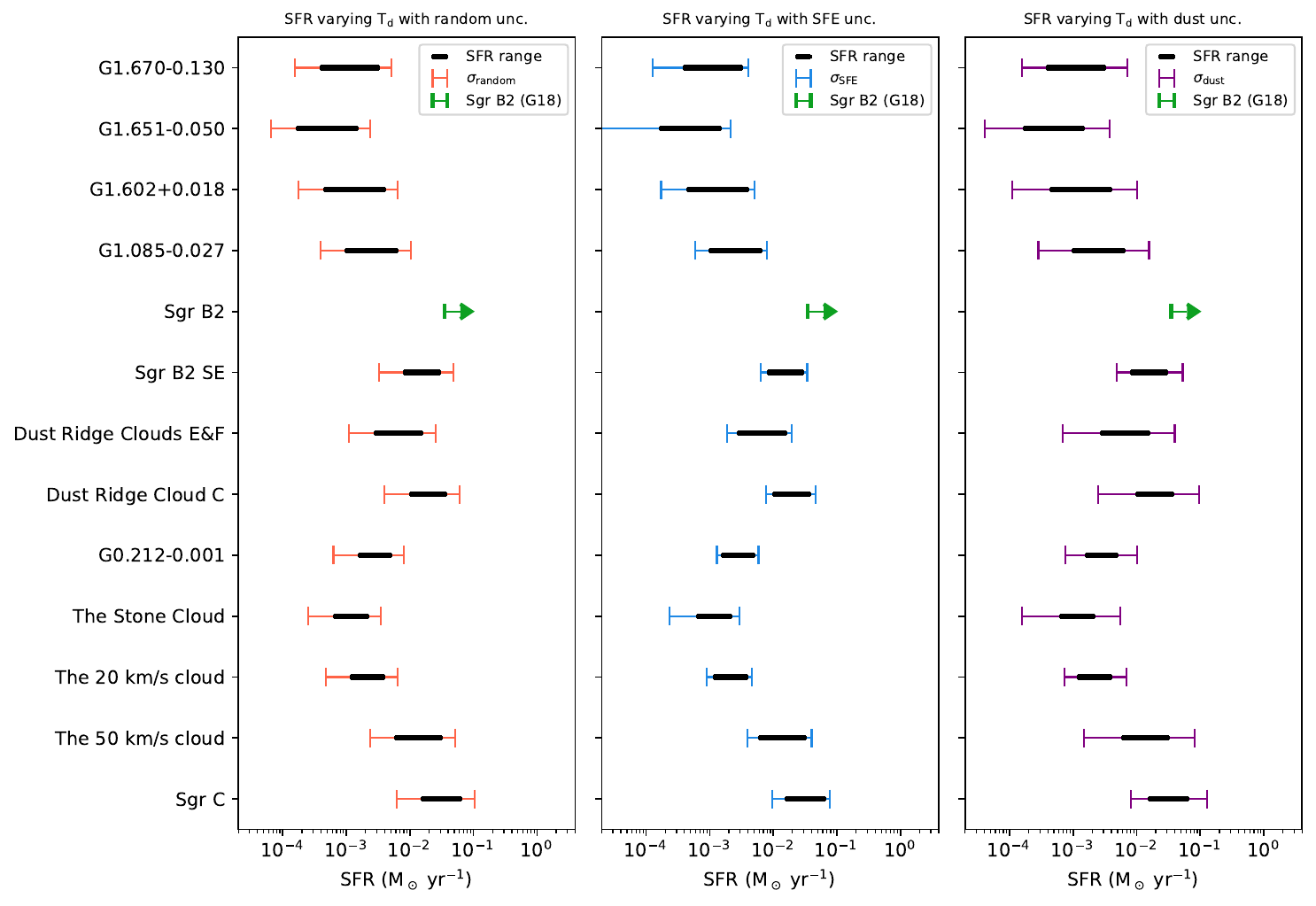}
\end{center}
\caption{The range of star formation rates from robustly star-forming sources and uncertainties derived using a variable dust temperature in calculating source masses, using only robust star formation indicators. In each panel, the black bar represents the star formation rate from equation \ref{eq:sfr_cloud} for physical properties calculated using a range of assumed dust temperatures from the Herschel-derived estimate reported in the catalog to a constant 50K. The error bars in each panel represent the uncertainty due to, from left to right, random uncertainties (including local noise in the measured submillimeter flux, assumed distance, and dust temperature fluctuations), the variation due to our choice of star formation efficiency, and lastly the uncertainty in the local dust properties of each source. The lower limit on the SFR from Sgr B2 is shown in green using the constraining values presented in \citet{ginsburg_distributed_2018}.}
\label{fig:sfr_bounds}
\end{figure*}

\begin{figure*}[h]
\begin{center}
\includegraphics[trim = 0mm 0mm 0mm 0mm, width = .99 \textwidth]{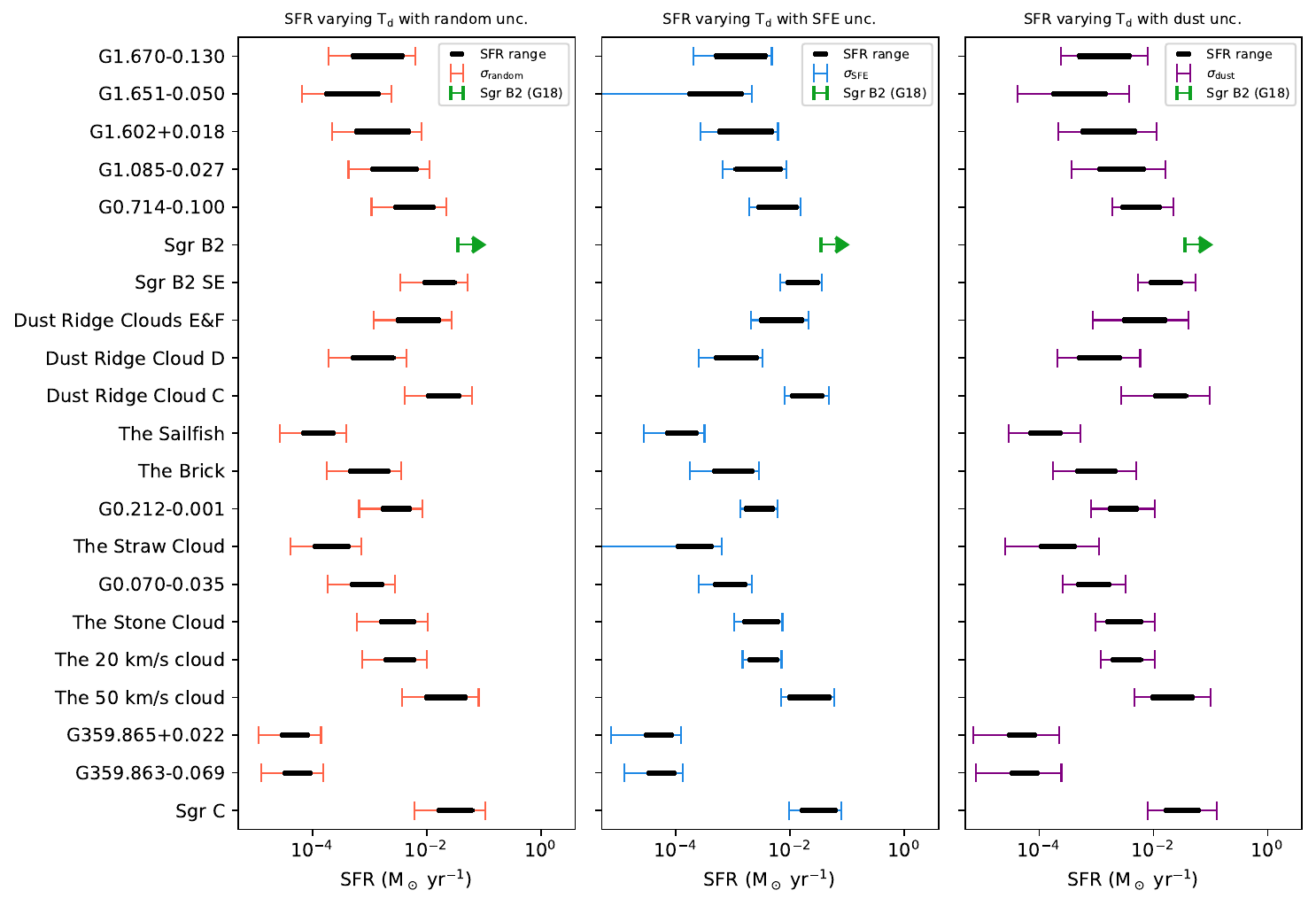}
\end{center}
\caption{A version of Figure \ref{fig:sfr_bounds} that also includes ambiguously star-forming sources. Again, in each panel, the black bar represents the star formation rate from equation \ref{eq:sfr_cloud} for physical properties calculated using a range of assumed dust temperatures from the Herschel-derived estimate reported in the catalog to a constant 50K. The error bars in each panel represent the uncertainty due to, from left to right, random uncertainties (including local noise in the measured submillimeter flux, assumed distance, and dust temperature fluctuations), the variation due to our choice of star formation efficiency, and lastly the uncertainty in the local dust properties of each source. The lower limit on the SFR from Sgr B2 is shown in green using the constraining values presented in \citet{ginsburg_distributed_2018}.}
\label{fig:sfr_bounds_amb}
\end{figure*}

\subsection{The Relationship Between Tracers of High-Mass Star Formation and the Total Star Formation Rate of CMZ Clouds}\label{sec:discussion:sfrs}

Given the completeness of the \textit{CMZoom} catalog to sites of incipient high-mass star formation, the star formation rates presented in this work effectively incorporate all possible deeply-embedded high-mass star formation in the CMZ. If we wish to understand this subset of the star-forming population in the context of the holistic star formation properties of the Galactic Center and calculate an incipient star formation rate for the CMZ as a whole, we must carefully consider our assumptions about the star formation activity and the protostellar population for which these tracers are not sensitive. 

Firstly, the protostellar ages of these star-forming objects are not well constrained. As dense gas in the CMZ's molecular clouds collapses, some local density maxima will achieve the conditions for gravitational instability and collapse to form clusters of stars. The high-mass protostars heat the gas and dust around them and add substantial energy into the surrounding ISM, leading to the emission that we use as a tracer of star formation in this work. Feedback from massive stars in these clusters will begin to erode the cold, submillimeter-bright envelope embedding the cluster as photodissociation regions emerge. Therefore, if we see signatures of active high-mass star formation such as methanol masers or compact dust emission in the far infrared associated with the densest submillimeter bright substructure within a cloud, we know that this star formation has not progressed long enough to destroy or significantly displace its natal envelope. While the amount of time needed for massive YSOs to destroy their envelope is not yet precisely understood and is likely to vary by environment, it is believed that it takes on the order of $\sim 4 \times 10^5$ years \citep[][]{davies_red_2011}. Therefore we anticipate that the sources identified in this work probe a range of protostellar ages less than $\sim 4\times 10^5$ years. 

The sample of previously catalogued YSOs used in this work \citep[][]{an_massive_2011, yusef-zadeh_star_2009, immer_recent_2012} contains many sources that do not correspond to \textit{CMZoom} catalog sources, likely either representing evolved stellar contaminants or more evolved young high-mass stars that have already destroyed the cold dense components of their envelopes. The presence of a YSO candidate in the absence of compact submillimeter emission, given the completeness of \textit{CMZoom}, implies that these sources are more advanced in their protostellar evolution. Therefore we expect that \textit{CMZoom}-associated star formation signatures are tracing a more specific, younger population of protostellar sources, which is how we define the term ``incipient'' star formation throughout this work.

As discussed briefly in Section \ref{sec:results}, the cloud star formation rates that we calculate according to equation \ref{eq:sfr_cloud} rely on ill-constrained assumptions about the star formation efficiency per free-fall time on these spatial scales and missing lower-mass star formation within the clouds. Given the diverse morphologies and star formation signatures of \textit{CMZoom} sources, we adopted a range for star formation efficiency per free-fall time, $0.1 \leq \epsilon_{\rm SF} \leq 0.4$ with a fiducial value of 0.25. It is not possible to make a more specific correction due to the IMF than the one used in equation \ref{eq:sfr_cloud} without a thorough analysis of the low-mass YSO population of the CMZ, which is beyond the capabilities of the present data set but may be feasible with future, higher resolution and sensitivity observations. There are reports of a top-heavy IMF in the known Galactic Center young star clusters- namely the Arches, Quintuplet, and Young Nuclear clusters \citep[e.g. ][]{hosek_unusual_2019, rui_quintuplet_2019}. While it is still unclear if an abnormal slope is characteristic of nascent CMZ star clusters or high-mass clusters in general, any variation in the typical IMF will influence the assumptions used in calculating these SFRs. A universally top-heavy IMF would imply that the values presented in this work would be overestimates of the true SFRs of CMZ clouds.

With the context of these significant uncertainties, we can estimate the total star formation rate of the Galactic Center, excluding the circumnuclear disk. Summing over the clouds' star formation rates derived in Section \ref{sec:results:sfrs}, using our assumed star formation efficiency $\epsilon_{\rm SF}=0.25\pm0.15$, the IMF correction described in Section \ref{sec:results:sfrs}, and assuming an elevated dust temperature of 50K, we find an incipient CMZ SFR of $\sim$0.08 M$_\odot$ yr$^{-1}$ across the 13 robustly star-forming \textit{CMZoom} clouds. This value includes the star formation rate for Sgr B2 calculated by \citet{ginsburg_distributed_2018} of 0.036 M$_\odot$ yr$^{-1}$, since the \textit{CMZoom} observations of the corresponding region are difficult to interpret due to the local noise properties (see section 5 of \citet{hatchfield_cmzoom_2020} for more details). However, if the dust temperatures on 0.1 pc scales near star-forming sources are closer to the Herschel-calculated dust temperatures, the masses used in equation \ref{eq:sfr_cloud} are considerably higher. Including the sources with ambiguous star formation tracers (detailed in Section \ref{sec:method:by-eye}), and with those lower dust temperatures reported in the catalog, we find a higher total CMZ SFR of $\sim$0.45 M$_\odot$ yr$^{-1}$ across 20 robustly and ambiguously star-forming \textit{CMZoom} clouds, which is remarkably similar to the empirical dense gas scaling relation-derived value for the CMZ's SFR ($\sim$ 0.46 M$_\odot$ yr$^{-1}$) presented in \citealt{lu_census_2019}.

\subsection{Is Star Formation in the Central Molecular Zone Episodic?}

Previous studies of the CMZ SFR using methods including YSO counting, HII region counting, and integrated light measurements find a CMZ SFR with mean value computed by \citet{henshaw_star_2022} of 0.07$^{+0.08}_{-0.02}$. As stated above, this is in line with our fiducial SFR value of $\sim$0.08 M$_\odot$ yr$^{-1}$. However, our fiducial value does not include any ambiguously star-forming sources, and assumes higher dust temperatures than those measured with Herschel, along with a star formation efficiency which is very ill-constrained for sources on these spatial scales. If we include the ambiguously star-forming sources, and a higher star formation efficiency of 0.4, equation \ref{eq:sfr_cloud} yields a much higher star formation rate of 0.45 M$_\odot$ yr$^{-1}$, which we can interpret as an upper limit (excluding the uncertainty from the dust temperatures). If these assumed higher values are reasonable and the majority of the ambiguously star-forming sources do host a high-mass protostellar activity, this implies the star formation rate of the CMZ is increasing. 

There is both theoretical and observational precedent for large fluctuations in the Galactic Center's SFR. Within such a model for the CMZ's evolution, the dearth of recent star formation is the result of episodic fluctuations, being preceded and followed by periods of enhanced star formation \citep[e.g.][]{kruijssen_what_2014, krumholz_dynamical_2015, armillotta_life_2019, orr_fiery_2021}. As described in more detail in Section 3.2 of \citet{henshaw_star_2022}, there is increasing observational evidence to suggest that the SFR of the CMZ has varied significantly on timescales of tens of Myr (given the stellar masses and ages of the Arches and Quintuplet clusters) and on timescales of hundreds of Myr. \citet{lu_star_2019} suggested that the SFR of the CMZ may show signs of imminently increasing from its present value. In recent work analyzing the stellar age distribution of the nuclear stellar disk, \citet{nogueras-lara_early_2020} showed that the SFR in the inner few hundred parsecs has reached higher values of $\sim$0.5 M$_\odot$ yr$^{-1}$ within the last Gyr, similar to the upper limit SFR of 0.45 M$_\odot$ yr$^{-1}$ calculated in this work. 

A recent census of the CMZ's high-mass star formation using VLA C-band observations of ultra-compact HII regions and methanol masers presented in \citealt{lu_census_2019} finds a lower limit SFR of $\sim$0.025 M$_\odot$ yr$^{-1}$ for the Galactic Center. The calculation of this value assumes a characteristic timescale of 0.3 Myr and a total stellar mass extrapolated from the IMF. The discrepancy between this and the value presented in our analysis may be due to a difference in the timescale and phase of star formation traced by both methods (the median freefall time for star-forming sources in the \textit{CMZoom} catalog is ~0.1 Myr). It is likely that much of the star-forming mass we characterize in this work represents more recent and future star formation, not necessarily associated with observed masers and observable ultra-compact HII regions. 

Alternatively, we can calculate a much more conservative estimate of the star formation rate to act as a lower limit on the CMZ's SFR by neglecting the IMF correction and only considering the mass derived from the observed, robustly star-forming objects. This lower limit value still depends on the choice of star formation efficiency and dust temperature, and as such must still be interpreted with caution. Neglecting the IMF correction, and with an assumed dust temperature of 50K and a low star formation efficiency of 10\%, we find a total SFR (excluding the CND) of $\sim$0.05 M$_\odot$ yr$^{-1}$, the majority of which is due to the estimate of the SFR of Sgr B2 ($\sim$0.036 M$_\odot$ yr$^{-1}$, from \citet{ginsburg_distributed_2018}). This lower limit lies much closer to previous estimates of the CMZ's SFR. The most significant sources of uncertainty in the measurements of cloud SFRs remain the dust opacity, dust temperature, and star formation efficiency described in Section \ref{sec:results} and shown in Figure \ref{fig:sfr_bounds}.

\subsection{The Nature of Atoll Structures in Dense Submillimeter Gas}\label{sec:discussion:atoll}

\begin{figure*}[htbp]
\begin{center}
\includegraphics[trim = 0mm 0mm 0mm 0mm,clip, width = .8 \textwidth]{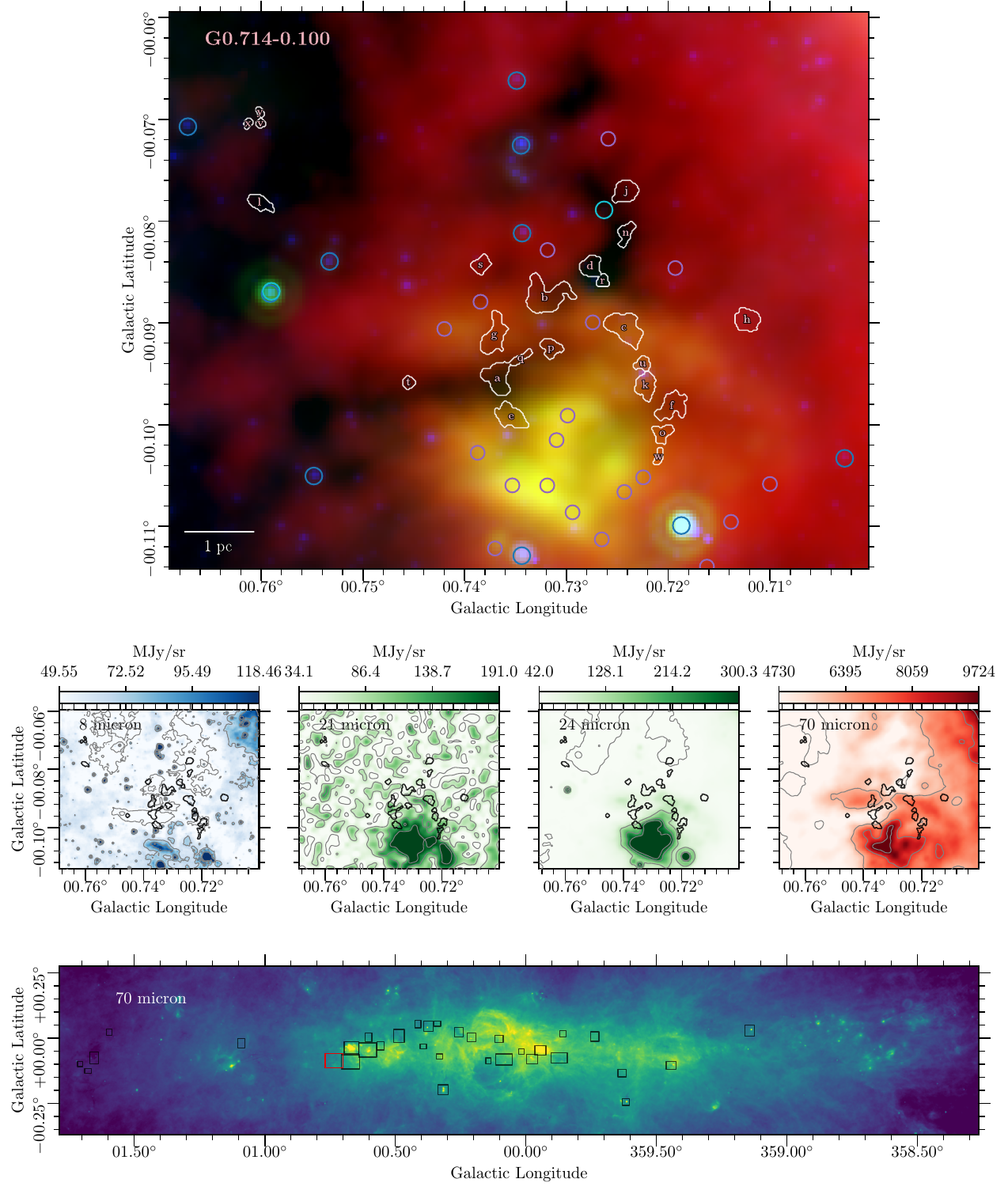}
\end{center}
\caption{An example of the ``atoll'' morphology of dust continuum sources observed by \textit{CMZoom}, shown in the three color images used for the by-eye classification procedure detailed in Section \ref{sec:method:by-eye}. These dust ring configurations are likely to represent remaining gas surrounding young HII regions. The three colors used in the top panel are 8$\mu$m (blue, GLIMPSE, from \citet{benjamin_glimpse_2003}, 24$\mu$m (green, MIPSGAL, from \citealt{carey_mipsgal_2009}), and 70$\mu$m (red, Hi-Gal, from \citet{molinari_hi-gal_2010}). These are each shown individually in the bottom four panels, along with the 21$\mu$m emission from MSX \citep[][]{egan_midcourse_2003}. Overlaid on the composite three-color image are white contours outlining the \textit{CMZoom} leaves, along with cyan circles demarcating YSO candidates from a compilation of those identified by \citet{yusef-zadeh_star_2009}, \citet{an_massive_2011}, and \citet{immer_recent_2012}, purple circles indicating the 70$\mu$m point sources catalogued by \citet{molinari_hi-gal_2010}, and darker blue circles representing the point sources identified by \citet{gutermuth_24_2015}. The radial size of these circles corresponds to the FWHM condition used do determine plausible association with \textit{CMZoom} leaves. The bottom panel shows a 70$\mu$m mosaic of the Galactic Center with each \textit{CMZoom} region shown as black box, with the specific region from the above panels highlighted in red.}
\label{fig:atoll_example}
\end{figure*}

Amidst the wealth of dense gas morphologies observed throughout the sample of CMZ clouds, several regions exhibit a common pattern of dense submillimeter sources clustered surrounding a more evolved source, reminiscent of an atoll of cold gas islands enclosing an far-infrared bright lagoon. These ``atoll sources'' occur in several clouds surveyed by \textit{CMZoom}, an example of which is shown in Figure \ref{fig:atoll_example}. Atoll sources are flagged by-eye, as their morphology varies significantly from region to region. It is not obvious that these sources represent a common physical phenomenon or several different phenomena which appear similar on the plane of the sky. Line-of-sight projection issues rule out any attempt to estimate a completeness to these types of sources, and it is possible that other sources in the catalog would appear as atoll sources from other viewing angles. It is possible that these sources only appear in proximity to their central ``lagoon'' of compact infrared emission due to plane-of-the-sky projection effects.

In several cases where higher-resolution ALMA data are available, these sources appear to be rings of dense gas surrounding an extended HII region, and therefore represent an interesting phase of molecular cloud evolution. At this point, stellar feedback from young stars is actively eroding away their natal envelopes, though it is unclear how long this phase lasts. Parsec-scale bubbles are found in simulations \citep[e.g.][]{rosen_blowing_2021} and observations \citep[e.g.][]{feddersen_expanding_2018} of regions of recent and ongoing star formation, and the lingering presence of a detectable cold dust component to their spectral energy distributions may help to constrain their protostellar age and class. Presently, our criteria for identifying these structures are qualitative and a common physical interpretation cannot be argued here, so we aim to simply note their presence and their potential significance. Future work will investigate the properties of these atoll sources and their central HII region lagoons in the combined context of their continuum and spectral properties, taking into account the submillimeter emission along with radio continuum observations using the VLA.

\section{Summary} \label{sec:summary}

The CMZ hosts a diverse population of giant molecular clouds with different morphologies and fragmented structures, and provides a unique opportunity to study the formation of stars in a more extreme environment than the Galactic Disk. The \textit{CMZoom} survey has allowed us to resolve the sub-parsec substructure of all dense material throughout the CMZ. By comparing the highly-complete catalog of dense submillimeter sources from this survey with previous wide surveys of signatures of high-mass star formation, we are able to characterize all possible sites of ongoing, deeply embedded high-mass star formation in the Galactic Center, excluding Sgr B2 and the circumnuclear disk. In this work, we have cross-referenced the \textit{CMZoom} catalog of submillimeter-bright substructure with catalogs of YSO candidates \citep[][]{yusef-zadeh_star_2009, an_massive_2011, immer_recent_2012} and catalogs of far infrared point sources \citep[][]{gutermuth_24_2015, molinari_hi-gal_2016}. We also catalogued compact emission localized with the \textit{CMZoom} catalog leaves in maps of 8\micron\, and 24\micron\, maps from Spitzer, 21\micron\, maps from MSX, 25 and 37\micron\, maps from SOFIA's FORCAST instrument, and 70\micron\, emission from Herschel by-eye. The key results from our analysis are:
\begin{itemize}
    \item Of the objects catalogued by \textit{CMZoom} (excluding the Sgr B2 complex and the CND region surrounding Sgr A*), 39 show robust signatures of ongoing high-mass star formation (a maser, previously identified YSO, or by-eye identified 24 and/or 70$\mu$m compact source), while 57 others show ambiguous signatures which may or may not be associated with active star formation. The remaining 103 leaves appear presently robustly quiescent, either representing a very early a stage of gravitational collapse without thermal emission in the far-infrared, or representing transient, turbulent density fluctuations in the the CMZ's molecular clouds.
    \item Assuming a range of star formation efficiencies and local dust temperatures, we calculate a cloud-by-cloud SFR for each region in the \textit{CMZoom} footprint, sensitive to all sites of recent ($\lesssim 4\times 10^5$ yr) and incipient high-mass star formation in the CMZ. Combining these cloud star formation rates, we estimate an incipient SFR for the entire CMZ of $\sim 0.08$ M$_\odot$ yr$^{-1}$. We calculate an upper limit CMZ SFR of $\sim 0.45$ M$_\odot$ yr$^{-1}$, including more ambiguously star-forming sources, using lower dust temperature estimates and a higher assumed star formation efficiency. These SFRs are higher than some previous measurements of the CMZ's SFR averaged over longer time scales (considerably higher in the lower dust temperature case), suggesting the possibility of an imminent increase in the Galactic Center's star formation rate.
    \item CMZ sources with higher volume densities are not more likely to host ongoing high-mass star formation. This implies either that denser material is not necessarily more likely to host high-mass star formation, that the densities measured do not accurately represent the densities of the pre-stellar envelope, or that our observational assumptions used to derive a 3D density from the objects' 1.3mm dust continuum emission are significantly flawed. Upcoming work constraining gas densities within the \textit{CMZoom} leaves will help resolve this ambiguity.
    \item The gas column density ratio (N$_{\rm SMA}$(H$_2$)/ N$_{\rm Herschel}$(H$_2)$) tends to be higher in regions where high-mass stars are actively forming, particularly above a threshold of N$_{\rm SMA}$(H$_2$)/N$_{\rm Herschel}$(H$_2)$ $\gtrsim 1.5$. It is unclear if this threshold is dependent on the physical scales probed by these particular observations, and thus it should be interpreted with caution.
    \item We identify the common morphology of ``atoll'' sources, ring-like structures of dense gas surrounding an evolved, far infrared-bright source, resembling a string of islands around a lagoon. Some of these sources correspond to the outskirts of known HII regions. Thus, we suspect these systems represent a particular stage of high-mass star formation, in which the HII region of a newborn high-mass star is in the process of destroying its natal envelope, and the continuum sources detected are the remaining pockets of dense gas which may or may not host ongoing or future star formation. These atoll sources may be valuable targets for follow-up study.  
\end{itemize}

While a great deal of uncertainty remains about the present, past, and future of the CMZ's star formation rate, the expansion and analysis of the \textit{CMZoom} catalog presented in and published with this work constrains both lower and upper limits on the incipient cloud-by-cloud and whole-CMZ star formation properties. Future observational work will aim to characterize the nature of the ambiguously star-forming sources in the catalog, further constrain dust and gas temperatures on more relevant spatial scales, and better resolve the low-mass component of the CMZ's YSO population. Simultaneously, further efforts are likely to allow better estimation of a relevant star formation efficiency. As our understanding improves, the results from this catalog will continue to constrain the CMZ's star formation properties, guiding future observations and interpretation of the Galactic Center's evolution. 

\begin{acknowledgements}
We would like to thank the anonymous reviewer for their insightful comments which substantially improved the quality of this work. HPH's research was supported by an appointment to the NASA Postdoctoral Program administered by Oak Ridge Associated Universities under contract with NASA. HPH was supported by JPL, which is run under contract by California Institute of Technology for NASA. HPH gratefully acknowledges support for this work from the SOFIA Archival Research Program (program ID 09\_0540). HPH also thanks the LSSTC Data Science Fellowship Program, which is funded by LSSTC, NSF Cybertraining grant \#1829740, the Brinson Foundation, and the Moore Foundation; his participation in the program has benefited this work. Additionally, HPH gratefully acknowledges support from the National Science Foundation under Award No. 1816715. EACM gratefully acknowledges support by the National Science Foundation under grant No. AST-1813765.
AG acknowledges support from NSF grants AST 2008101 and CAREER 2142300.
CB gratefully  acknowledges  funding  from  National  Science  Foundation  under  Award  Nos. 1816715, 2108938, 2206510, and CAREER 2145689, as well as from the National Aeronatics and Space Administration through the Astrophysics Data Analaysis Program under Award No. 21-ADAP21-0179 and through the SOFIA archival research program under Award No.  09$\_$0540.  
ATB would like to acknowledge funding from the European Research Council (ERC) under the European Union’s Horizon 2020 research and innovation programme (grant agreement No.726384/Empire).
JMDK gratefully acknowledges funding from the European Research Council (ERC) under the European Union's Horizon 2020 research and innovation programme via the ERC Starting Grant MUSTANG (grant agreement number 714907). COOL Research DAO is a Decentralized Autonomous Organization supporting research in astrophysics aimed at uncovering our cosmic origins.
LCH was supported by the National Science Foundation of China (11721303, 11991052, 12011540375, 12233001) and the China Manned Space Project (CMS-CSST-2021-A04, CMS-CSST-2021-A06).
Herschel is an ESA space observatory with science instruments provided by European-led Principal Investigator consortia and with important participation from NASA.
\software{This research made significant use of Astropy \citep{collaboration_astropy_2022}
This research also utilized astrodendro, a Python package to compute dendrograms of Astronomical data (\href{http://www.dendrograms.org/}{http://www.dendrograms.org/}) as well as SAO Image DS9 \citep{joye_new_2003}, as well as the python module IMF created by Adam Ginsburg (\href{https://github.com/keflavich/imf}{https://github.com/keflavich/imf}). }
\end{acknowledgements}

\bibliography{cmzoom_lib_8-23.bib}{}
\bibliographystyle{aasjournal}

\appendix

\section{Gallery of Three-color Images Used in Star Formation Tracer Search }\label{app:rgb_gallery}

This appendix contains the complete gallery of three-color images used to assess the star formation status of each leaf in the \textit{CMZoom} robust catalog in Figures 10 - 33. The red colorscale shows the 70$\mu$m emission map from Herschel \citep{molinari_hi-gal_2010,molinari_source_2011}. The green colorscale shows the combined 21 and 24$\mu$m emission from MSX and Spitzer respectively \citep{benjamin_glimpse_2003}. The blue colorscale shows the 8$\mu$m map from Spitzer \citep{benjamin_glimpse_2003}. Each color is shown in logscale, scaled between a local 5\% and 95\% boundary value for each box. Overlaid on the composite three-color image are white contours representing the \textit{CMZoom} leaves, along with cyan circles representing YSO candidates from a compilation of those identified by \citealt{yusef-zadeh_star_2009}, \citealt{an_massive_2011}, and \citealt{immer_recent_2012}, purple circles representing the 70$\mu$m point sources cataloged by \citealt{molinari_hi-gal_2016}, and darker blue circles representing the point sources identified by \citealt{gutermuth_24_2015}. The radial size of these circles corresponds to the FWHM condition used to determine plausible association with nearby \textit{CMZoom} leaves. Region numbers used in this section do not correspond to the mask numbers referenced in the tables throughout the rest of this work.

\begin{figure*}
\begin{center}
\includegraphics[trim = 0mm 0mm 0mm 0mm, clip, width = .90 \textwidth]{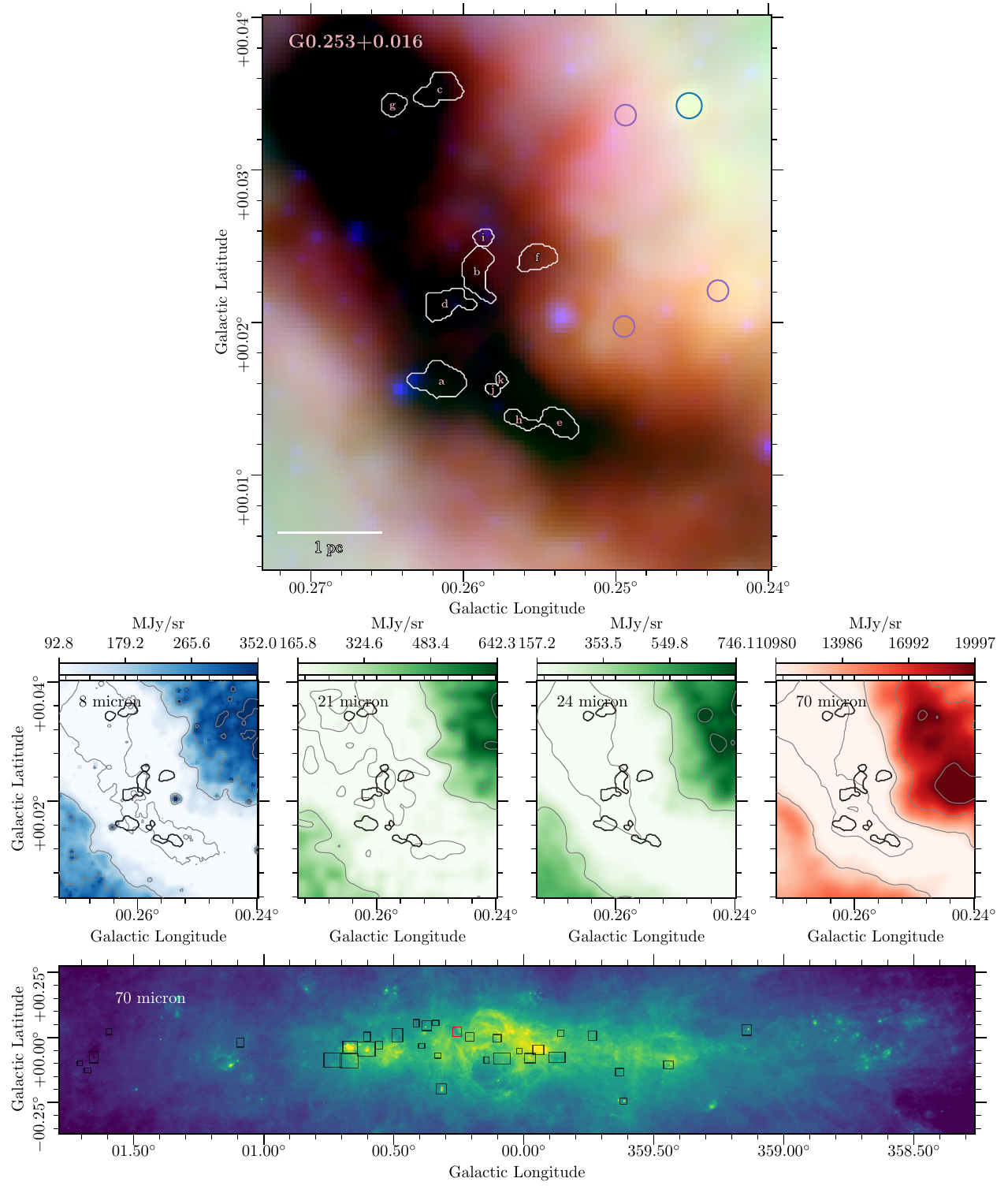}\label{fig:rgb_1}
\end{center}
\caption{Region 1, Cloud ID G0.253+0.016, also known as The Brick. The red colorscale shows the 70$\mu$m emission map from Herschel \citep{molinari_hi-gal_2010,molinari_source_2011}. The green colorscale shows the combined 21 and 24$\mu$m emission from MSX and Spitzer respectively \citep{benjamin_glimpse_2003}. The blue colorscale shows the 8$\mu$m map from Spitzer \citep{benjamin_glimpse_2003}. Each color is shown in logscale, scaled between a local 5\% and 95\% boundary value for each box. Overlaid on the composite three-color image are white contours representing the \textit{CMZoom} leaves, along with cyan circles representing YSO candidates from a compilation of those identified by \citealt{yusef-zadeh_star_2009}, \citealt{an_massive_2011}, and \citealt{immer_recent_2012}, purple circles representing the 70$\mu$m point sources cataloged by \citealt{molinari_hi-gal_2016}, and darker blue circles representing the point sources identified by \citealt{gutermuth_24_2015}. The radial size of these circles corresponds to the FWHM condition used to determine plausible association with nearby \textit{CMZoom} leaves.}
\end{figure*}

\begin{figure*}
\begin{center}
\includegraphics[trim = 0mm 0mm 0mm 0mm, clip, width = .90 \textwidth]{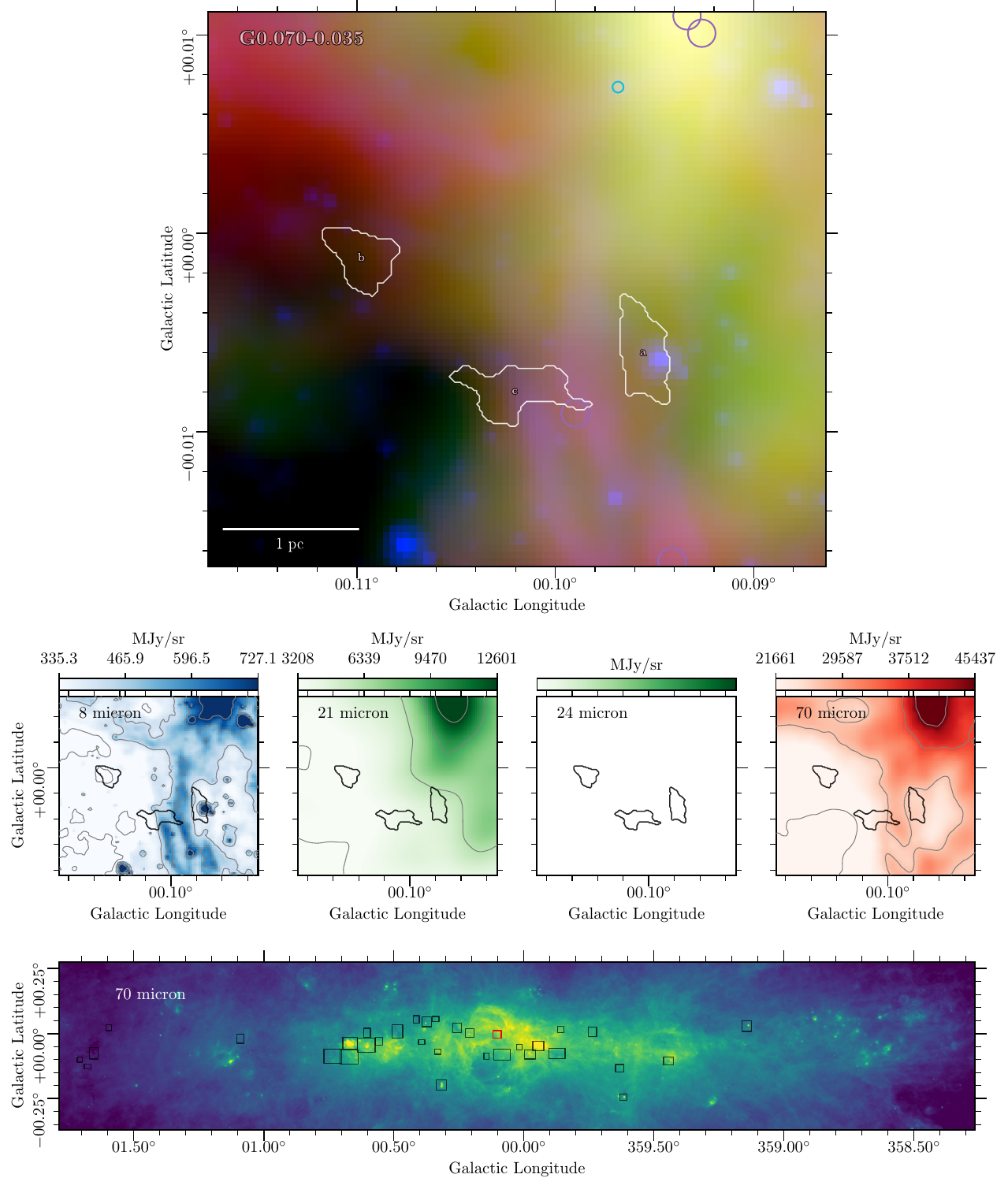}
\end{center}
\caption{Region 2, Cloud ID G0.070-0.035. The red colorscale shows the 70$\mu$m emission map from Herschel \citep{molinari_hi-gal_2010,molinari_source_2011}. The green colorscale shows the combined 21 and 24$\mu$m emission from MSX and Spitzer respectively \citep{benjamin_glimpse_2003}. The blue colorscale shows the 8$\mu$m map from Spitzer \citep{benjamin_glimpse_2003}. Each color is shown in logscale, scaled between a local 5\% and 95\% boundary value for each box. Overlaid on the composite three-color image are white contours representing the \textit{CMZoom} leaves, along with cyan circles representing YSO candidates from a compilation of those identified by \citealt{yusef-zadeh_star_2009}, \citealt{an_massive_2011}, and \citealt{immer_recent_2012}, purple circles representing the 70$\mu$m point sources cataloged by \citealt{molinari_hi-gal_2016}, and darker blue circles representing the point sources identified by \citealt{gutermuth_24_2015}. The radial size of these circles corresponds to the FWHM condition used to determine plausible association with nearby \textit{CMZoom} leaves.}
\label{fig:rgb_2}
\end{figure*}

\begin{figure*}
\begin{center}
\includegraphics[trim = 0mm 0mm 0mm 0mm, clip, width = .90 \textwidth]{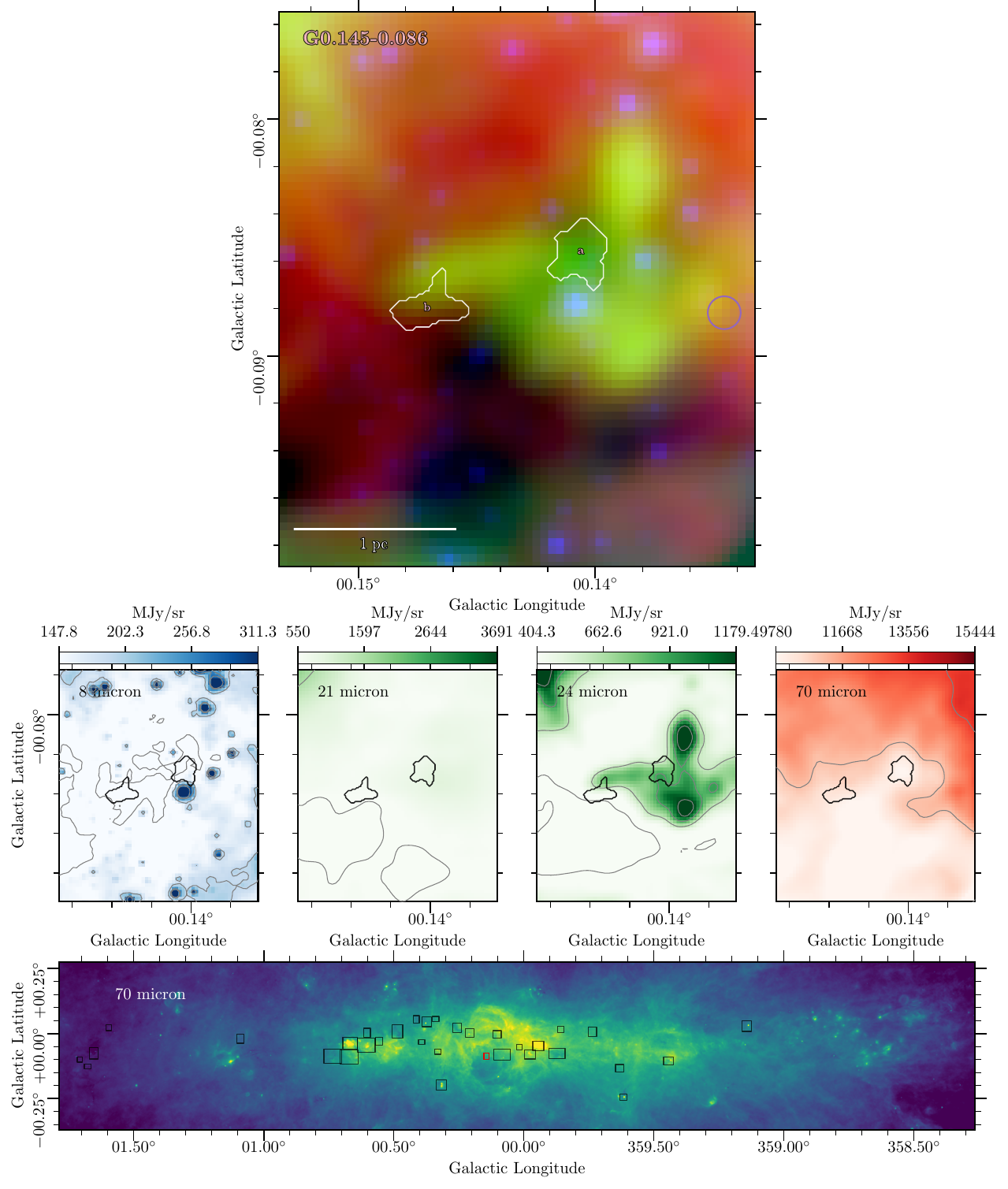}
\end{center}
\caption{Region 3, Cloud ID G0.145-0.086, also known as the Straw Cloud. The red colorscale shows the 70$\mu$m emission map from Herschel \citep{molinari_hi-gal_2010,molinari_source_2011}. The green colorscale shows the combined 21 and 24$\mu$m emission from MSX and Spitzer respectively \citep{benjamin_glimpse_2003}. The blue colorscale shows the 8$\mu$m map from Spitzer \citep{benjamin_glimpse_2003}. Each color is shown in logscale, scaled between a local 5\% and 95\% boundary value for each box. Overlaid on the composite three-color image are white contours representing the \textit{CMZoom} leaves, along with cyan circles representing YSO candidates from a compilation of those identified by \citealt{yusef-zadeh_star_2009}, \citealt{an_massive_2011}, and \citealt{immer_recent_2012}, purple circles representing the 70$\mu$m point sources cataloged by \citealt{molinari_hi-gal_2016}, and darker blue circles representing the point sources identified by \citealt{gutermuth_24_2015}. The radial size of these circles corresponds to the FWHM condition used to determine plausible association with nearby \textit{CMZoom} leaves.}
\label{fig:rgb_3}
\end{figure*}

\begin{figure*}
\begin{center}
\includegraphics[trim = 0mm 0mm 0mm 0mm, clip, width = .90 \textwidth]{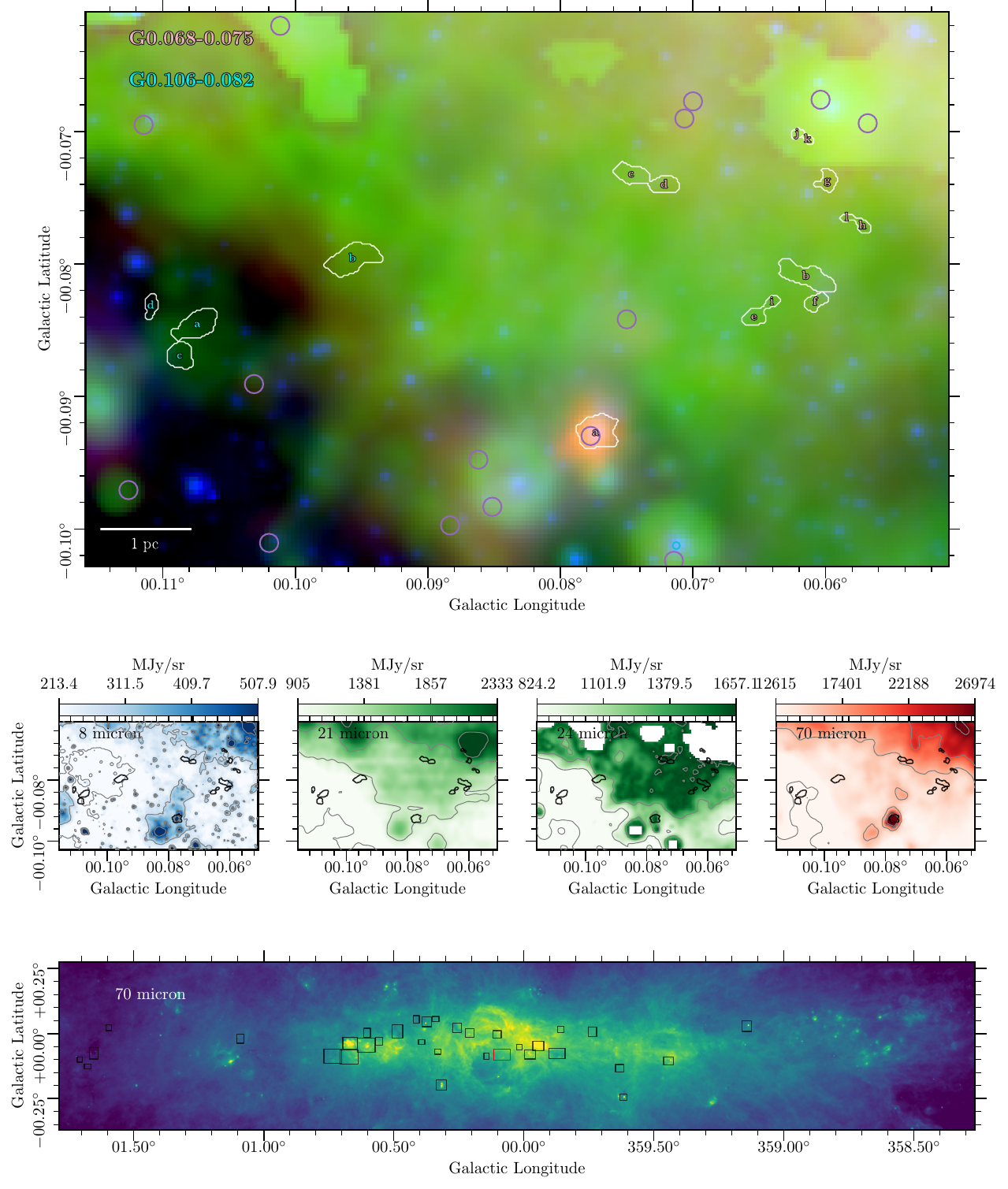}
\end{center}
\caption{Region 4, including Cloud IDs G0.068-0.075 and G0.106-0.082, also known as the Sticks Cloud and the Stone Clouds respectively. The red colorscale shows the 70$\mu$m emission map from Herschel \citep{molinari_hi-gal_2010,molinari_source_2011}. The green colorscale shows the combined 21 and 24$\mu$m emission from MSX and Spitzer respectively \citep{benjamin_glimpse_2003}. The blue colorscale shows the 8$\mu$m map from Spitzer \citep{benjamin_glimpse_2003}. Each color is shown in logscale, scaled between a local 5\% and 95\% boundary value for each box. Overlaid on the composite three-color image are white contours representing the \textit{CMZoom} leaves, along with cyan circles representing YSO candidates from a compilation of those identified by \citealt{yusef-zadeh_star_2009}, \citealt{an_massive_2011}, and \citealt{immer_recent_2012}, purple circles representing the 70$\mu$m point sources cataloged by \citealt{molinari_hi-gal_2016}, and darker blue circles representing the point sources identified by \citealt{gutermuth_24_2015}. The radial size of these circles corresponds to the FWHM condition used to determine plausible association with nearby \textit{CMZoom} leaves.}
\label{fig:rgb_4}
\end{figure*}

\begin{figure*}
\begin{center}
\includegraphics[trim = 0mm 0mm 0mm 0mm, clip, width = .90 \textwidth]{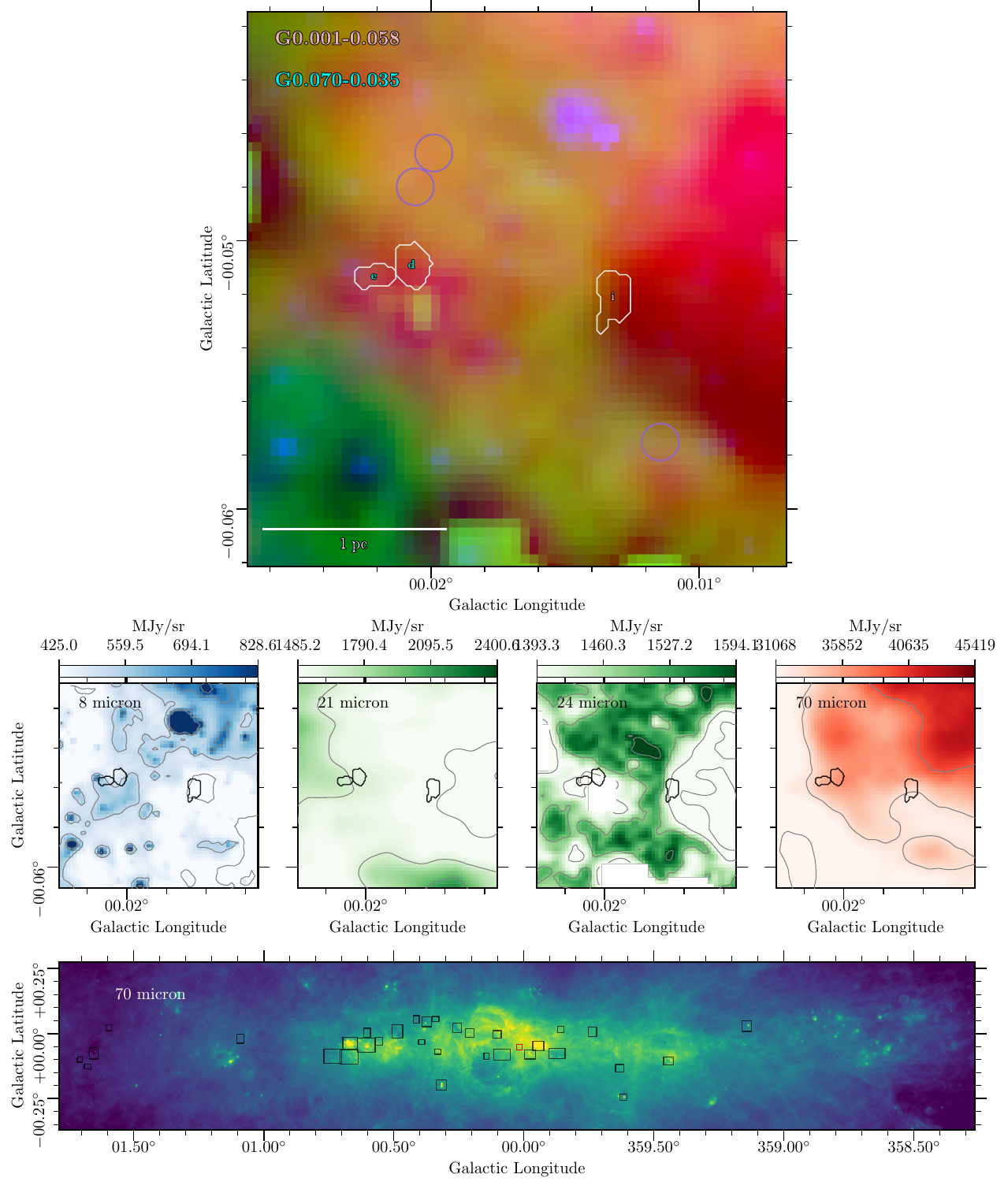}
\end{center}
\caption{Region 5, Cloud ID G0.070-0.035 and G0.001-0.058 (also known as the 20 km s$^{-1}$ Cloud). The red colorscale shows the 70$\mu$m emission map from Herschel \citep{molinari_hi-gal_2010,molinari_source_2011}. The green colorscale shows the combined 21 and 24$\mu$m emission from MSX and Spitzer respectively \citep{benjamin_glimpse_2003}. The blue colorscale shows the 8$\mu$m map from Spitzer \citep{benjamin_glimpse_2003}. Each color is shown in logscale, scaled between a local 5\% and 95\% boundary value for each box. Overlaid on the composite three-color image are white contours representing the \textit{CMZoom} leaves, along with cyan circles representing YSO candidates from a compilation of those identified by \citealt{yusef-zadeh_star_2009}, \citealt{an_massive_2011}, and \citealt{immer_recent_2012}, purple circles representing the 70$\mu$m point sources cataloged by \citealt{molinari_hi-gal_2016}, and darker blue circles representing the point sources identified by \citealt{gutermuth_24_2015}. The radial size of these circles corresponds to the FWHM condition used to determine plausible association with nearby \textit{CMZoom} leaves.}
\label{fig:rgb_5}
\end{figure*}

\begin{figure*}
\begin{center}
\includegraphics[trim = 0mm 0mm 0mm 0mm, clip, width = .90 \textwidth]{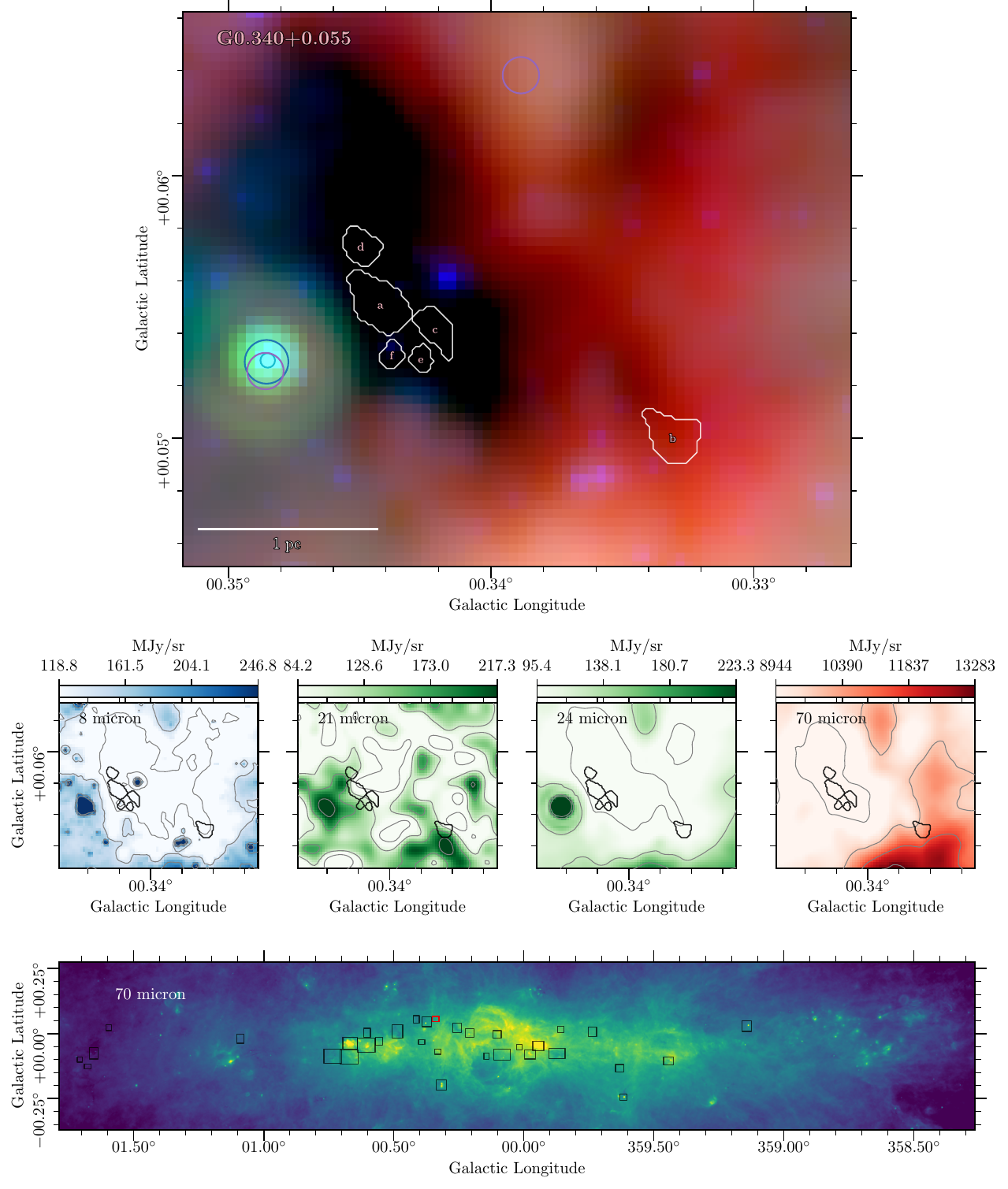}
\end{center}
\caption{Region 6, Cloud ID G0.340+0.055, also known as Dust Ridge Cloud B. The red colorscale shows the 70$\mu$m emission map from Herschel \citep{molinari_hi-gal_2010,molinari_source_2011}. The green colorscale shows the combined 21 and 24$\mu$m emission from MSX and Spitzer respectively \citep{benjamin_glimpse_2003}. The blue colorscale shows the 8$\mu$m map from Spitzer \citep{benjamin_glimpse_2003}. Each color is shown in logscale, scaled between a local 5\% and 95\% boundary value for each box. Overlaid on the composite three-color image are white contours representing the \textit{CMZoom} leaves, along with cyan circles representing YSO candidates from a compilation of those identified by \citealt{yusef-zadeh_star_2009}, \citealt{an_massive_2011}, and \citealt{immer_recent_2012}, purple circles representing the 70$\mu$m point sources cataloged by \citealt{molinari_hi-gal_2016}, and darker blue circles representing the point sources identified by \citealt{gutermuth_24_2015}. The radial size of these circles corresponds to the FWHM condition used to determine plausible association with nearby \textit{CMZoom} leaves.}
\label{fig:rgb_6}
\end{figure*}

\begin{figure*}
\begin{center}
\includegraphics[trim = 0mm 0mm 0mm 0mm, clip, width = .90 \textwidth]{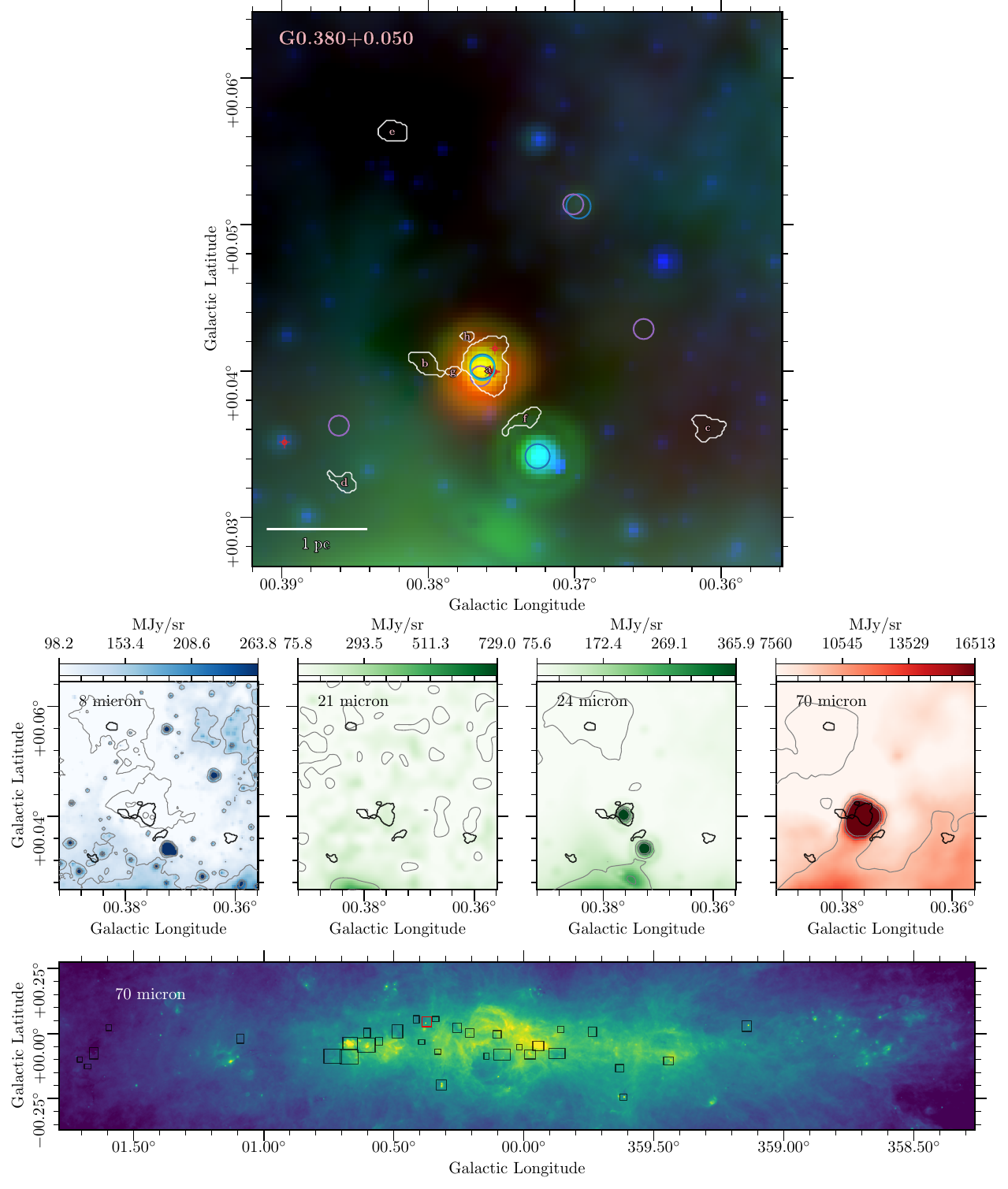}
\end{center}
\caption{Region 7, Cloud ID G0.380+0.050, also known as Dust Ridge Cloud C. The red colorscale shows the 70$\mu$m emission map from Herschel \citep{molinari_hi-gal_2010,molinari_source_2011}. The green colorscale shows the combined 21 and 24$\mu$m emission from MSX and Spitzer respectively \citep{benjamin_glimpse_2003}. The blue colorscale shows the 8$\mu$m map from Spitzer \citep{benjamin_glimpse_2003}. Each color is shown in logscale, scaled between a local 5\% and 95\% boundary value for each box. Overlaid on the composite three-color image are white contours representing the \textit{CMZoom} leaves, along with cyan circles representing YSO candidates from a compilation of those identified by \citealt{yusef-zadeh_star_2009}, \citealt{an_massive_2011}, and \citealt{immer_recent_2012}, purple circles representing the 70$\mu$m point sources cataloged by \citealt{molinari_hi-gal_2016}, and darker blue circles representing the point sources identified by \citealt{gutermuth_24_2015}. The radial size of these circles corresponds to the FWHM condition used to determine plausible association with nearby \textit{CMZoom} leaves.}
\label{fig:rgb_7}
\end{figure*}

\begin{figure*}
\begin{center}
\includegraphics[trim = 0mm 0mm 0mm 0mm, clip, width = .90 \textwidth]{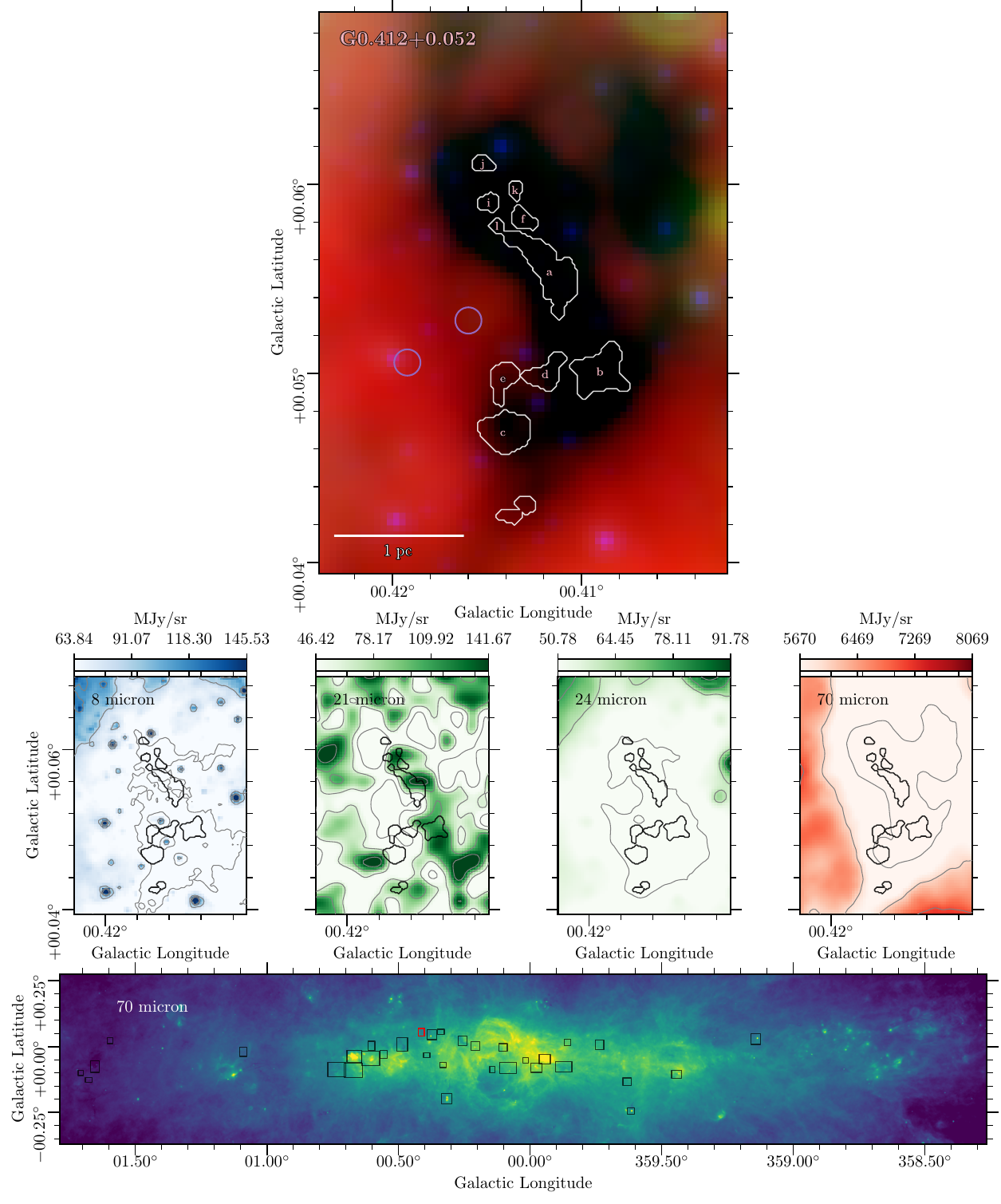}
\end{center}
\caption{Region 8, Cloud ID G0.412+0.052, also known as Dust Ridge Cloud D. The red colorscale shows the 70$\mu$m emission map from Herschel \citep{molinari_hi-gal_2010,molinari_source_2011}. The green colorscale shows the combined 21 and 24$\mu$m emission from MSX and Spitzer respectively \citep{benjamin_glimpse_2003}. The blue colorscale shows the 8$\mu$m map from Spitzer \citep{benjamin_glimpse_2003}. Each color is shown in logscale, scaled between a local 5\% and 95\% boundary value for each box. Overlaid on the composite three-color image are white contours representing the \textit{CMZoom} leaves, along with cyan circles representing YSO candidates from a compilation of those identified by \citealt{yusef-zadeh_star_2009}, \citealt{an_massive_2011}, and \citealt{immer_recent_2012}, purple circles representing the 70$\mu$m point sources cataloged by \citealt{molinari_hi-gal_2016}, and darker blue circles representing the point sources identified by \citealt{gutermuth_24_2015}. The radial size of these circles corresponds to the FWHM condition used to determine plausible association with nearby \textit{CMZoom} leaves.}
\label{fig:rgb_8}
\end{figure*}

\begin{figure*}
\begin{center}
\includegraphics[trim = 0mm 0mm 0mm 0mm, clip, width = .90 \textwidth]{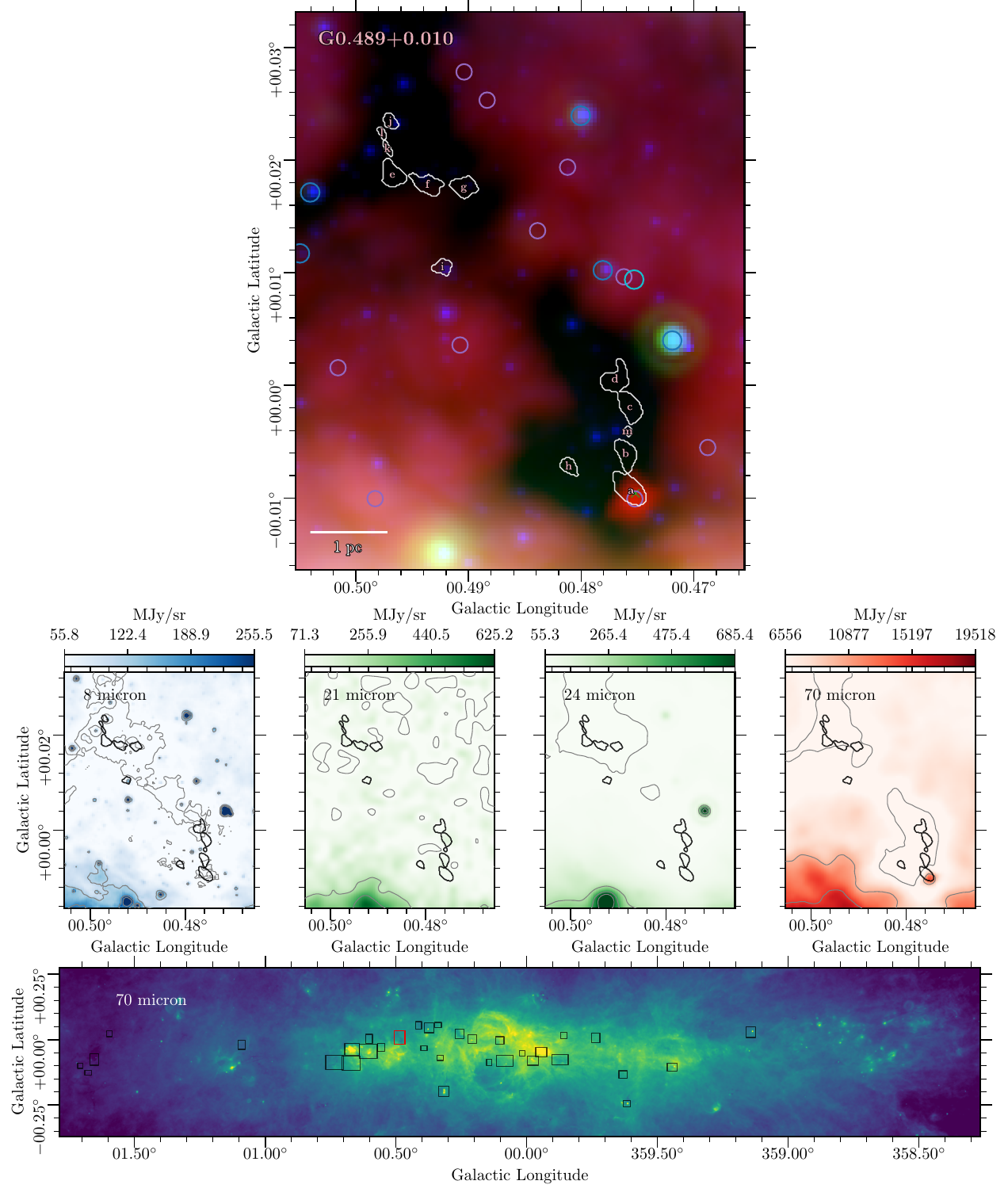}
\end{center}
\caption{Region 9, Cloud ID G0.489+0.010, otherwise known as Dust Ridge Clouds E and F. The red colorscale shows the 70$\mu$m emission map from Herschel \citep{molinari_hi-gal_2010,molinari_source_2011}. The green colorscale shows the combined 21 and 24$\mu$m emission from MSX and Spitzer respectively \citep{benjamin_glimpse_2003}. The blue colorscale shows the 8$\mu$m map from Spitzer \citep{benjamin_glimpse_2003}. Each color is shown in logscale, scaled between a local 5\% and 95\% boundary value for each box. Overlaid on the composite three-color image are white contours representing the \textit{CMZoom} leaves, along with cyan circles representing YSO candidates from a compilation of those identified by \citealt{yusef-zadeh_star_2009}, \citealt{an_massive_2011}, and \citealt{immer_recent_2012}, purple circles representing the 70$\mu$m point sources cataloged by \citealt{molinari_hi-gal_2016}, and darker blue circles representing the point sources identified by \citealt{gutermuth_24_2015}. The radial size of these circles corresponds to the FWHM condition used to determine plausible association with nearby \textit{CMZoom} leaves.}
\label{fig:rgb_9}
\end{figure*}

\begin{figure*}
\begin{center}
\includegraphics[trim = 0mm 0mm 0mm 0mm, clip, width = .90 \textwidth]{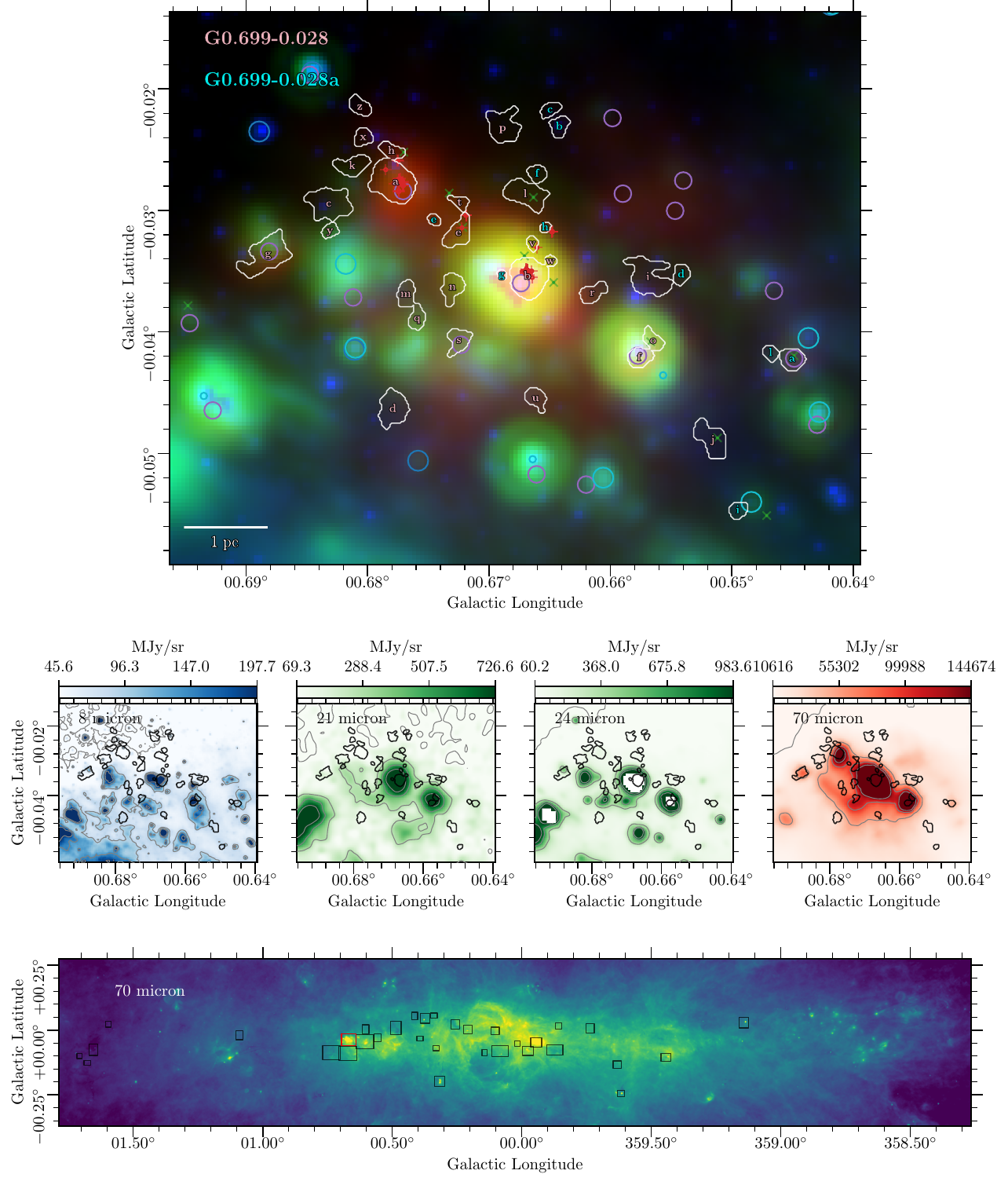}
\end{center}
\caption{Region 10, Cloud ID G0.699-0.028, also known as Sgr B2. The red colorscale shows the 70$\mu$m emission map from Herschel \citep{molinari_hi-gal_2010,molinari_source_2011}. The green colorscale shows the combined 21 and 24$\mu$m emission from MSX and Spitzer respectively \citep{benjamin_glimpse_2003}. The blue colorscale shows the 8$\mu$m map from Spitzer \citep{benjamin_glimpse_2003}. Each color is shown in logscale, scaled between a local 5\% and 95\% boundary value for each box. Overlaid on the composite three-color image are white contours representing the \textit{CMZoom} leaves, along with cyan circles representing YSO candidates from a compilation of those identified by \citealt{yusef-zadeh_star_2009}, \citealt{an_massive_2011}, and \citealt{immer_recent_2012}, purple circles representing the 70$\mu$m point sources cataloged by \citealt{molinari_hi-gal_2016}, and darker blue circles representing the point sources identified by \citealt{gutermuth_24_2015}. The radial size of these circles corresponds to the FWHM condition used to determine plausible association with nearby \textit{CMZoom} leaves.}
\label{fig:rgb_10}
\end{figure*}

\begin{figure*}
\begin{center}
\includegraphics[trim = 0mm 0mm 0mm 0mm, clip, width = .90 \textwidth]{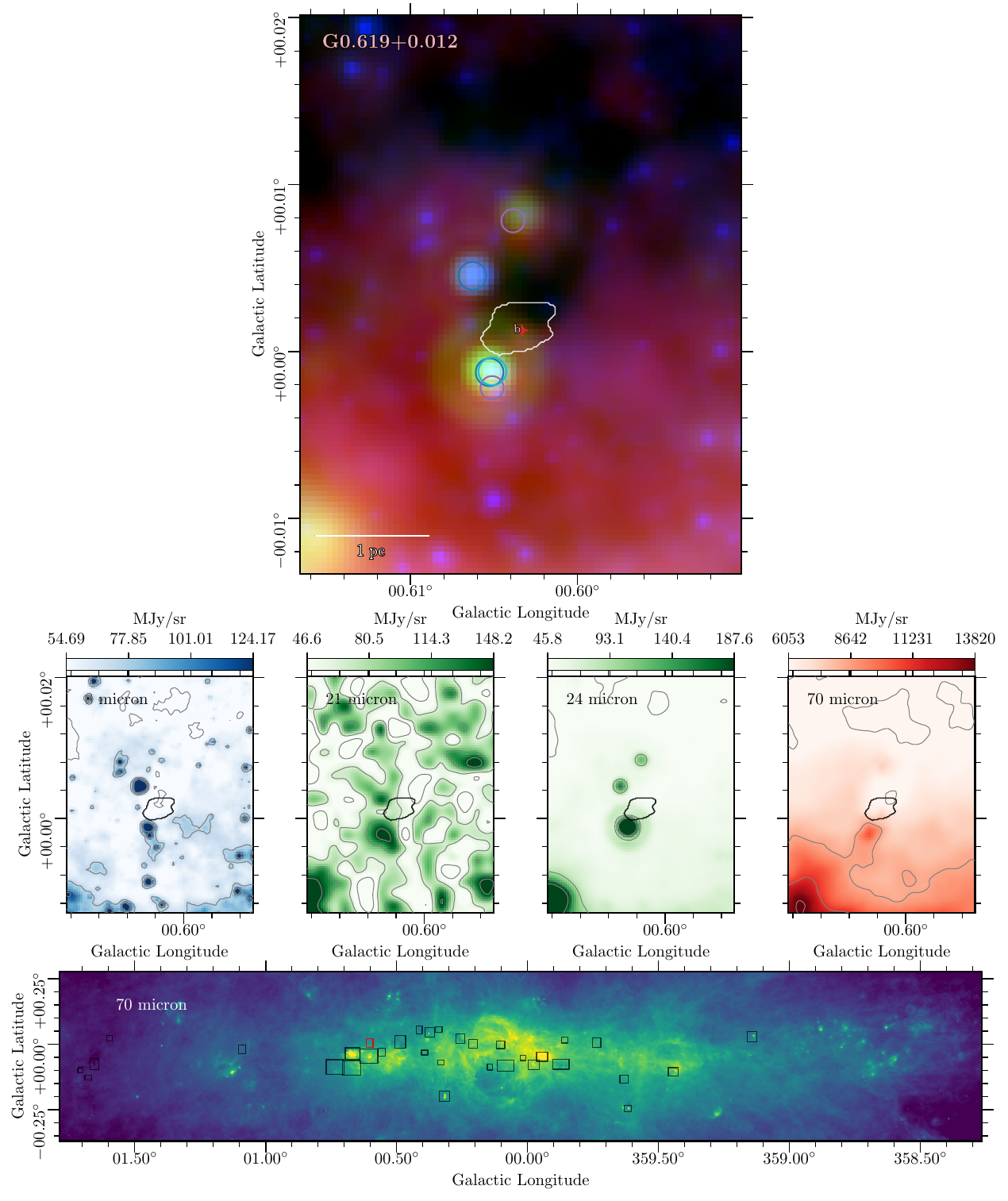}
\end{center}
\caption{Region 11, Cloud ID G0.619+0.012, also known as Sgr B2 SE. The red colorscale shows the 70$\mu$m emission map from Herschel \citep{molinari_hi-gal_2010,molinari_source_2011}. The green colorscale shows the combined 21 and 24$\mu$m emission from MSX and Spitzer respectively \citep{benjamin_glimpse_2003}. The blue colorscale shows the 8$\mu$m map from Spitzer \citep{benjamin_glimpse_2003}. Each color is shown in logscale, scaled between a local 5\% and 95\% boundary value for each box. Overlaid on the composite three-color image are white contours representing the \textit{CMZoom} leaves, along with cyan circles representing YSO candidates from a compilation of those identified by \citealt{yusef-zadeh_star_2009}, \citealt{an_massive_2011}, and \citealt{immer_recent_2012}, purple circles representing the 70$\mu$m point sources cataloged by \citealt{molinari_hi-gal_2016}, and darker blue circles representing the point sources identified by \citealt{gutermuth_24_2015}. The radial size of these circles corresponds to the FWHM condition used to determine plausible association with nearby \textit{CMZoom} leaves.}
\label{fig:rgb_11}
\end{figure*}

\begin{figure*}
\begin{center}
\includegraphics[trim = 0mm 0mm 0mm 0mm, clip, width = .90 \textwidth]{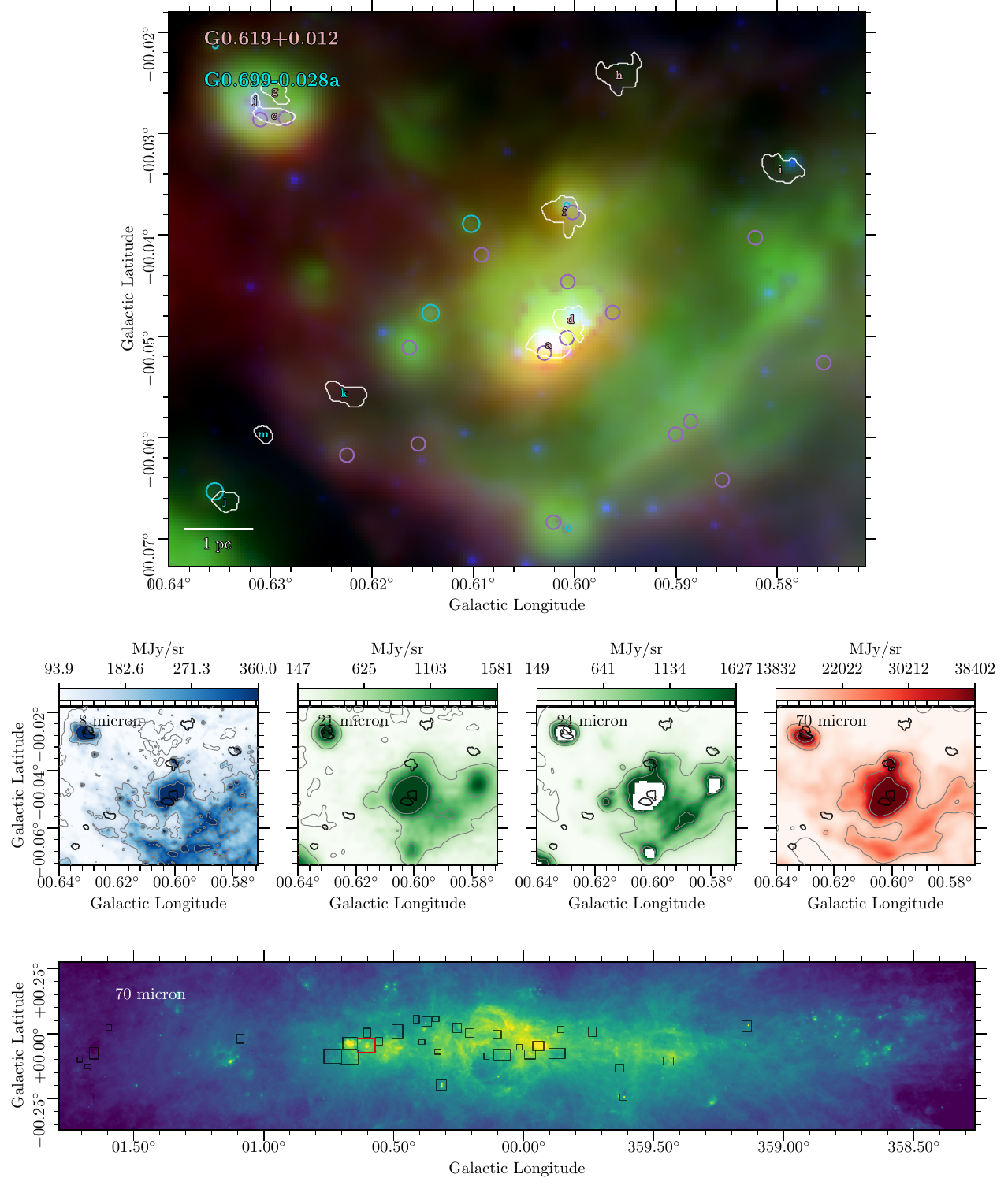}
\end{center}
\caption{Region 12, Cloud IDs G0.699-0.028 and G0.619+0.012, also known as Sgr B2 and Sgr B2 SE respectively. The red colorscale shows the 70$\mu$m emission map from Herschel \citep{molinari_hi-gal_2010,molinari_source_2011}. The green colorscale shows the combined 21 and 24$\mu$m emission from MSX and Spitzer respectively \citep{benjamin_glimpse_2003}. The blue colorscale shows the 8$\mu$m map from Spitzer \citep{benjamin_glimpse_2003}. Each color is shown in logscale, scaled between a local 5\% and 95\% boundary value for each box. Overlaid on the composite three-color image are white contours representing the \textit{CMZoom} leaves, along with cyan circles representing YSO candidates from a compilation of those identified by \citealt{yusef-zadeh_star_2009}, \citealt{an_massive_2011}, and \citealt{immer_recent_2012}, purple circles representing the 70$\mu$m point sources cataloged by \citealt{molinari_hi-gal_2016}, and darker blue circles representing the point sources identified by \citealt{gutermuth_24_2015}. The radial size of these circles corresponds to the FWHM condition used to determine plausible association with nearby \textit{CMZoom} leaves.}
\label{fig:rgb_12}
\end{figure*}

\begin{figure*}
\begin{center}
\includegraphics[trim = 0mm 0mm 0mm 0mm, clip, width = .90 \textwidth]{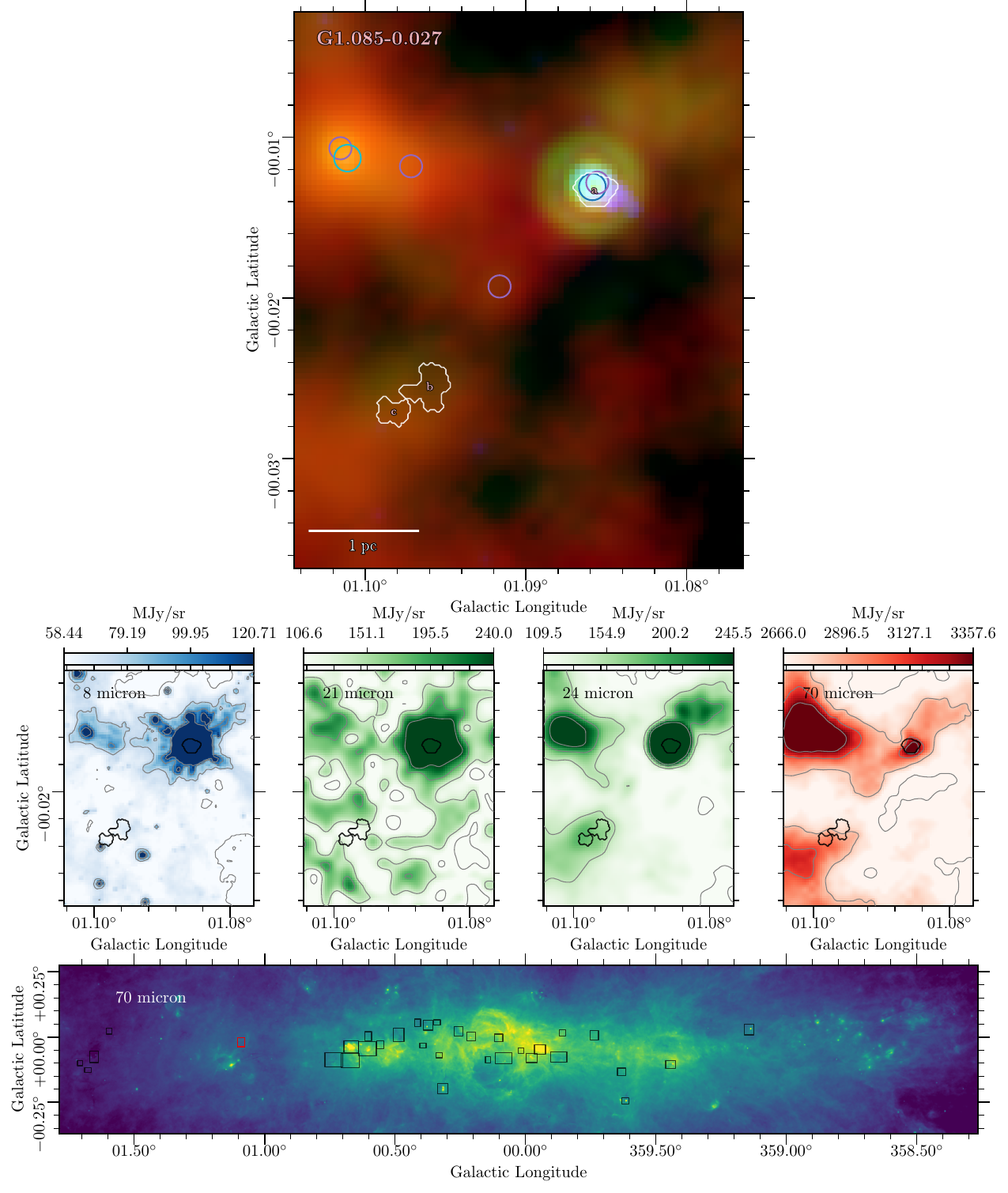}
\end{center}
\caption{Region 31, Cloud ID G1.085-0.027. The red colorscale shows the 70$\mu$m emission map from Herschel \citep{molinari_hi-gal_2010,molinari_source_2011}. The green colorscale shows the combined 21 and 24$\mu$m emission from MSX and Spitzer respectively \citep{benjamin_glimpse_2003}. The blue colorscale shows the 8$\mu$m map from Spitzer \citep{benjamin_glimpse_2003}. Each color is shown in logscale, scaled between a local 5\% and 95\% boundary value for each box. Overlaid on the composite three-color image are white contours representing the \textit{CMZoom} leaves, along with cyan circles representing YSO candidates from a compilation of those identified by \citealt{yusef-zadeh_star_2009}, \citealt{an_massive_2011}, and \citealt{immer_recent_2012}, purple circles representing the 70$\mu$m point sources cataloged by \citealt{molinari_hi-gal_2016}, and darker blue circles representing the point sources identified by \citealt{gutermuth_24_2015}. The radial size of these circles corresponds to the FWHM condition used to determine plausible association with nearby \textit{CMZoom} leaves. }
\label{fig:rgb_13}
\end{figure*}

\begin{figure*}
\begin{center}
\includegraphics[trim = 0mm 0mm 0mm 0mm, clip, width = .90 \textwidth]{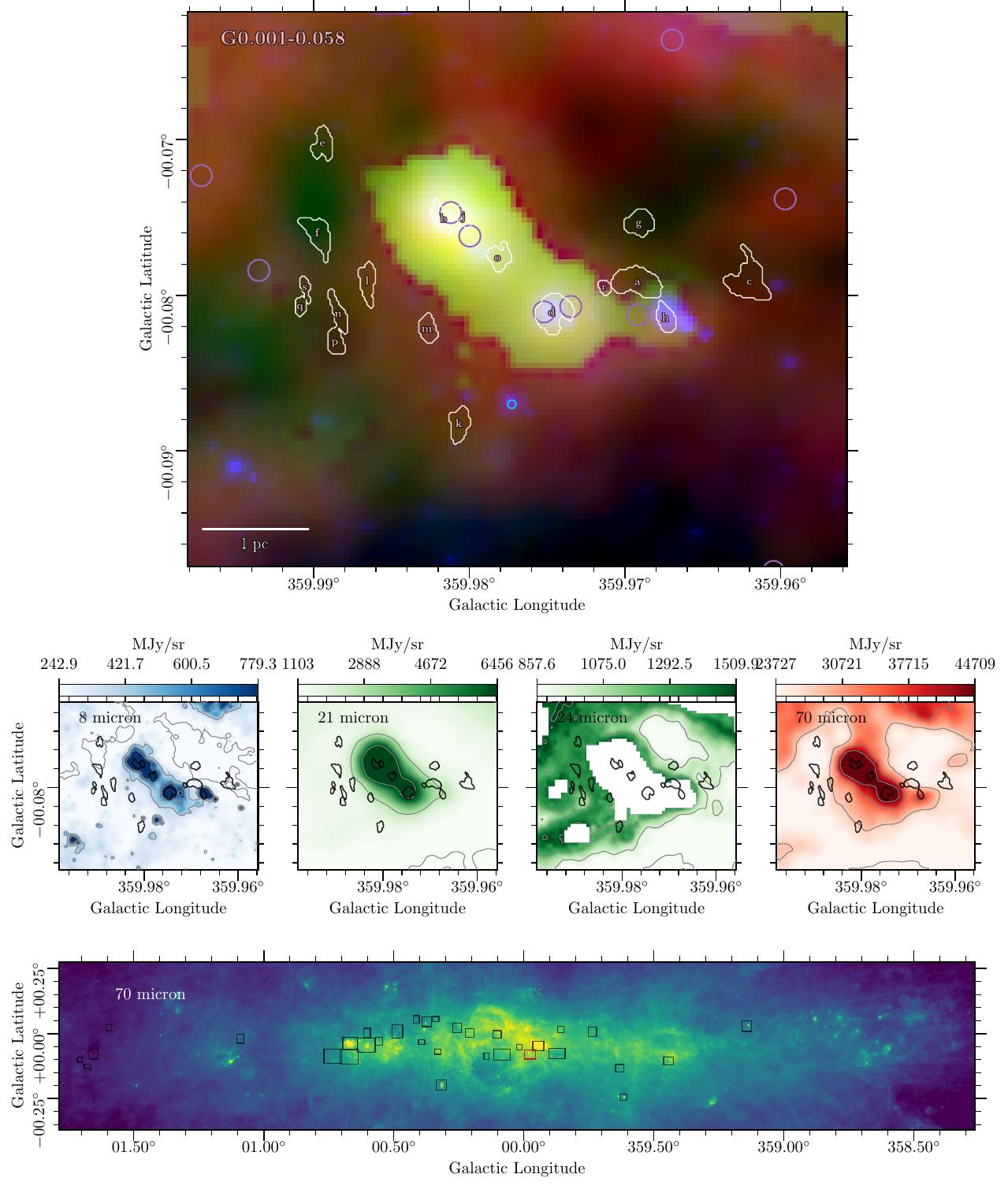}
\end{center}
\caption{Region 14, Cloud ID G0.001-0.058, otherwise known as the 20 km s$^{-1}$ Cloud. The red colorscale shows the 70$\mu$m emission map from Herschel \citep{molinari_hi-gal_2010,molinari_source_2011}. The green colorscale shows the combined 21 and 24$\mu$m emission from MSX and Spitzer respectively \citep{benjamin_glimpse_2003}. The blue colorscale shows the 8$\mu$m map from Spitzer \citep{benjamin_glimpse_2003}. Each color is shown in logscale, scaled between a local 5\% and 95\% boundary value for each box. Overlaid on the composite three-color image are white contours representing the \textit{CMZoom} leaves, along with cyan circles representing YSO candidates from a compilation of those identified by \citealt{yusef-zadeh_star_2009}, \citealt{an_massive_2011}, and \citealt{immer_recent_2012}, purple circles representing the 70$\mu$m point sources cataloged by \citealt{molinari_hi-gal_2016}, and darker blue circles representing the point sources identified by \citealt{gutermuth_24_2015}. The radial size of these circles corresponds to the FWHM condition used to determine plausible association with nearby \textit{CMZoom} leaves.}
\label{fig:rgb_14}
\end{figure*}

\begin{figure*}
\begin{center}
\includegraphics[trim = 0mm 0mm 0mm 0mm, clip, width = .90 \textwidth]{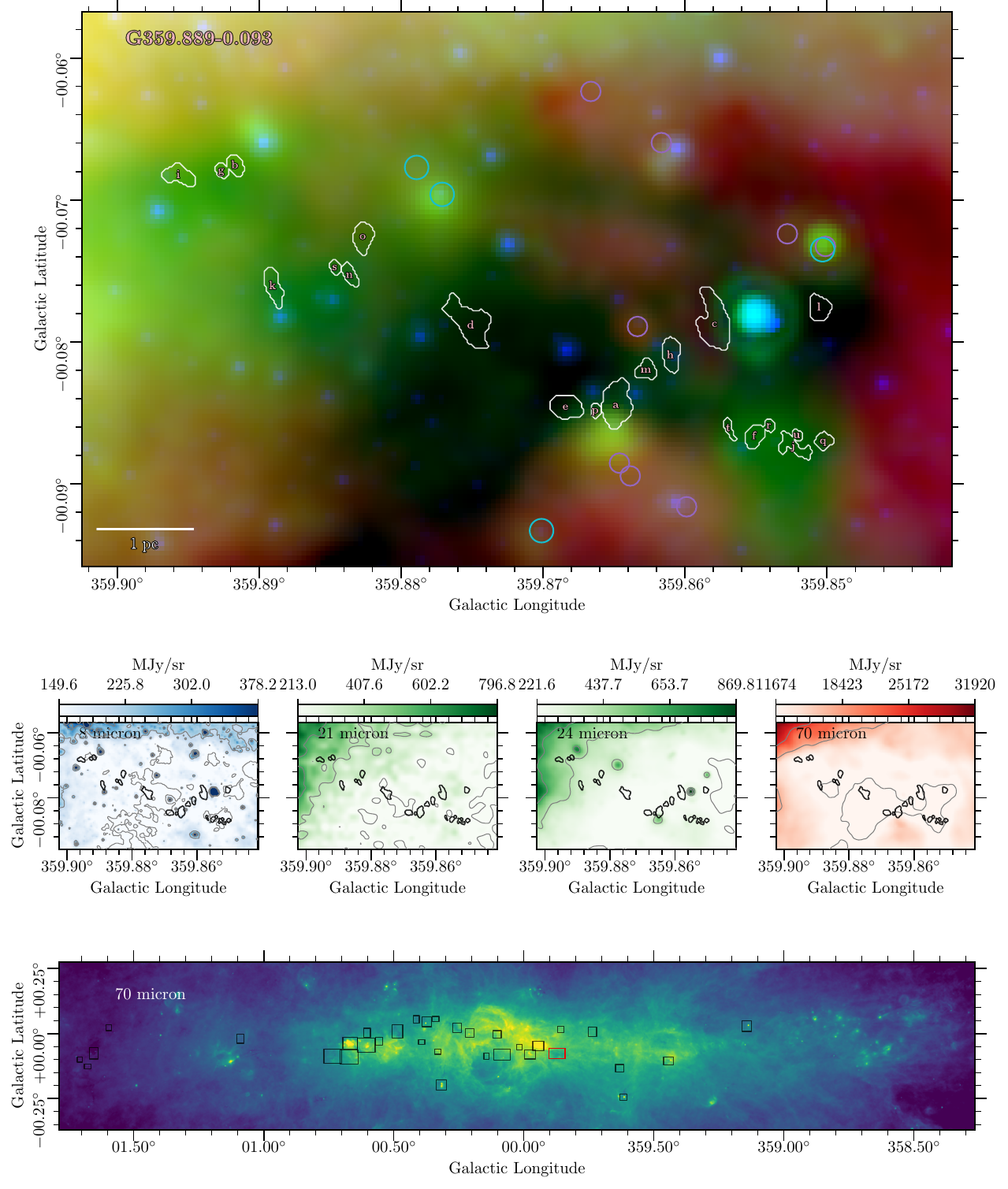}
\end{center}
\caption{Region 15, Cloud ID G359.889-0.093, also known as the 50 km s$^{-1}$ Cloud. The red colorscale shows the 70$\mu$m emission map from Herschel \citep{molinari_hi-gal_2010,molinari_source_2011}. The green colorscale shows the combined 21 and 24$\mu$m emission from MSX and Spitzer respectively \citep{benjamin_glimpse_2003}. The blue colorscale shows the 8$\mu$m map from Spitzer \citep{benjamin_glimpse_2003}. Each color is shown in logscale, scaled between a local 5\% and 95\% boundary value for each box. Overlaid on the composite three-color image are white contours representing the \textit{CMZoom} leaves, along with cyan circles representing YSO candidates from a compilation of those identified by \citealt{yusef-zadeh_star_2009}, \citealt{an_massive_2011}, and \citealt{immer_recent_2012}, purple circles representing the 70$\mu$m point sources cataloged by \citealt{molinari_hi-gal_2016}, and darker blue circles representing the point sources identified by \citealt{gutermuth_24_2015}. The radial size of these circles corresponds to the FWHM condition used to determine plausible association with nearby \textit{CMZoom} leaves.}
\label{fig:rgb_15}
\end{figure*}

\begin{figure*}
\begin{center}
\includegraphics[trim = 0mm 0mm 0mm 0mm, clip, width = .90 \textwidth]{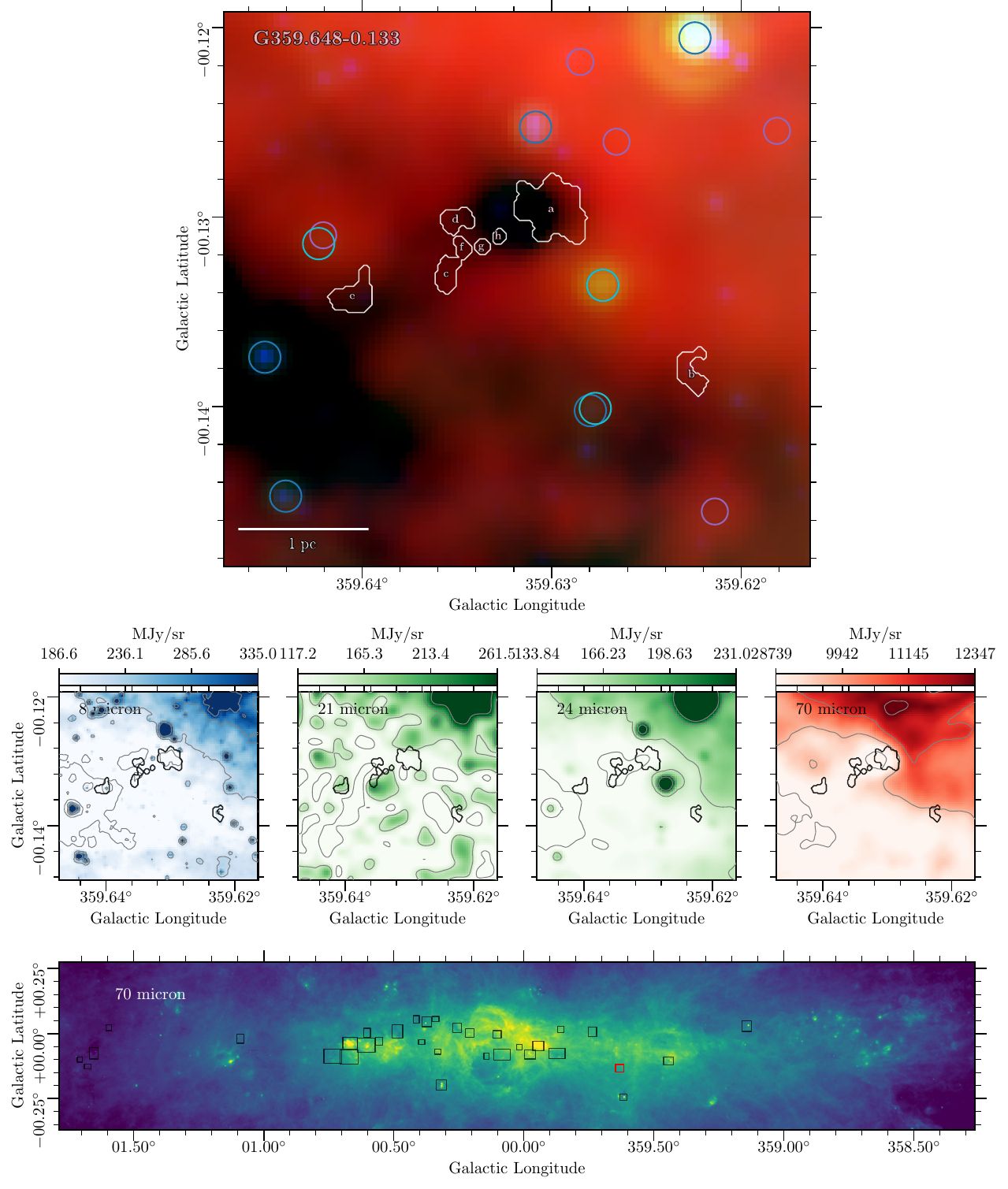}
\end{center}
\caption{Region 16, Cloud ID G359.648-0.133. The red colorscale shows the 70$\mu$m emission map from Herschel \citep{molinari_hi-gal_2010,molinari_source_2011}. The green colorscale shows the combined 21 and 24$\mu$m emission from MSX and Spitzer respectively \citep{benjamin_glimpse_2003}. The blue colorscale shows the 8$\mu$m map from Spitzer \citep{benjamin_glimpse_2003}. Each color is shown in logscale, scaled between a local 5\% and 95\% boundary value for each box. Overlaid on the composite three-color image are white contours representing the \textit{CMZoom} leaves, along with cyan circles representing YSO candidates from a compilation of those identified by \citealt{yusef-zadeh_star_2009}, \citealt{an_massive_2011}, and \citealt{immer_recent_2012}, purple circles representing the 70$\mu$m point sources cataloged by \citealt{molinari_hi-gal_2016}, and darker blue circles representing the point sources identified by \citealt{gutermuth_24_2015}. The radial size of these circles corresponds to the FWHM condition used to determine plausible association with nearby \textit{CMZoom} leaves.}
\label{fig:rgb_16}
\end{figure*}

\begin{figure*}
\begin{center}
\includegraphics[trim = 0mm 0mm 0mm 0mm, clip, width = .90 \textwidth]{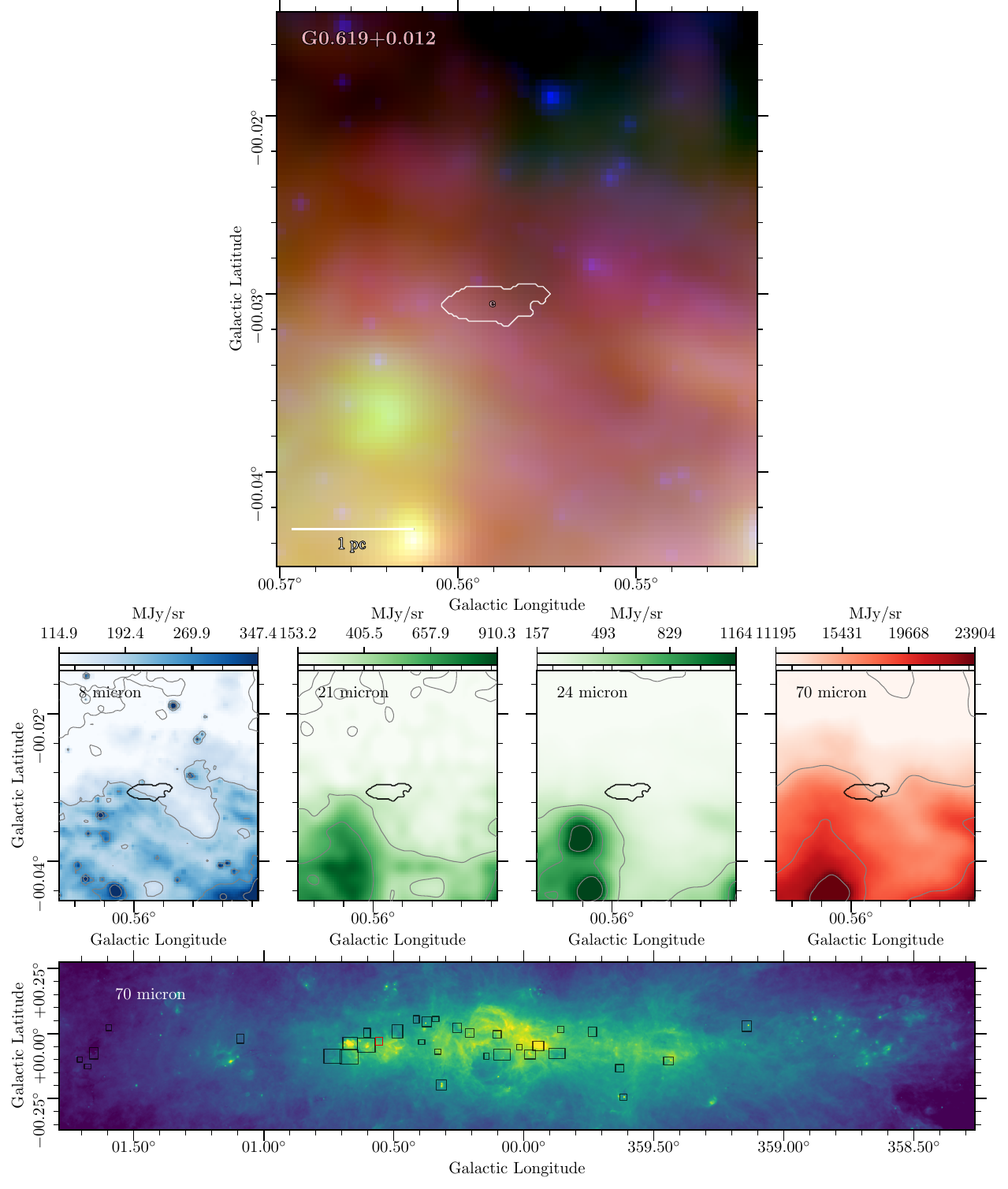}
\end{center}
\caption{Region 17, the remaining portion of Cloud ID G0.619+0.012, part of the region also known as Sgr B2 SE. The red colorscale shows the 70$\mu$m emission map from Herschel \citep{molinari_hi-gal_2010,molinari_source_2011}. The green colorscale shows the combined 21 and 24$\mu$m emission from MSX and Spitzer respectively \citep{benjamin_glimpse_2003}. The blue colorscale shows the 8$\mu$m map from Spitzer \citep{benjamin_glimpse_2003}. Each color is shown in logscale, scaled between a local 5\% and 95\% boundary value for each box. Overlaid on the composite three-color image are white contours representing the \textit{CMZoom} leaves, along with cyan circles representing YSO candidates from a compilation of those identified by \citealt{yusef-zadeh_star_2009}, \citealt{an_massive_2011}, and \citealt{immer_recent_2012}, purple circles representing the 70$\mu$m point sources cataloged by \citealt{molinari_hi-gal_2016}, and darker blue circles representing the point sources identified by \citealt{gutermuth_24_2015}. The radial size of these circles corresponds to the FWHM condition used to determine plausible association with nearby \textit{CMZoom} leaves.}
\label{fig:rgb_17}
\end{figure*}

\begin{figure*}
\begin{center}
\includegraphics[trim = 0mm 0mm 0mm 0mm, clip, width = .90 \textwidth]{msx_mips_17-full.pdf}
\end{center}
\caption{Region 18, Cloud ID G359.484-0.132, also known as Sgr C. The red colorscale shows the 70$\mu$m emission map from Herschel \citep{molinari_hi-gal_2010,molinari_source_2011}. The green colorscale shows the combined 21 and 24$\mu$m emission from MSX and Spitzer respectively \citep{benjamin_glimpse_2003}. The blue colorscale shows the 8$\mu$m map from Spitzer \citep{benjamin_glimpse_2003}. Each color is shown in logscale, scaled between a local 5\% and 95\% boundary value for each box. Overlaid on the composite three-color image are white contours representing the \textit{CMZoom} leaves, along with cyan circles representing YSO candidates from a compilation of those identified by \citealt{yusef-zadeh_star_2009}, \citealt{an_massive_2011}, and \citealt{immer_recent_2012}, purple circles representing the 70$\mu$m point sources cataloged by \citealt{molinari_hi-gal_2016}, and darker blue circles representing the point sources identified by \citealt{gutermuth_24_2015}. The radial size of these circles corresponds to the FWHM condition used to determine plausible association with nearby \textit{CMZoom} leaves.}
\label{fig:rgb_18}
\end{figure*}

\begin{figure*}
\begin{center}
\includegraphics[trim = 0mm 0mm 0mm 0mm, clip, width = .90 \textwidth]{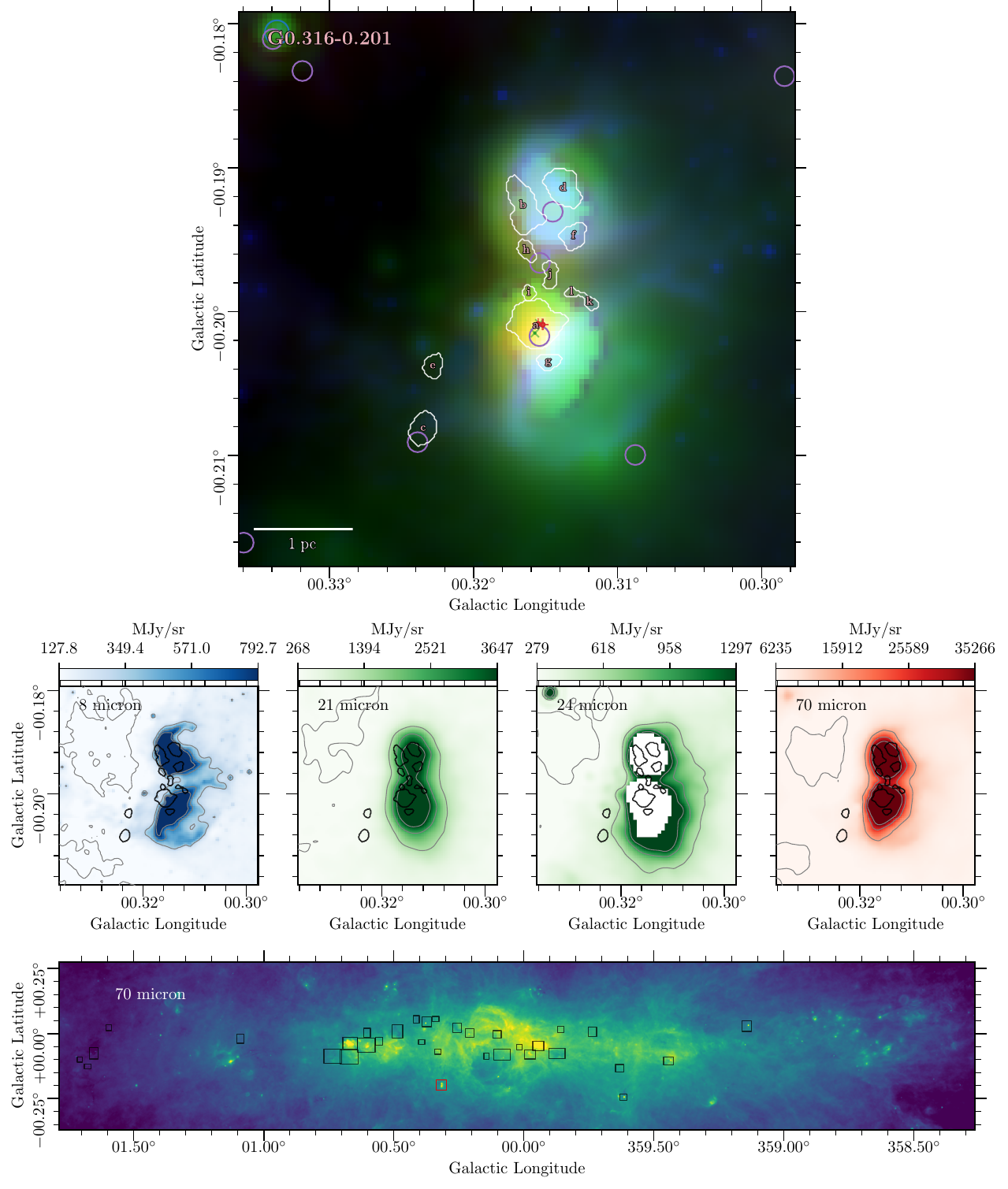}
\end{center}
\caption{Region 19, Cloud ID G0.316-0.201. The red colorscale shows the 70$\mu$m emission map from Herschel \citep{molinari_hi-gal_2010,molinari_source_2011}. The green colorscale shows the combined 21 and 24$\mu$m emission from MSX and Spitzer respectively \citep{benjamin_glimpse_2003}. The blue colorscale shows the 8$\mu$m map from Spitzer \citep{benjamin_glimpse_2003}. Each color is shown in logscale, scaled between a local 5\% and 95\% boundary value for each box. Overlaid on the composite three-color image are white contours representing the \textit{CMZoom} leaves, along with cyan circles representing YSO candidates from a compilation of those identified by \citealt{yusef-zadeh_star_2009}, \citealt{an_massive_2011}, and \citealt{immer_recent_2012}, purple circles representing the 70$\mu$m point sources cataloged by \citealt{molinari_hi-gal_2016}, and darker blue circles representing the point sources identified by \citealt{gutermuth_24_2015}. The radial size of these circles corresponds to the FWHM condition used to determine plausible association with nearby \textit{CMZoom} leaves.}
\label{fig:rgb_19}
\end{figure*}

\begin{figure*}
\begin{center}
\includegraphics[trim = 0mm 0mm 0mm 0mm, clip, width = .90 \textwidth]{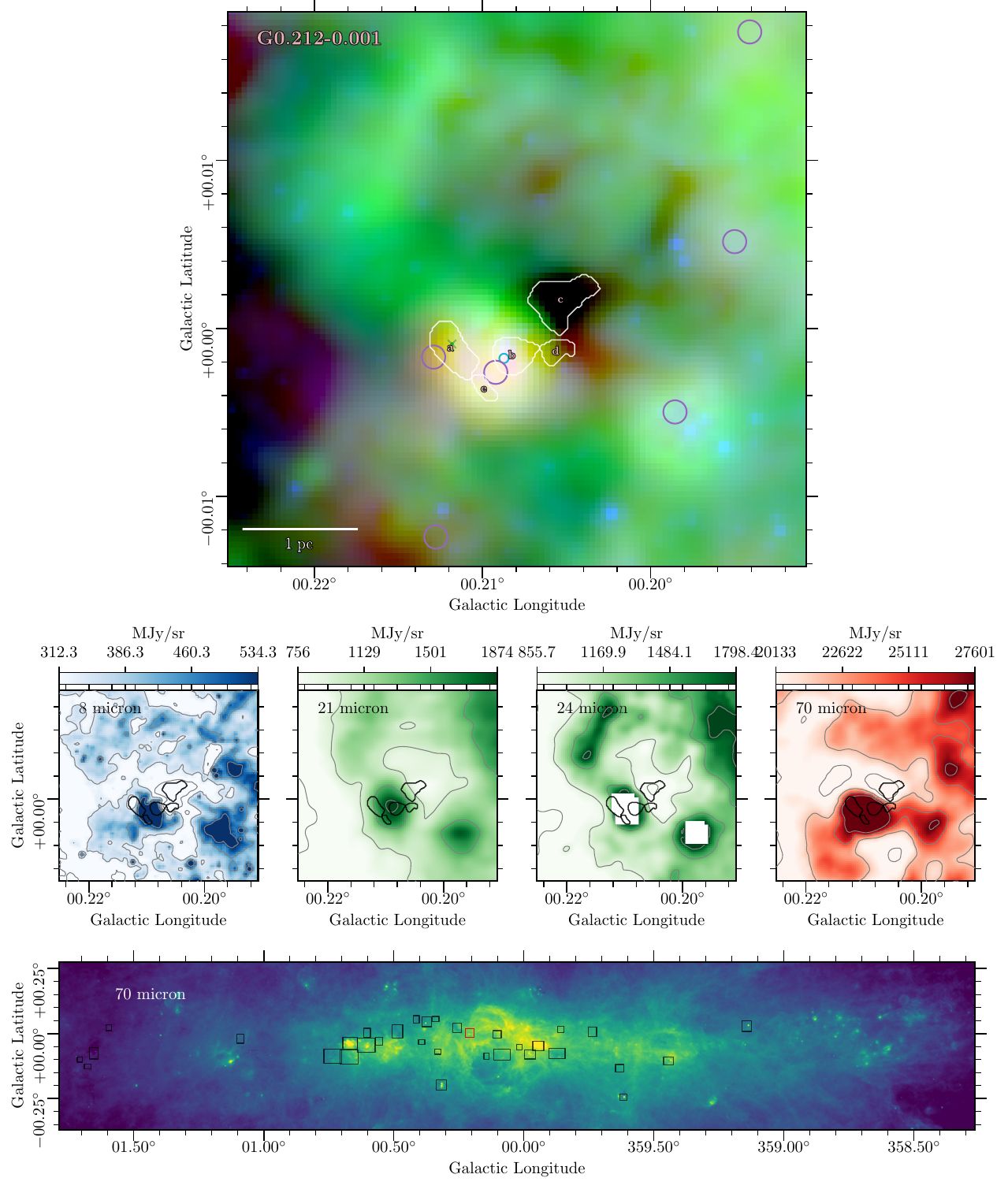}
\end{center}
\caption{Region 20, Cloud ID G0.212-0.001. The red colorscale shows the 70$\mu$m emission map from Herschel \citep{molinari_hi-gal_2010,molinari_source_2011}. The green colorscale shows the combined 21 and 24$\mu$m emission from MSX and Spitzer respectively \citep{benjamin_glimpse_2003}. The blue colorscale shows the 8$\mu$m map from Spitzer \citep{benjamin_glimpse_2003}. Each color is shown in logscale, scaled between a local 5\% and 95\% boundary value for each box. Overlaid on the composite three-color image are white contours representing the \textit{CMZoom} leaves, along with cyan circles representing YSO candidates from a compilation of those identified by \citealt{yusef-zadeh_star_2009}, \citealt{an_massive_2011}, and \citealt{immer_recent_2012}, purple circles representing the 70$\mu$m point sources cataloged by \citealt{molinari_hi-gal_2016}, and darker blue circles representing the point sources identified by \citealt{gutermuth_24_2015}. The radial size of these circles corresponds to the FWHM condition used to determine plausible association with nearby \textit{CMZoom} leaves.}
\label{fig:rgb_20}
\end{figure*}

\begin{figure*}
\begin{center}
\includegraphics[trim = 0mm 0mm 0mm 0mm, clip, width = .90 \textwidth]{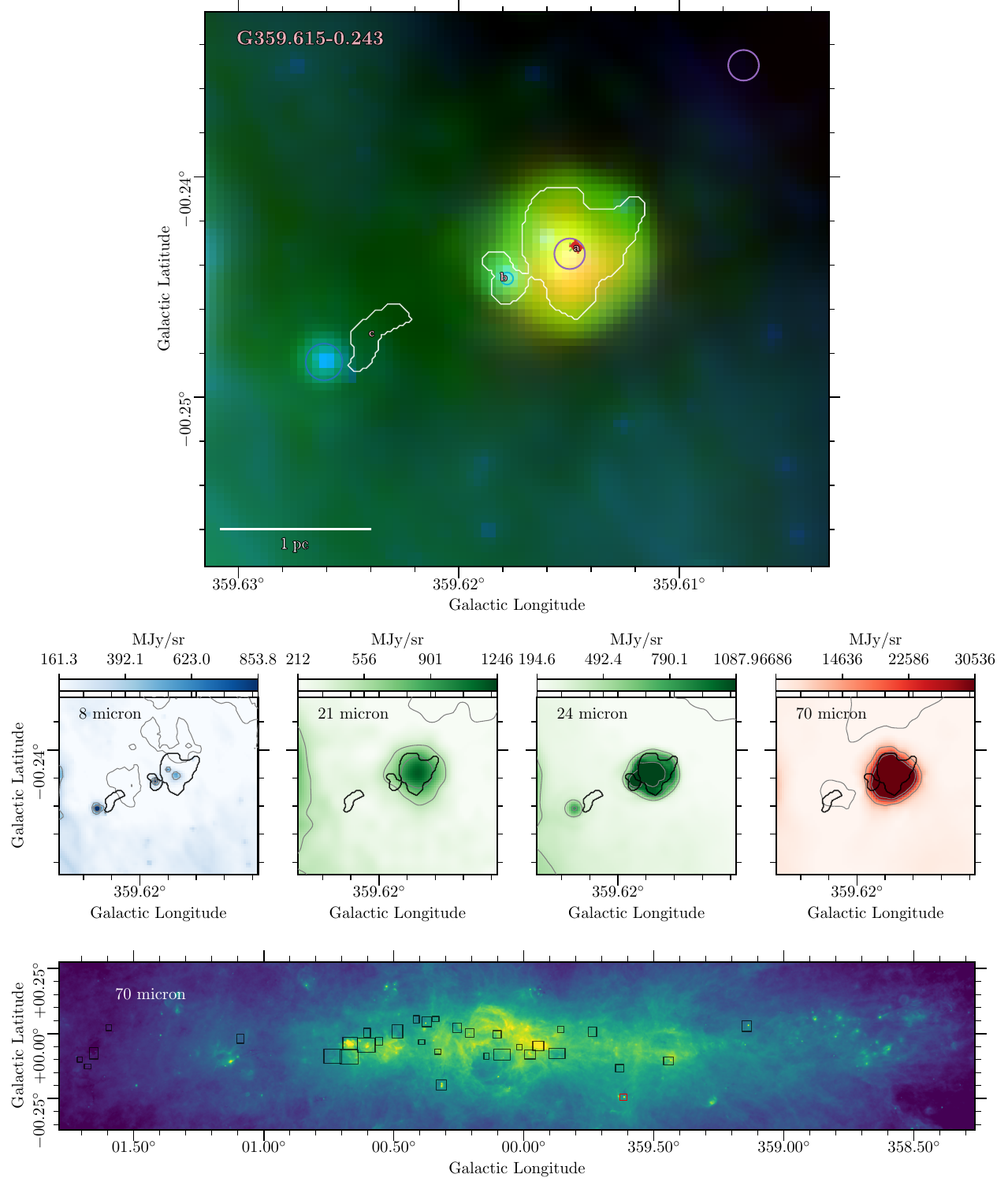}
\end{center}
\caption{Region 21, Cloud ID G359.615-0.243. The red colorscale shows the 70$\mu$m emission map from Herschel \citep{molinari_hi-gal_2010,molinari_source_2011}. The green colorscale shows the combined 21 and 24$\mu$m emission from MSX and Spitzer respectively \citep{benjamin_glimpse_2003}. The blue colorscale shows the 8$\mu$m map from Spitzer \citep{benjamin_glimpse_2003}. Each color is shown in logscale, scaled between a local 5\% and 95\% boundary value for each box. Overlaid on the composite three-color image are white contours representing the \textit{CMZoom} leaves, along with cyan circles representing YSO candidates from a compilation of those identified by \citealt{yusef-zadeh_star_2009}, \citealt{an_massive_2011}, and \citealt{immer_recent_2012}, purple circles representing the 70$\mu$m point sources cataloged by \citealt{molinari_hi-gal_2016}, and darker blue circles representing the point sources identified by \citealt{gutermuth_24_2015}. The radial size of these circles corresponds to the FWHM condition used to determine plausible association with nearby \textit{CMZoom} leaves.}
\label{fig:rgb_21}
\end{figure*}

\begin{figure*}
\begin{center}
\includegraphics[trim = 0mm 0mm 0mm 0mm, clip, width = .90 \textwidth]{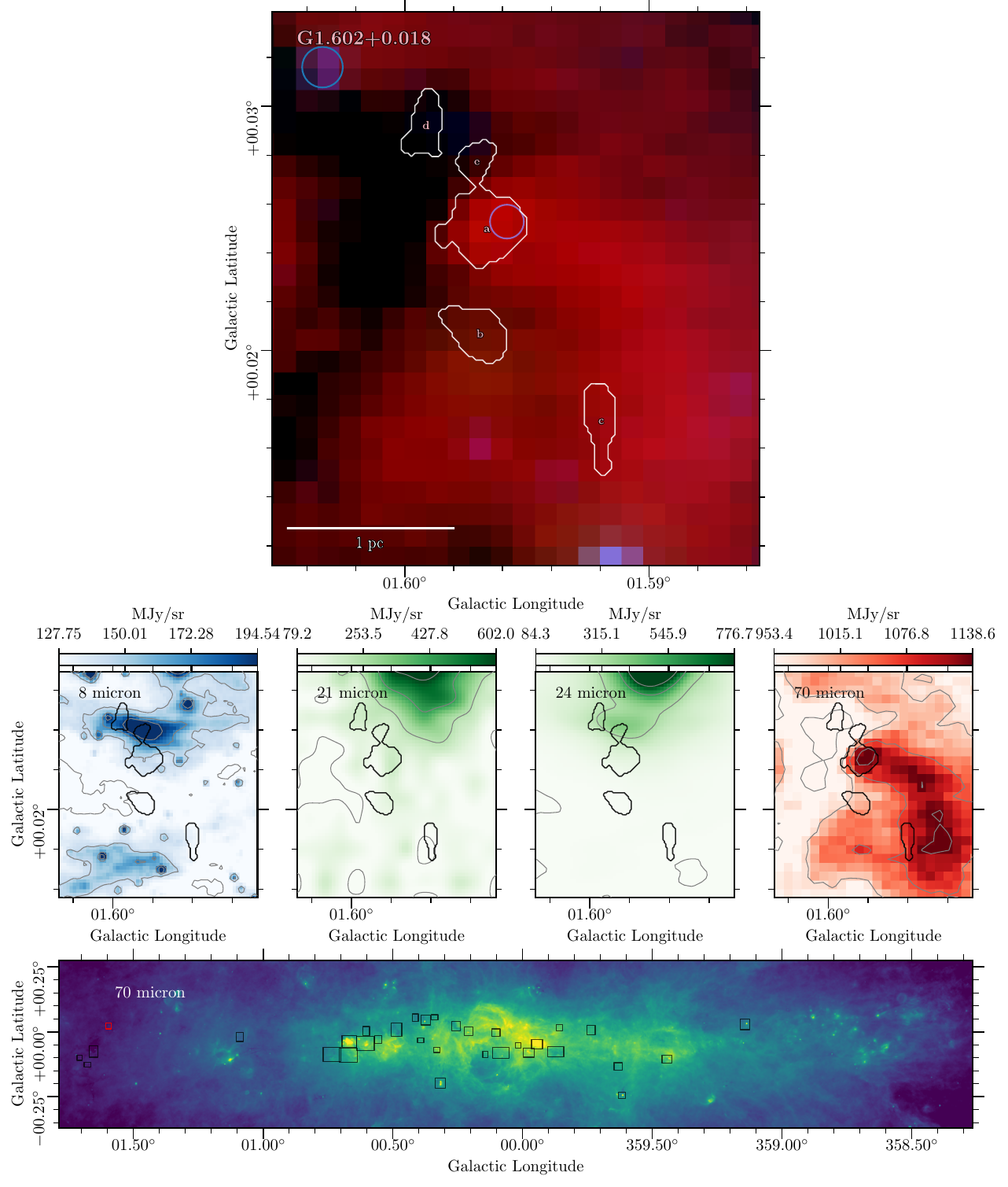}
\end{center}
\caption{Region 22, Cloud ID G1.602+0.018. The red colorscale shows the 70$\mu$m emission map from Herschel \citep{molinari_hi-gal_2010,molinari_source_2011}. The green colorscale shows the combined 21 and 24$\mu$m emission from MSX and Spitzer respectively \citep{benjamin_glimpse_2003}. The blue colorscale shows the 8$\mu$m map from Spitzer \citep{benjamin_glimpse_2003}. Each color is shown in logscale, scaled between a local 5\% and 95\% boundary value for each box. Overlaid on the composite three-color image are white contours representing the \textit{CMZoom} leaves, along with cyan circles representing YSO candidates from a compilation of those identified by \citealt{yusef-zadeh_star_2009}, \citealt{an_massive_2011}, and \citealt{immer_recent_2012}, purple circles representing the 70$\mu$m point sources cataloged by \citealt{molinari_hi-gal_2016}, and darker blue circles representing the point sources identified by \citealt{gutermuth_24_2015}. The radial size of these circles corresponds to the FWHM condition used to determine plausible association with nearby \textit{CMZoom} leaves.}
\label{fig:rgb_22}
\end{figure*}

\begin{figure*}
\begin{center}
\includegraphics[trim = 0mm 0mm 0mm 0mm, clip, width = .90 \textwidth]{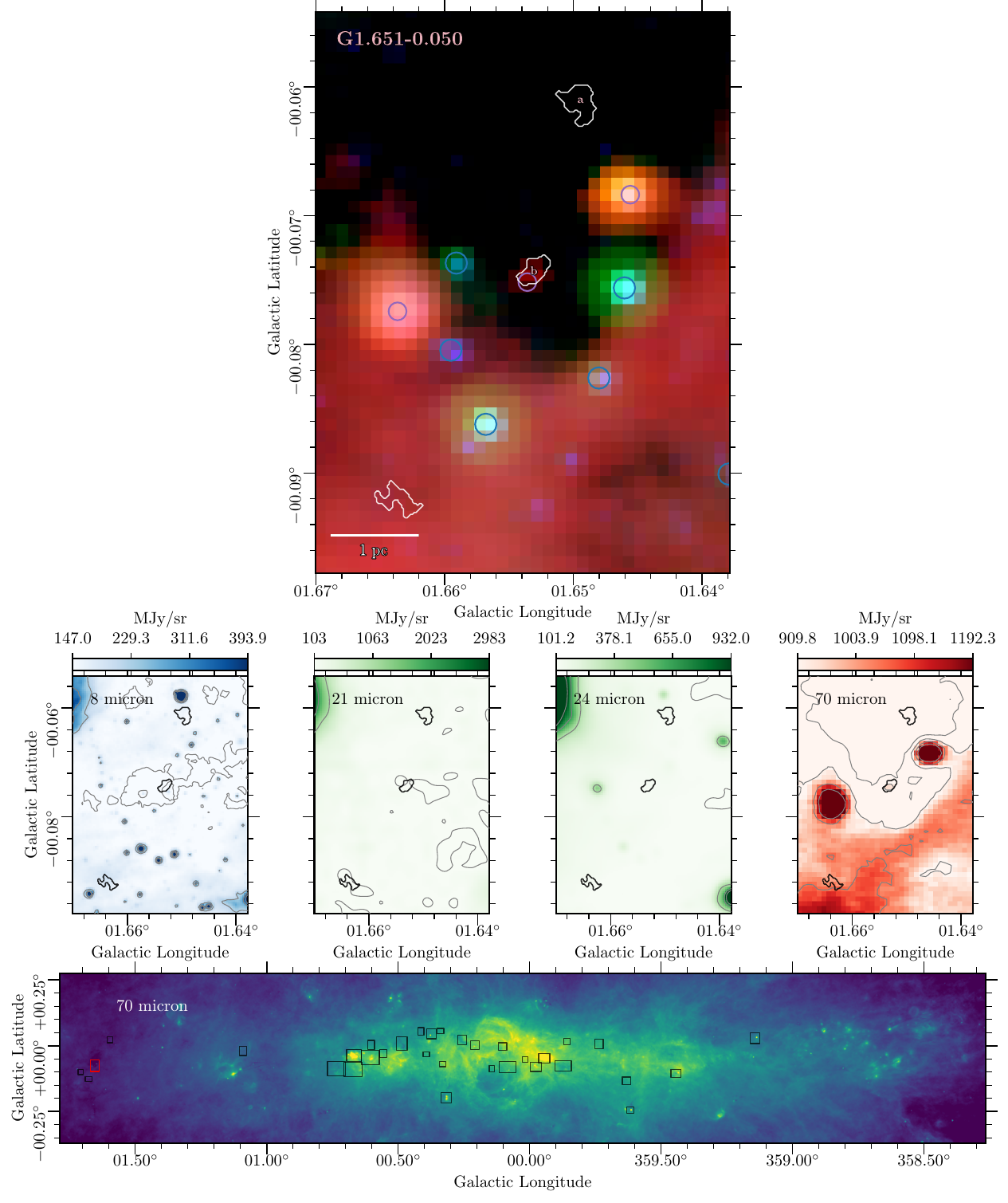}
\end{center}
\caption{Region 23, Cloud ID G1.651-0.050. The red colorscale shows the 70$\mu$m emission map from Herschel \citep{molinari_hi-gal_2010,molinari_source_2011}. The green colorscale shows the combined 21 and 24$\mu$m emission from MSX and Spitzer respectively \citep{benjamin_glimpse_2003}. The blue colorscale shows the 8$\mu$m map from Spitzer \citep{benjamin_glimpse_2003}. Each color is shown in logscale, scaled between a local 5\% and 95\% boundary value for each box. Overlaid on the composite three-color image are white contours representing the \textit{CMZoom} leaves, along with cyan circles representing YSO candidates from a compilation of those identified by \citealt{yusef-zadeh_star_2009}, \citealt{an_massive_2011}, and \citealt{immer_recent_2012}, purple circles representing the 70$\mu$m point sources cataloged by \citealt{molinari_hi-gal_2016}, and darker blue circles representing the point sources identified by \citealt{gutermuth_24_2015}. The radial size of these circles corresponds to the FWHM condition used to determine plausible association with nearby \textit{CMZoom} leaves.}
\label{fig:rgb_23}
\end{figure*}

\begin{figure*}
\begin{center}
\includegraphics[trim = 0mm 0mm 0mm 0mm, clip, width = .90 \textwidth]{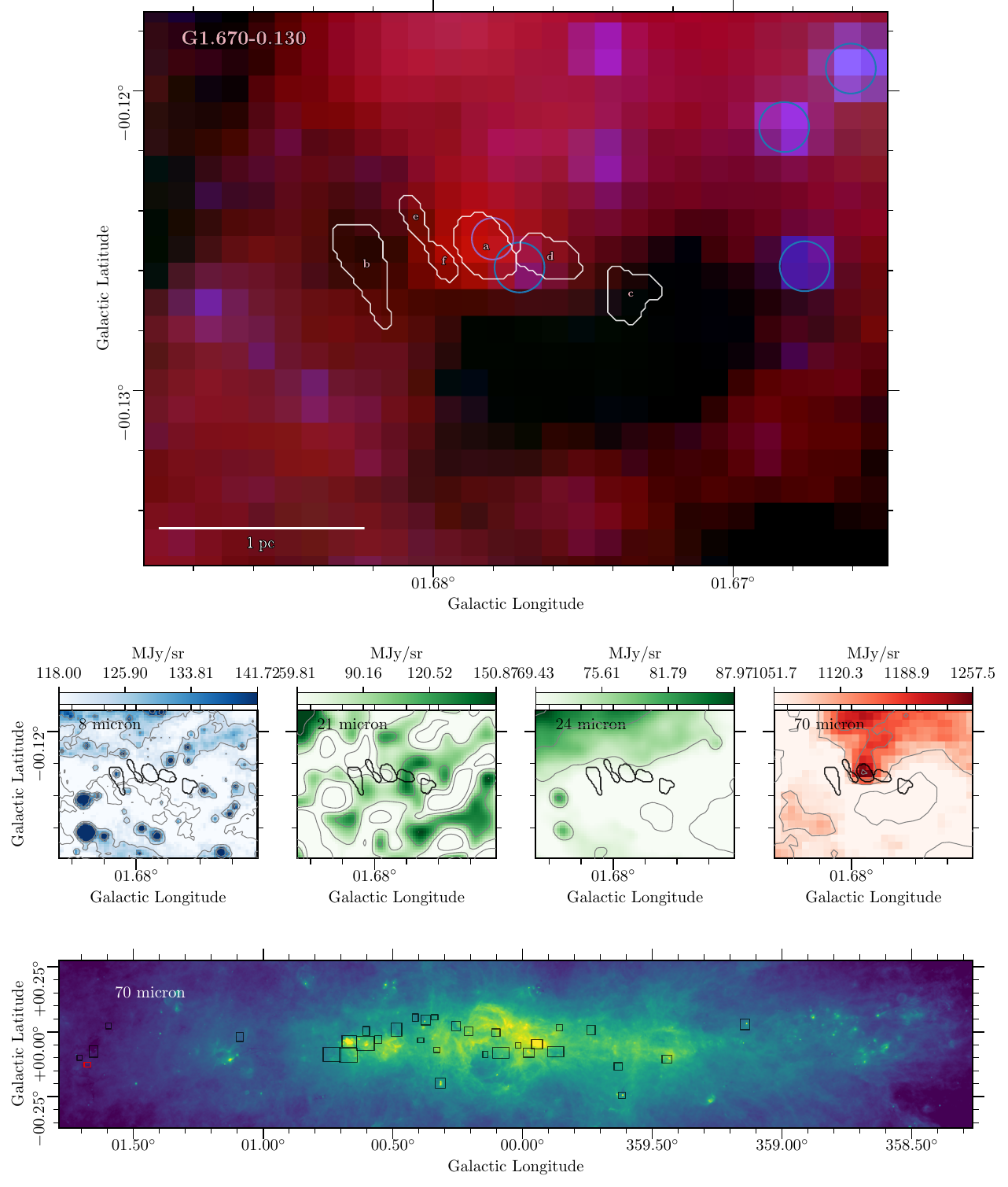}
\end{center}
\caption{Region 24, Cloud ID G1.670-0.130. The red colorscale shows the 70$\mu$m emission map from Herschel \citep{molinari_hi-gal_2010,molinari_source_2011}. The green colorscale shows the combined 21 and 24$\mu$m emission from MSX and Spitzer respectively \citep{benjamin_glimpse_2003}. The blue colorscale shows the 8$\mu$m map from Spitzer \citep{benjamin_glimpse_2003}. Each color is shown in logscale, scaled between a local 5\% and 95\% boundary value for each box. Overlaid on the composite three-color image are white contours representing the \textit{CMZoom} leaves, along with cyan circles representing YSO candidates from a compilation of those identified by \citealt{yusef-zadeh_star_2009}, \citealt{an_massive_2011}, and \citealt{immer_recent_2012}, purple circles representing the 70$\mu$m point sources cataloged by \citealt{molinari_hi-gal_2016}, and darker blue circles representing the point sources identified by \citealt{gutermuth_24_2015}. The radial size of these circles corresponds to the FWHM condition used to determine plausible association with nearby \textit{CMZoom} leaves.}
\label{fig:rgb_24}
\end{figure*}

\begin{figure*}
\begin{center}
\includegraphics[trim = 0mm 0mm 0mm 0mm,clip, width = .90 \textwidth]{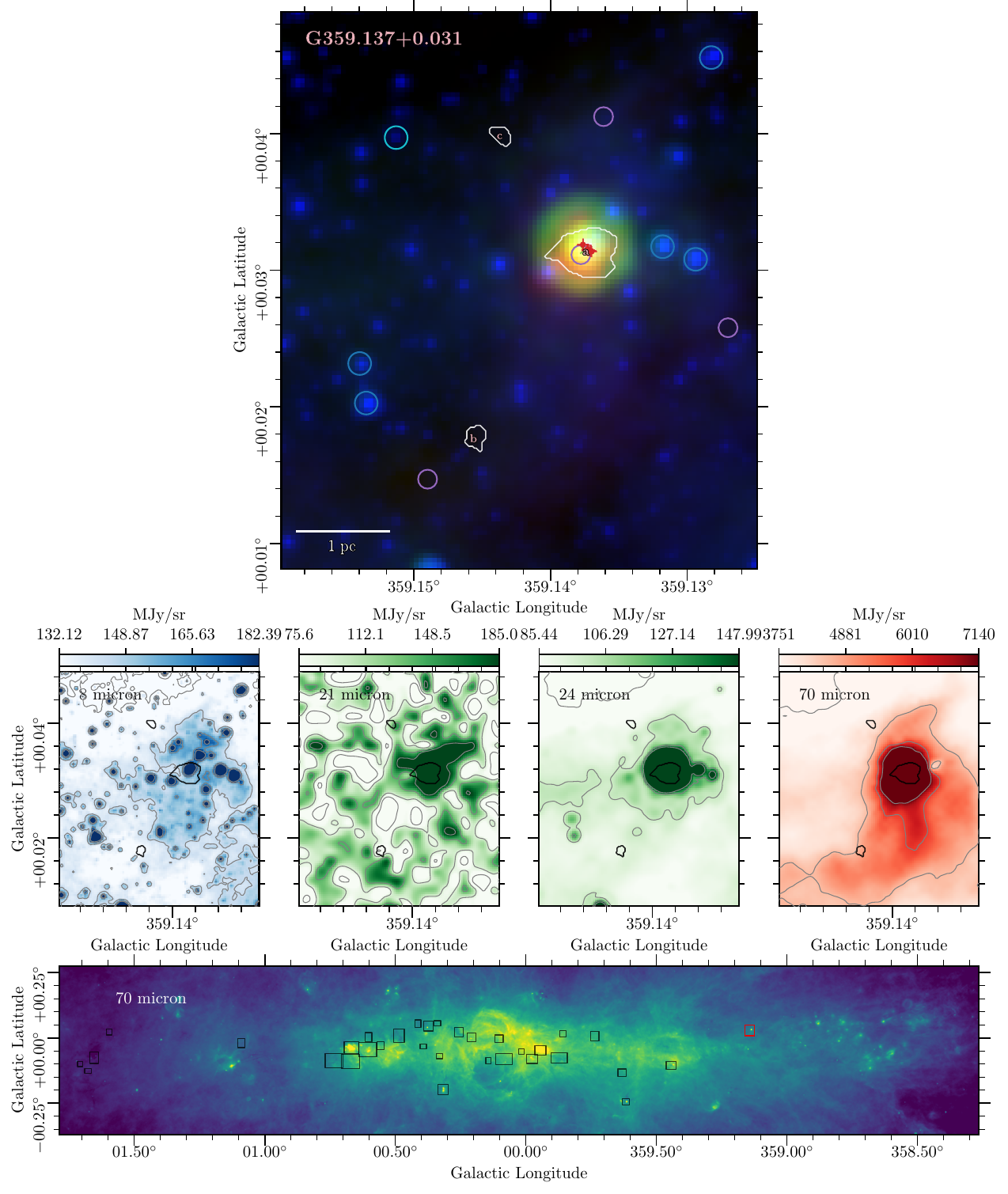}
\end{center}
\caption{Region 25, Cloud ID G0.359.137+0.031. The red colorscale shows the 70$\mu$m emission map from Herschel \citep{molinari_hi-gal_2010,molinari_source_2011}. The green colorscale shows the combined 21 and 24$\mu$m emission from MSX and Spitzer respectively \citep{benjamin_glimpse_2003}. The blue colorscale shows the 8$\mu$m map from Spitzer \citep{benjamin_glimpse_2003}. Each color is shown in logscale, scaled between a local 5\% and 95\% boundary value for each box. Overlaid on the composite three-color image are white contours representing the \textit{CMZoom} leaves, along with cyan circles representing YSO candidates from a compilation of those identified by \citealt{yusef-zadeh_star_2009}, \citealt{an_massive_2011}, and \citealt{immer_recent_2012}, purple circles representing the 70$\mu$m point sources cataloged by \citealt{molinari_hi-gal_2016}, and darker blue circles representing the point sources identified by \citealt{gutermuth_24_2015}. The radial size of these circles corresponds to the FWHM condition used to determine plausible association with nearby \textit{CMZoom} leaves.}
\label{fig:rgb_25}
\end{figure*}

\begin{figure*}
\begin{center}
\includegraphics[trim = 0mm 0mm 0mm 0mm,clip, width = .90 \textwidth]{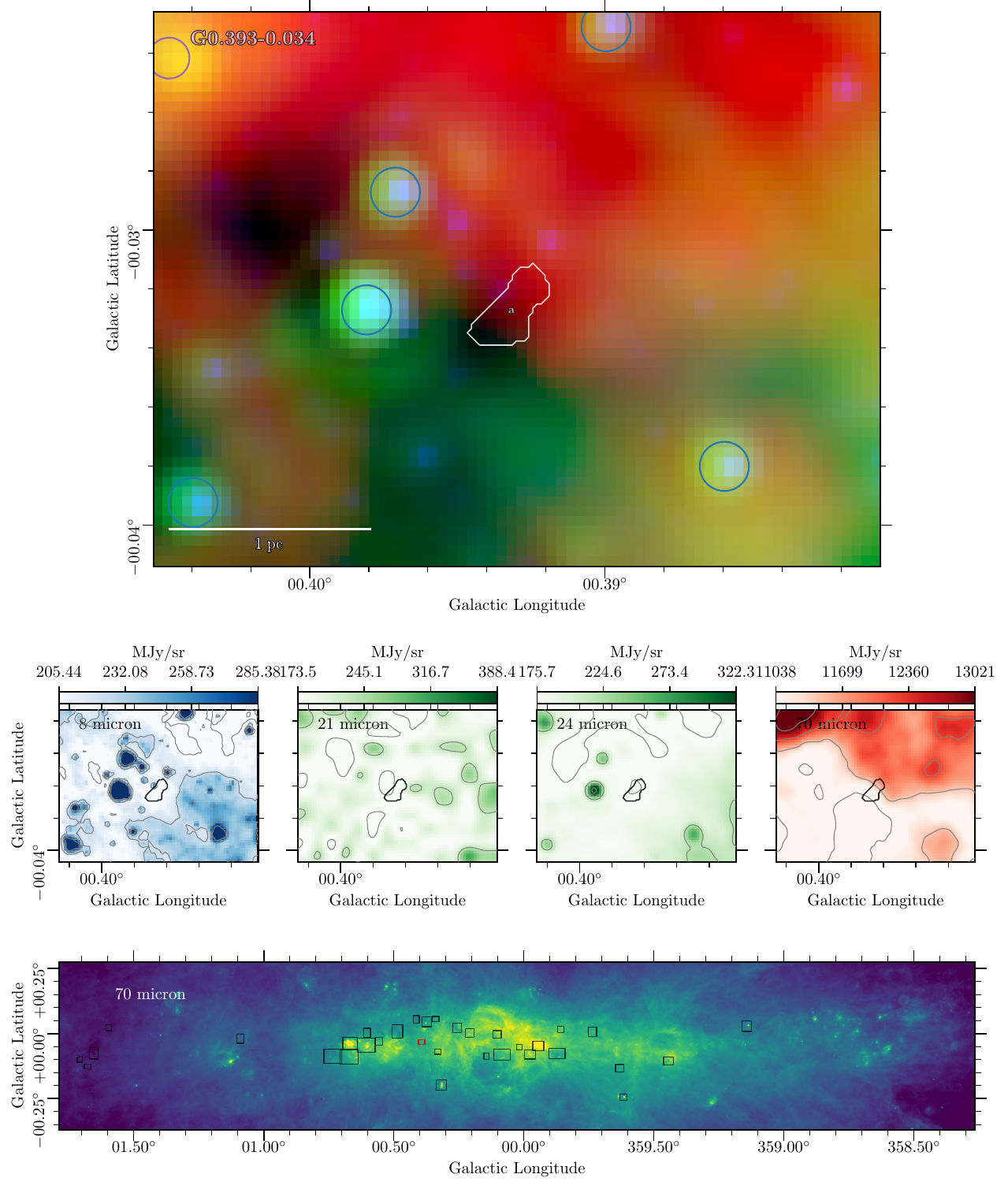}
\end{center}
\caption{Region 26, Cloud ID G0.393-0.034. The red colorscale shows the 70$\mu$m emission map from Herschel \citep{molinari_hi-gal_2010,molinari_source_2011}. The green colorscale shows the combined 21 and 24$\mu$m emission from MSX and Spitzer respectively \citep{benjamin_glimpse_2003}. The blue colorscale shows the 8$\mu$m map from Spitzer \citep{benjamin_glimpse_2003}. Each color is shown in logscale, scaled between a local 5\% and 95\% boundary value for each box. Overlaid on the composite three-color image are white contours representing the \textit{CMZoom} leaves, along with cyan circles representing YSO candidates from a compilation of those identified by \citealt{yusef-zadeh_star_2009}, \citealt{an_massive_2011}, and \citealt{immer_recent_2012}, purple circles representing the 70$\mu$m point sources cataloged by \citealt{molinari_hi-gal_2016}, and darker blue circles representing the point sources identified by \citealt{gutermuth_24_2015}. The radial size of these circles corresponds to the FWHM condition used to determine plausible association with nearby \textit{CMZoom} leaves. }
\label{fig:rgb_26}
\end{figure*}

\begin{figure*}
\begin{center}
\includegraphics[trim = 0mm 0mm 0mm 0mm,clip, width = .90 \textwidth]{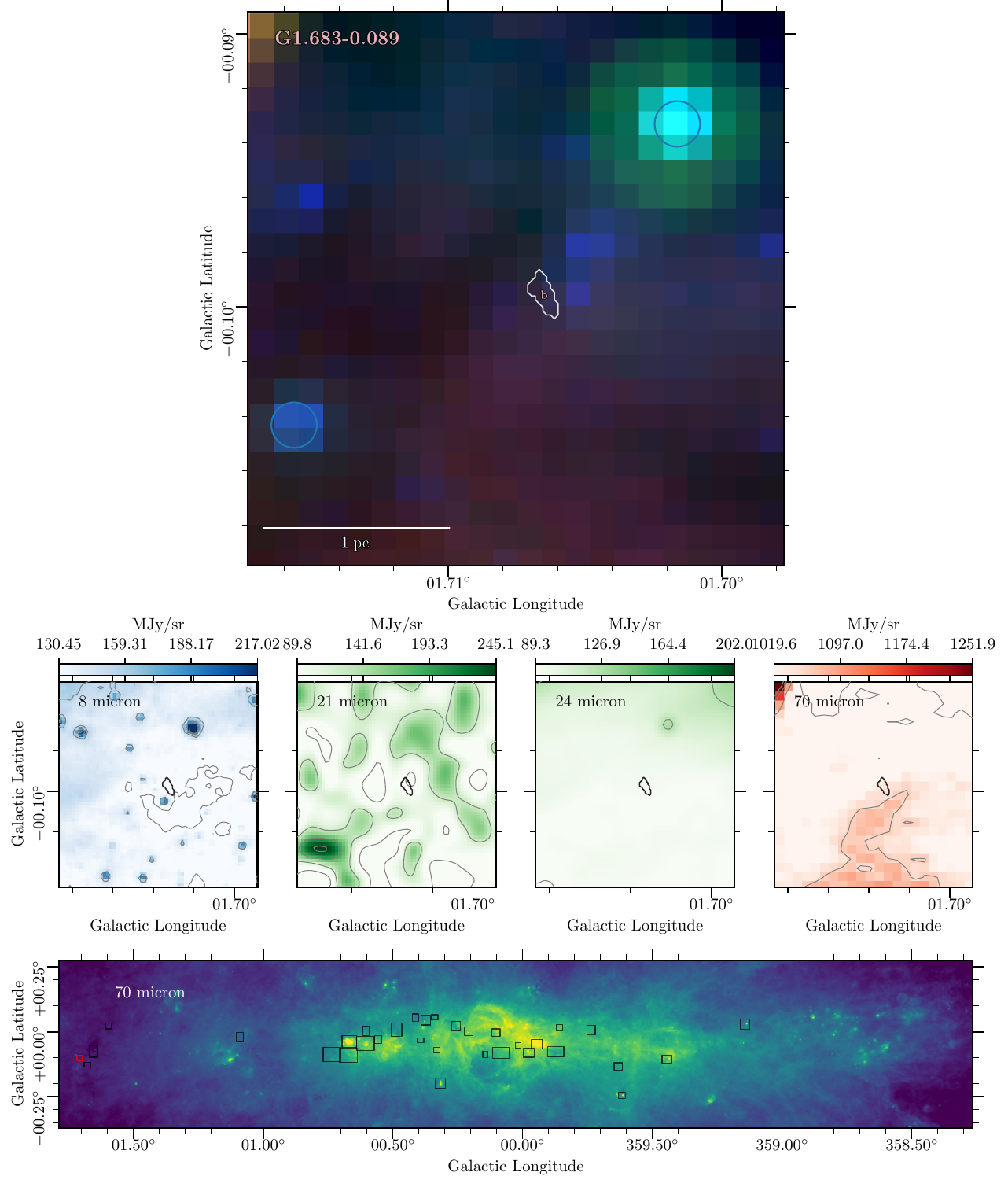}
\end{center}
\caption{Region 27, Cloud ID G1.683-0.089. The red colorscale shows the 70$\mu$m emission map from Herschel \citep{molinari_hi-gal_2010,molinari_source_2011}. The green colorscale shows the combined 21 and 24$\mu$m emission from MSX and Spitzer respectively \citep{benjamin_glimpse_2003}. The blue colorscale shows the 8$\mu$m map from Spitzer \citep{benjamin_glimpse_2003}. Each color is shown in logscale, scaled between a local 5\% and 95\% boundary value for each box. Overlaid on the composite three-color image are white contours representing the \textit{CMZoom} leaves, along with cyan circles representing YSO candidates from a compilation of those identified by \citealt{yusef-zadeh_star_2009}, \citealt{an_massive_2011}, and \citealt{immer_recent_2012}, purple circles representing the 70$\mu$m point sources cataloged by \citealt{molinari_hi-gal_2016}, and darker blue circles representing the point sources identified by \citealt{gutermuth_24_2015}. The radial size of these circles corresponds to the FWHM condition used to determine plausible association with nearby \textit{CMZoom} leaves.}
\label{fig:rgb_27}
\end{figure*}

\begin{figure*}
\begin{center}
\includegraphics[trim = 0mm 0mm 0mm 0mm,clip, width = .90 \textwidth]{msx_mips_27-full.pdf}
\end{center}
\caption{Region 28, Cloud ID G0.714-0.100. The red colorscale shows the 70$\mu$m emission map from Herschel \citep{molinari_hi-gal_2010,molinari_source_2011}. The green colorscale shows the combined 21 and 24$\mu$m emission from MSX and Spitzer respectively \citep{benjamin_glimpse_2003}. The blue colorscale shows the 8$\mu$m map from Spitzer \citep{benjamin_glimpse_2003}. Each color is shown in logscale, scaled between a local 5\% and 95\% boundary value for each box. Overlaid on the composite three-color image are white contours representing the \textit{CMZoom} leaves, along with cyan circles representing YSO candidates from a compilation of those identified by \citealt{yusef-zadeh_star_2009}, \citealt{an_massive_2011}, and \citealt{immer_recent_2012}, purple circles representing the 70$\mu$m point sources cataloged by \citealt{molinari_hi-gal_2016}, and darker blue circles representing the point sources identified by \citealt{gutermuth_24_2015}. The radial size of these circles corresponds to the FWHM condition used to determine plausible association with nearby \textit{CMZoom} leaves.}
\label{fig:rgb_28}
\end{figure*}

\begin{figure*}
\begin{center}
\includegraphics[trim = 0mm 0mm 0mm 0mm,clip, width = .90 \textwidth]{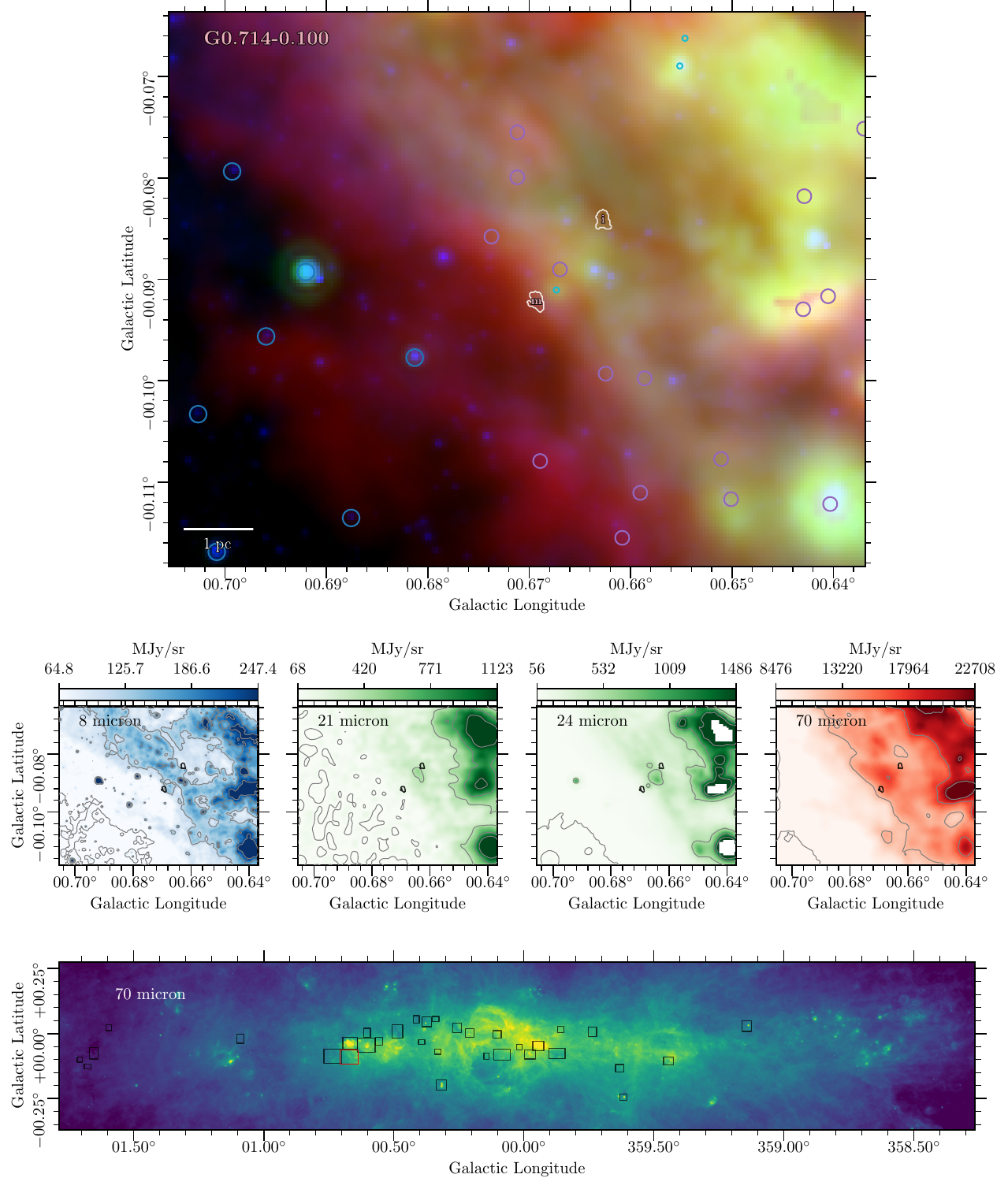}
\end{center}
\caption{Region 29, Cloud ID G0.714-0.100. The red colorscale shows the 70$\mu$m emission map from Herschel \citep{molinari_hi-gal_2010,molinari_source_2011}. The green colorscale shows the combined 21 and 24$\mu$m emission from MSX and Spitzer respectively \citep{benjamin_glimpse_2003}. The blue colorscale shows the 8$\mu$m map from Spitzer \citep{benjamin_glimpse_2003}. Each color is shown in logscale, scaled between a local 5\% and 95\% boundary value for each box. Overlaid on the composite three-color image are white contours representing the \textit{CMZoom} leaves, along with cyan circles representing YSO candidates from a compilation of those identified by \citealt{yusef-zadeh_star_2009}, \citealt{an_massive_2011}, and \citealt{immer_recent_2012}, purple circles representing the 70$\mu$m point sources cataloged by \citealt{molinari_hi-gal_2016}, and darker blue circles representing the point sources identified by \citealt{gutermuth_24_2015}. The radial size of these circles corresponds to the FWHM condition used to determine plausible association with nearby \textit{CMZoom} leaves.}
\label{fig:rgb_29}
\end{figure*}

\begin{figure*}
\begin{center}
\includegraphics[trim = 0mm 0mm 0mm 0mm, clip, width = .90 \textwidth]{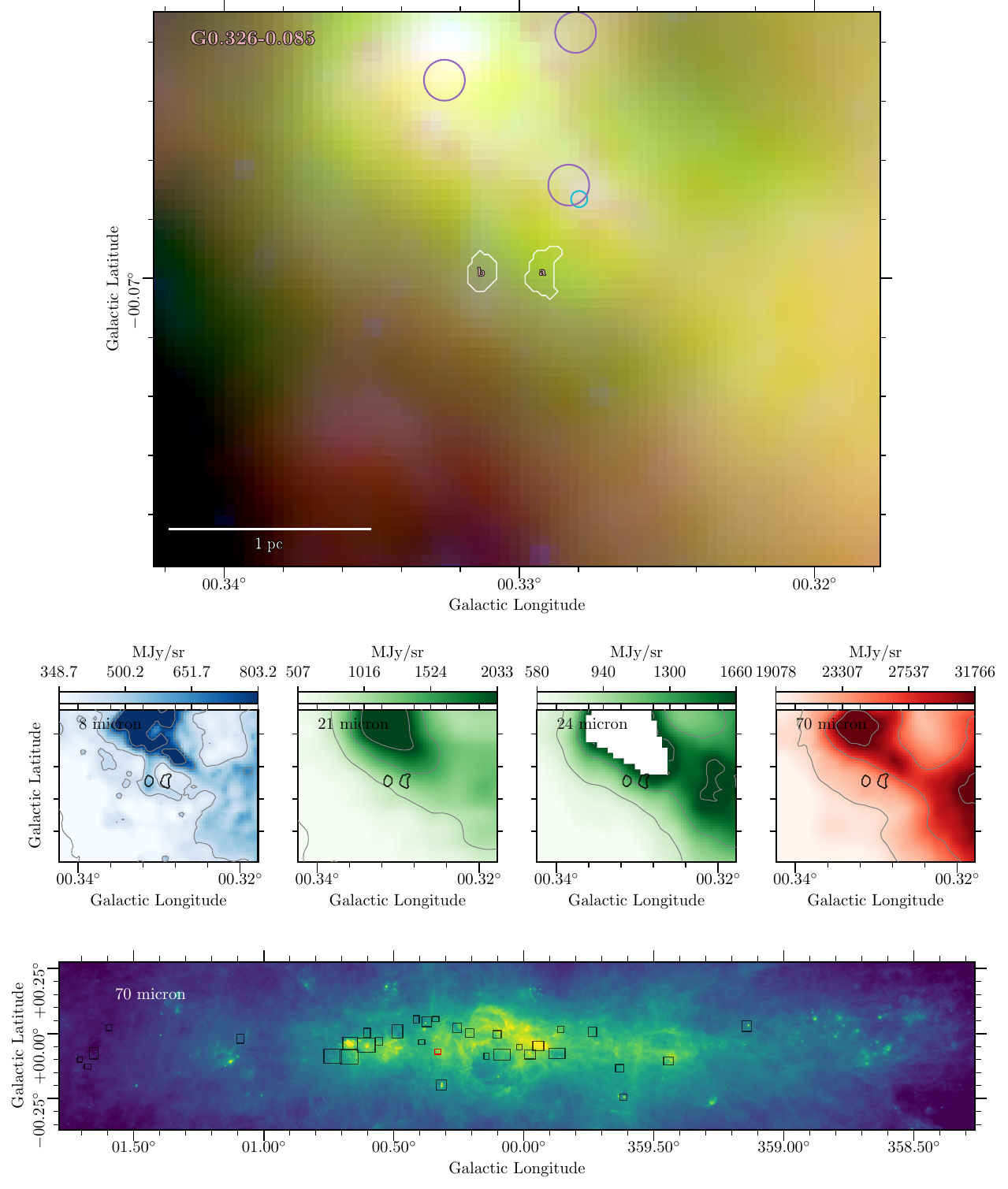}
\end{center}
\caption{Region 31, Cloud ID G0.326-0.085, also known as the Sailfish. The red colorscale shows the 70$\mu$m emission map from Herschel \citep{molinari_hi-gal_2010,molinari_source_2011}. The green colorscale shows the combined 21 and 24$\mu$m emission from MSX and Spitzer respectively \citep{benjamin_glimpse_2003}. The blue colorscale shows the 8$\mu$m map from Spitzer \citep{benjamin_glimpse_2003}. Each color is shown in logscale, scaled between a local 5\% and 95\% boundary value for each box. Overlaid on the composite three-color image are white contours representing the \textit{CMZoom} leaves, along with cyan circles representing YSO candidates from a compilation of those identified by \citealt{yusef-zadeh_star_2009}, \citealt{an_massive_2011}, and \citealt{immer_recent_2012}, purple circles representing the 70$\mu$m point sources cataloged by \citealt{molinari_hi-gal_2016}, and darker blue circles representing the point sources identified by \citealt{gutermuth_24_2015}. The radial size of these circles corresponds to the FWHM condition used to determine plausible association with nearby \textit{CMZoom} leaves.}
\label{fig:rgb_30}
\end{figure*}
\newpage

\begin{figure*}
\begin{center}
\includegraphics[trim = 0mm 0mm 0mm 0mm,clip, width = .90 \textwidth]{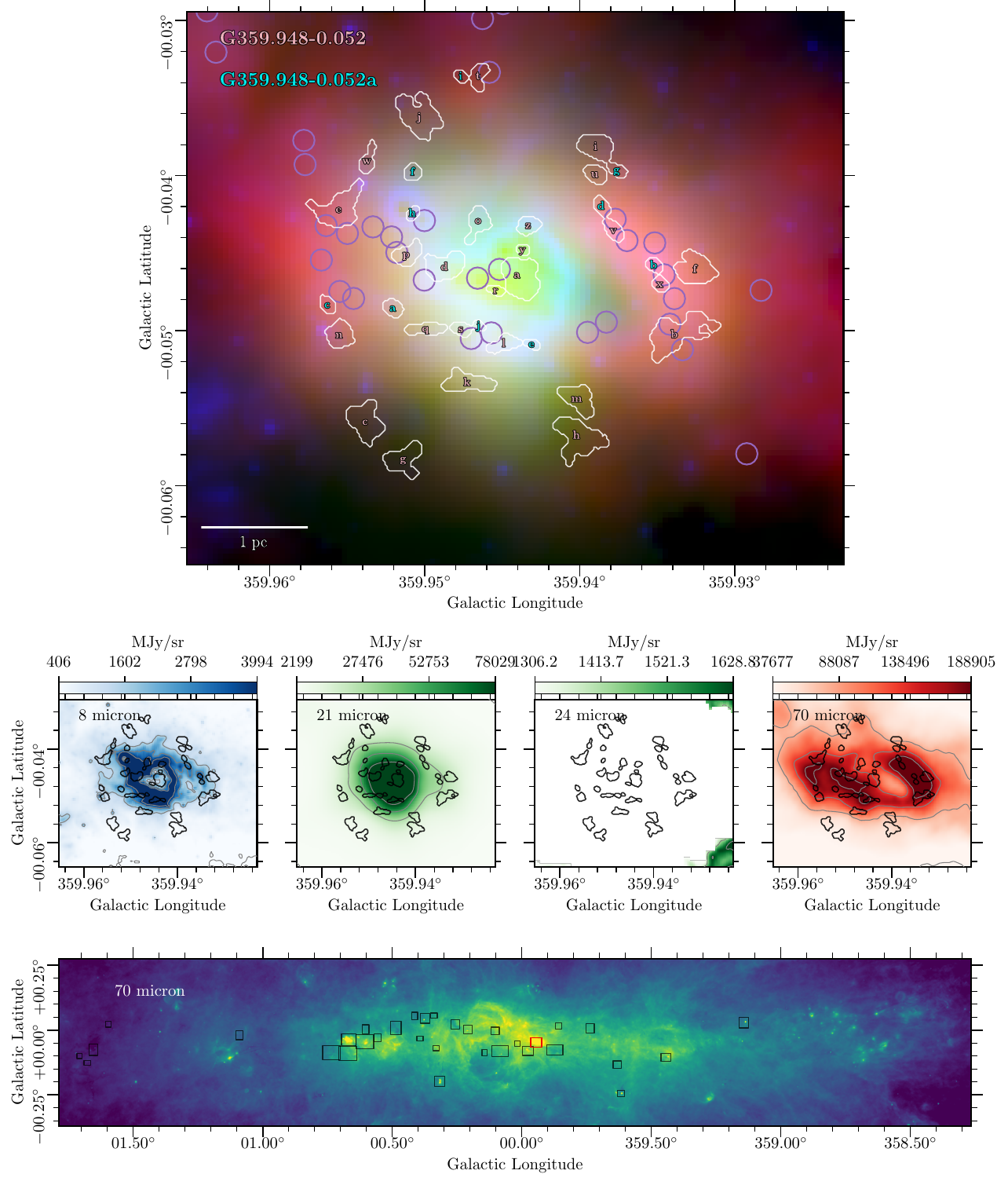}
\end{center}
\caption{Region 31, Cloud ID G359.948-0.052. This region contains the circumnuclear disk and Sgr A*, though leaves in this cloud were excluded from the analysis presented in this work due to significant contamination from non-continuum sources. The red colorscale shows the 70$\mu$m emission map from Herschel \citep{molinari_hi-gal_2010,molinari_source_2011}. The green colorscale shows the combined 21 and 24$\mu$m emission from MSX and Spitzer respectively \citep{benjamin_glimpse_2003}. The blue colorscale shows the 8$\mu$m map from Spitzer \citep{benjamin_glimpse_2003}. Each color is shown in logscale, scaled between a local 5\% and 95\% boundary value for each box. Overlaid on the composite three-color image are white contours representing the \textit{CMZoom} leaves, along with cyan circles representing YSO candidates from a compilation of those identified by \citealt{yusef-zadeh_star_2009}, \citealt{an_massive_2011}, and \citealt{immer_recent_2012}, purple circles representing the 70$\mu$m point sources cataloged by \citealt{molinari_hi-gal_2016}, and darker blue circles representing the point sources identified by \citealt{gutermuth_24_2015}. The radial size of these circles corresponds to the FWHM condition used to determine plausible association with nearby \textit{CMZoom} leaves.}
\label{fig:rgb_31}
\end{figure*}

\begin{figure*}
\begin{center}
\includegraphics[trim = 0mm 0mm 0mm 0mm,clip, width = .90 \textwidth]{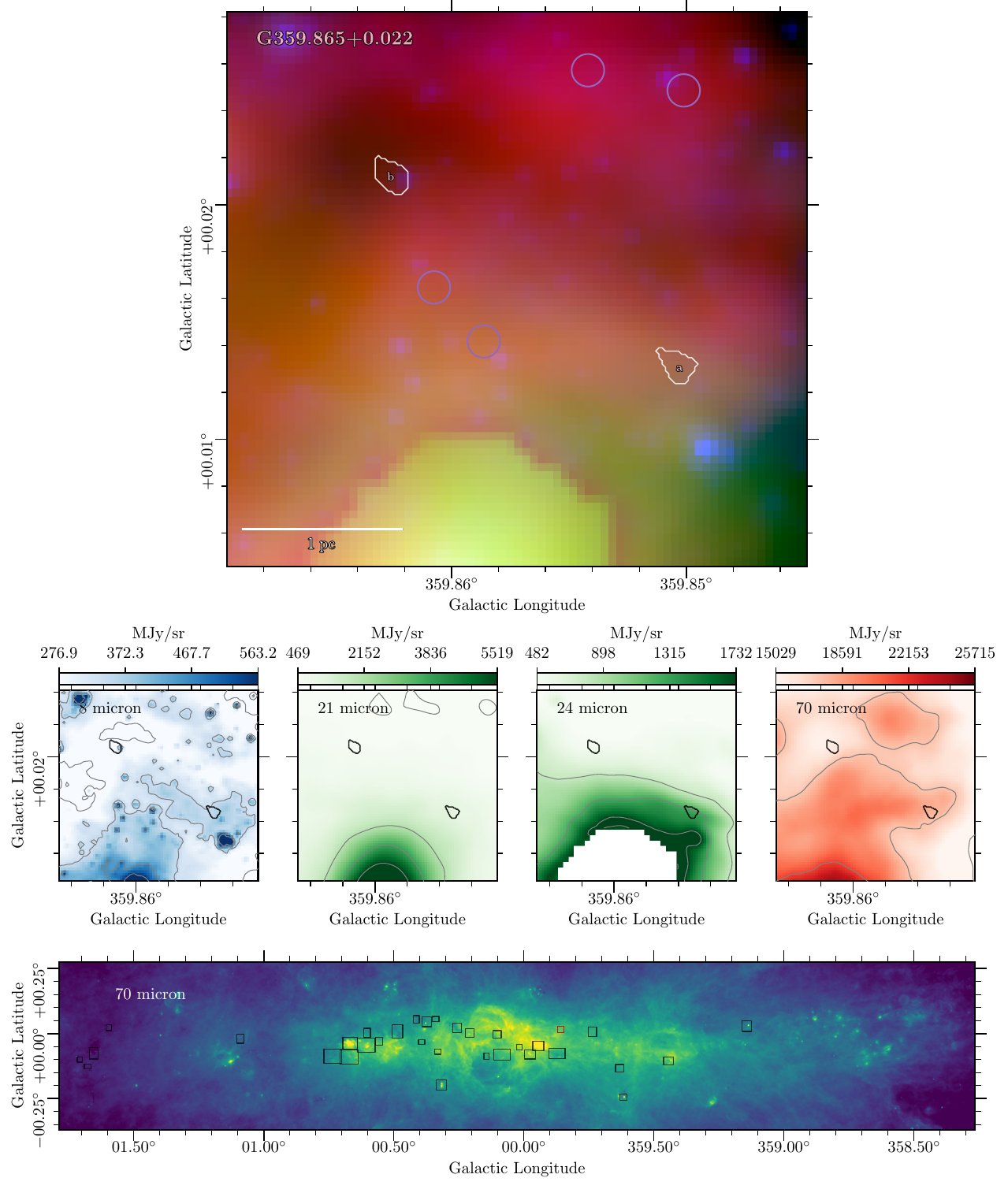}
\end{center}
\caption{Region 32, Cloud ID G359.865+0.022. The red colorscale shows the 70$\mu$m emission map from Herschel \citep{molinari_hi-gal_2010,molinari_source_2011}. The green colorscale shows the combined 21 and 24$\mu$m emission from MSX and Spitzer respectively \citep{benjamin_glimpse_2003}. The blue colorscale shows the 8$\mu$m map from Spitzer \citep{benjamin_glimpse_2003}. Each color is shown in logscale, scaled between a local 5\% and 95\% boundary value for each box. Overlaid on the composite three-color image are white contours representing the \textit{CMZoom} leaves, along with cyan circles representing YSO candidates from a compilation of those identified by \citealt{yusef-zadeh_star_2009}, \citealt{an_massive_2011}, and \citealt{immer_recent_2012}, purple circles representing the 70$\mu$m point sources cataloged by \citealt{molinari_hi-gal_2016}, and darker blue circles representing the point sources identified by \citealt{gutermuth_24_2015}. The radial size of these circles corresponds to the FWHM condition used to determine plausible association with nearby \textit{CMZoom} leaves.}
\label{fig:rgb_32}
\end{figure*}

\begin{figure*}
\begin{center}
\includegraphics[trim = 0mm 0mm 0mm 0mm,clip, width = .90 \textwidth]{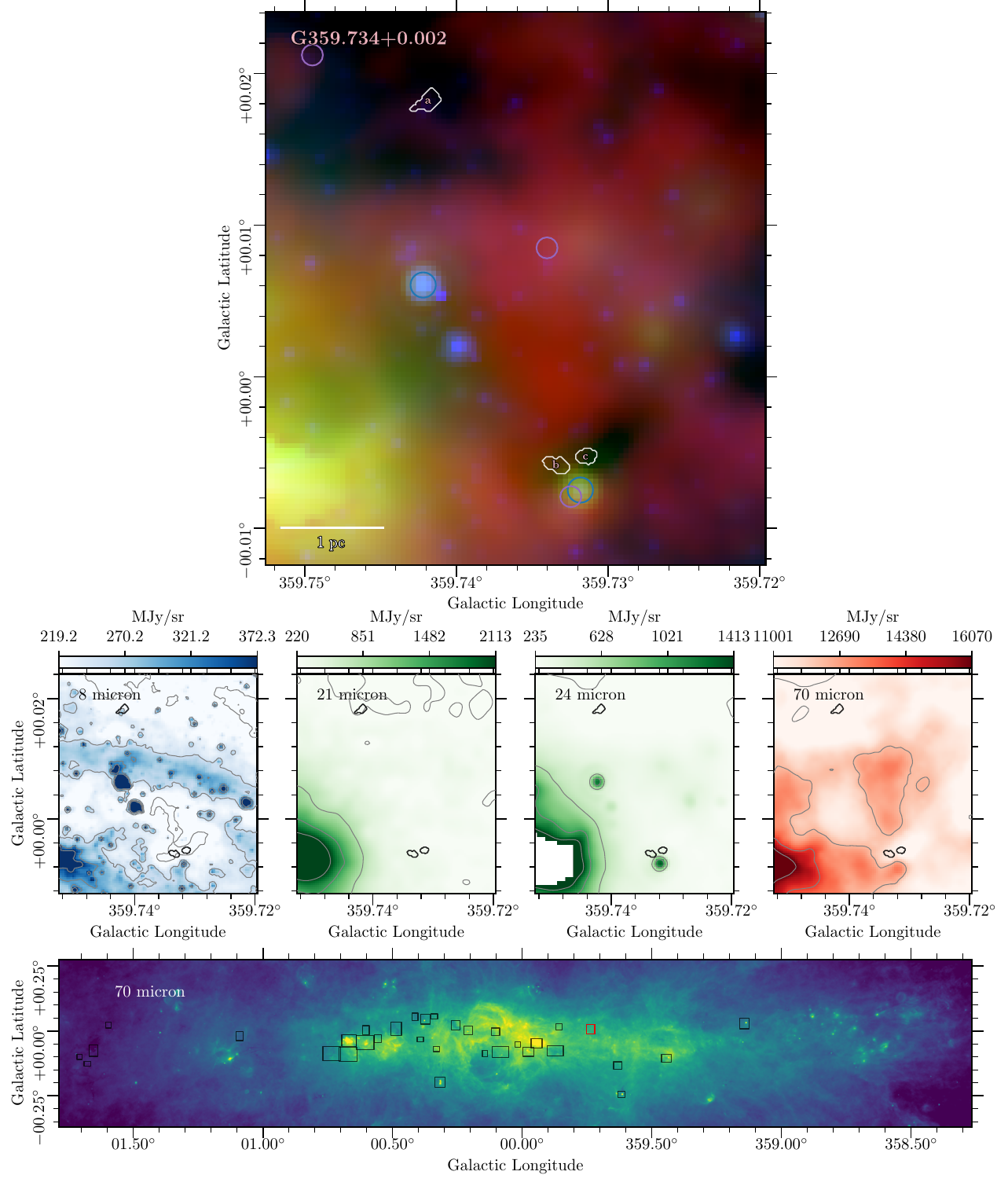}
\label{fig:rgb_33}
\end{center}
\caption{Region 33, Cloud ID G359.734+0.002. The red colorscale shows the 70$\mu$m emission map from Herschel \citep{molinari_hi-gal_2010,molinari_source_2011}. The green colorscale shows the combined 21 and 24$\mu$m emission from MSX and Spitzer respectively \citep{benjamin_glimpse_2003}. The blue colorscale shows the 8$\mu$m map from Spitzer \citep{benjamin_glimpse_2003}. Each color is shown in logscale, scaled between a local 5\% and 95\% boundary value for each box. Overlaid on the composite three-color image are white contours representing the \textit{CMZoom} leaves, along with cyan circles representing YSO candidates from a compilation of those identified by \citealt{yusef-zadeh_star_2009}, \citealt{an_massive_2011}, and \citealt{immer_recent_2012}, purple circles representing the 70$\mu$m point sources cataloged by \citealt{molinari_hi-gal_2016}, and darker blue circles representing the point sources identified by \citealt{gutermuth_24_2015}. The radial size of these circles corresponds to the FWHM condition used to determine plausible association with nearby \textit{CMZoom} leaves.}
\end{figure*}

\end{document}